\documentclass[12pt,a4paper]{article}%

\usepackage{adjustbox}
\usepackage{lmodern}
\usepackage[T1]{fontenc}
\usepackage[utf8]{inputenc}
\usepackage{parskip}
\usepackage{amssymb}
\usepackage{amsfonts}
\usepackage{xfrac}
\usepackage{amsmath}
\usepackage{mathtools}
\usepackage{enumitem}

\usepackage[UKenglish]{isodate}
\usepackage[page]{appendix} 
\usepackage{cases}
\usepackage{diagbox}
\usepackage{rotating}
\usepackage{bm}
\usepackage[margin=1.25in]{geometry}
\usepackage{setspace}
\usepackage[authoryear]{natbib}
\usepackage{setspace}
\usepackage{booktabs}
\usepackage[table,usenames,dvipsnames]{xcolor}

\usepackage{graphicx}
\usepackage{tikz}
\usepackage{pgfplots}
\usepackage[font=small,labelfont=sf]{caption}

\usepackage{array}
\newcolumntype{L}{>{$}l<{$}}

\usepackage[colorlinks=false,
            urlcolor=blue,
            citecolor=blue,
            linkcolor=blue,
            pdfauthor={Alexander and Davide},
            pdftitle={A title},
            bookmarks,
            bookmarksopen=true,
            bookmarksnumbered=true,
            pdfstartview={FitH},
            ]{hyperref}

\newcommand{\bk}{\color{black}}

\newcommand{\Tr}{{\textnormal{\textsf{T}}}}
\newcommand{\Ex}{\textnormal{\textsf{E}}}
\newcommand{\PP}{\mathbb{P}}

\newcommand{\var}{\textnormal{\textsf{var}}}
\newcommand{\cov}{\textnormal{\textsf{cov}}}

\newcommand{\st}{\star}

\newcommand{\eps}{\varepsilon}
\DeclareMathOperator*{\argmin}{\textnormal{\textsf{arg\,min}}}
\DeclareMathOperator*{\ssup}{\textnormal{\textsf{sup}}}
\DeclareMathOperator*{\iinf}{\textnormal{\textsf{inf}}}
\DeclareMathOperator*{\llim}{\textnormal{\textsf{lim}}}

\DeclareMathOperator*{\mmax}{\textnormal{\textsf{max}}}
\DeclareMathOperator*{\mmin}{\textnormal{\textsf{min}}}
\DeclareMathOperator*{\fra}{\mathfrak{a}}

\usepackage{accents}
\newcommand\ubar[1]{%
  \underaccent{\bar}{#1}}
\newtheorem{remark}{Remark}

\newtheorem{assumption}{Assumption}
\newtheorem{lemma}{Lemma}
\newtheorem{proposition}{Proposition}
\newtheorem{corollary}{Corollary}
 
\pgfplotsset{compat=newest}

\usepackage[table]{xcolor}
\definecolor{blk}{HTML}{FFF3BF} 

\definecolor{zcol}{HTML}{D6EAF8} 
\definecolor{brcol}{HTML}{D5F5E3} 

\usepgfplotslibrary{groupplots}
\pgfplotsset{compat=newest}
\usepgfplotslibrary{fillbetween}
\begin{document}

\title{\vspace*{-1.85em}Estimation and inference in models with multiple behavioural equilibria\thanks{We are grateful for the helpful suggestions and comments received at seminars at Erasmus University Rotterdam, Rijksuniversiteit Groningen, Collegio Sant’Anna (University of Pisa), Department of Econometrics and Statistics (Universität zu Köln), and the CFE 2025 Conference (Birkbeck, University of London). We would like to thank Mariia Artemova, Giulio Bottazzi, John Cochrane, Pietro Dindo, Christian Francq, Rutger-Jan Lange, Sven Otto, Dario Palumbo, Andreas Pick, Andrea Torsello and Dominik Wied for helpful comments and discussions. The second author acknowledges financial support under the National Recovery and
Resilience Plan (NRRP), Mission 4, Component 2, Investment 1.1, Call for tender No. 104 published
on 2.2.2022 by the Italian Ministry of University and Research (MUR), funded by the European Union
– NextGenerationEU– Project Title ”Market Learning and Robust Predictions (MALERP)” – CUP J53D23004190006. }}

\author{Alexander Mayer\footnote{{\it Corresponding author}, email: \href{mailto:a.s.mayer@ese.eur.nl}{a.s.mayer@ese.eur.nl}}\\
{\it Erasmus University \& Tinbergen Institute}\\
  Rotterdam, The Netherlands\\\vspace{-1.5em}
  \and
Davide Raggi\footnote{Email: \href{mailto:davide.raggi@unive.it}{davide.raggi@unive.it}}\\
{\it Università Ca' Foscari}\\
Venezia, Italy}
\date{\today} 
\maketitle 
\thispagestyle{empty}
\vspace*{-.6cm}
\begin{abstract} We develop estimation and inference methods for a  macroeconomic model with potentially multiple behavioural equilibria, where agents form expectations using a constant-gain learning rule. We discuss identification, estimation,  and inference for the structural parameters and propose uniform confidence bands for the equilibria. When equilibrium solutions are repeated, mixed convergence rates and non-standard limit distributions emerge. Monte Carlo simulations and an empirical application illustrate the finite-sample performance of our methods.

  \vspace{1em}\noindent{\bf Keywords:} adaptive learning, ergodicity, consistency, identification, asymptotic distribution.
\end{abstract}

\section{Introduction}

The rational expectations hypothesis provides the standard benchmark for incorporating expectations into macroeconomic models \citep{Muth1961,Lucas1972}. A large theoretical and empirical literature, however, documents systematic departures from full-information rationality, particularly in the formation of inflation expectations \citetext{see, among others, \citealp{evans:01}, \citealp{CoibionGorodnichenko2012}, \citealp{mn:2016}, and \citealp{bachmann:22}}. Several alternatives to rational expectations have consequently been proposed. While some approaches emphasize information frictions, another prominent route treats agents as statisticians who form expectations using simple, possibly misspecified, forecasting models and update their beliefs as new data become available \citep{marsar:89,evans:01}.

This is also the approach taken in this paper, which closely follows the specifications studied in \citet{HommesSorger1998}, \citet{Lansing2009}, \citet{hz14}, \citet{HommesMavromatisOzdenZhu2023}, and \cite{hajdini2026}. In particular, we consider a three-equation New Keynesian model 
in which agents update their beliefs by recursively estimating the parameters of first-order autoregressions, AR(1), that they (incorrectly) perceive to be the correct model of inflation and the output gap. 
As \citet{hz14} and \citet{HommesMavromatisOzdenZhu2023} show, this framework provides a parsimonious account of several features of observed inflation dynamics (see also \citealp{davide:25} and \citealp{kosh:forth}). 
Importantly, because agents remain even in the long run boundedly rational without correctly perceiving all stochastic aspects of the model, multiple so-called {\it behavioural equilibria} can coexist: self-confirming inflation paths that are fixed points of the perceived law of motion, rather than rational-expectations solutions of the full model. Despite the growing econometric literature on adaptive-learning models, little is known about frequentist estimation and inference in nonlinear models with multiple behavioural equilibria; see, e.g., \citet{chev:10}, \citet{adam:16}, \citet{chev:17}, \citet{chrismass:18,chrismass:19}, \citet{mayer:22,mayer:23}, \citet{christiano:24}, and \citet{mm:25}.

To close this gap, we {\it first} provide a statistical analysis of a multivariate New Keynesian model in which agents form expectations using constant-gain learning. Drawing on recent results for nonlinear time series models (\citealp{chot:19}), we derive primitive sufficient conditions for ($V$-uniform) ergodicity in terms of the joint spectral radius governing the dynamics and show that the effect of initializations vanishes geometrically fast. {\it Secondly}, building on identification arguments for observation-driven models \citep[e.g.][]{francq:04,blasques:22}, we establish identification of all parameters of the multivariate model under economically plausible sign restrictions: in particular, the learning parameter is recovered from the location of the complex poles of the agents' forecast, viewed as a function of the current innovation. We then propose a quasi maximum-likelihood estimator akin to \cite{linde2005}, for which strong consistency is established. {\it Thirdly}, we specialize to the single-equation New Keynesian Phillips curve (NKPC) with an exogenous marginal cost process considered by, among others, \citet{hz14}. For this stylized specification, nested by the general model, we derive a distributional theory, including asymptotic normality as well as inference procedures for the structural parameters and the potentially multiple behavioural equilibria. Here, several less-standard results emerge that connect to the broader econometric literature. For example, akin to \citet{hansen:1996,hansen:17}, we show how to conduct inference when certain parameters (viz., the learning gain) are not {\it jointly} identified. In this case, following \cite{saikkonen:95} and \cite{seo:11}, we can still verify $\sqrt n$-consistency of the remaining coefficients. Our analysis of the (multiple) equilibria is also conceptually related to \citet{kasy:15}, who develops a nonparametric procedure for inference on the number of roots of an unknown function. In contrast, we work in a parametric setup where the equilibria are the roots of a polynomial function that is known up to a finite dimensional parameter. This allows us to study the asymptotic distribution of equilibrium locations, its number, and uniform inference on the complete function. When inference concerns the (multiple) equilibria, convergence can be slow, akin to the second-order identification mechanism in \citet{dovonon:18}. {\it Finally}, we verify the theoretical findings in finite samples, using both simulated and real-world data. The empirical application to US data illustrates how our methods can be used to characterize uncertainty about equilibrium locations and learning dynamics in practice.

The remainder of this paper is organized as follows: Section \ref{sec:model} introduces the model. The time series properties of the resulting data generating process (dgp) are established in Section \ref{sec:prop}. Section \ref{sec:cons} discusses identification and point estimation based  on the quasi maximum likelihood principle. Section~\ref{sec:distnlse} derives the distributional theory in the special case of the single-equation NKPC. Finite sample evidence in the form of a Monte Carlo experiment and an estimation exercise using US data are presented in Sections~\ref{sec:mc} and  \ref{sec:emp}, respectively, before Section~\ref{sec:conc} concludes. All proofs are delegated to the appendix.

\section{Model}\label{sec:model}
Consider a small New Keynesian economy that jointly describes inflation ($\pi$), output gap ($y$), and nominal interest rates ($i$), via
\begin{align}
    \pi_t &= \delta\pi_{t+1}^e+\psi y_t+u_{\pi,t}, \label{eq:pc}\\
    y_t &= \mu y_{t+1}^e - \varphi(i_t-\pi_{t+1}^e)+u_{y,t}, \label{eq:is} \\ i_t &=  \rho_i i_{t-1} + (1-\rho_i)\phi^\Tr x_t + u_{i,t},
 \label{eq:taylor}
 \end{align}
where Eqs.~\eqref{eq:pc}-\eqref{eq:taylor} are, respectively, the Phillips curve, the IS curve, and a Taylor rule. Furthermore,  $x_{t}^e\coloneqq(\pi_{t}^e,y_{t}^e)^\Tr$ denote agents' not necessarily rational expectations of $x_{t} \coloneqq (\pi_{t},y_{t})^\Tr$. The relevant economic parameter is $\psi \in {\mathbb R}$, usually called the slope, that measures how inflation responds to changes in economic `{slack}' (see, e.g., \citealp{Woodford2003} for a textbook reference or \citealp{MavroeidisPlagborgMollerStock2014} for a more empirical treatment),  
whereas $\phi \coloneqq (\phi_\pi,\phi_y)^\Tr$ and $|\rho_i| < 1$ measures inertia in the monetary policy (see, e.g. \citealp[Section II.C]{clarida2000monetary}). The shocks $u_t\coloneqq (u_{\pi,t},u_{y,t},u_{i,t})^\Tr$
are generated by a VAR(1)
\begin{equation}
    u_t  = R u_{t-1}+\eps_{t},
\label{eq:var-shock}
\end{equation}
where $R$ is a $3\times3$ coefficient matrix and $\eps_{t}$ is a $3 \times 1$ vector of innovations. 

Following \cite{mns17} and \cite{gabaix20}, we deviate from the standard model with $\mu=1$  and allow \(\mu\in[0,1]\), so that expected future output is discounted. Since some of our later
results focus, for reasons of analytical tractability, on the
single-equation NKPC, this specification also serves a purpose specific to
this paper. In particular, at $\mu=\varphi=\rho_i=0$ we recover, after allowing for an intercept for the $y$-process and imposing restrictions on the shock dynamics~\eqref{eq:var-shock},
\begin{equation}\label{eq:seNKPC} 
\pi_t =  \delta\pi_{t+1}^e+\psi y_t+u_{t}, \qquad y_t = a+\rho y_{t-1}+\eps_{t},  
\end{equation}
which is the model studied by, among others, \citet{Lansing2009} and
\citet[Section~4]{hz14}: there, $y_t$ is a marginal cost driver (e.g. an output gap measure or real unit labour costs) which is assumed exogenous
to the system, while the innovations $v_t \coloneqq (u_t,\eps_t)^\Tr$ are mean-zero with $\var[v_t] = {\sf diag}(\sigma_u^2,\sigma_\eps^2)$.

Under full information rational expectations (RE), agents have complete knowledge of the model structure and set $x_{t+1}^e = \Ex[x_{t+1} \mid \mathcal{F}_t]$, where $\mathcal{F}_t \coloneqq \sigma(\{\eps_j: j \leq t\})$. For example, in the scalar NKPC, it can easily be shown that the RE equilibrium is  $\pi_t= \frac{a\delta \psi}{(1-\delta\rho)(1-\delta)}+ \frac{\psi}{(1-\delta\rho)}y_t + u_{t}.$ The RE paradigm, although providing an important benchmark, has been challenged over the years. One popular alternative, proposed by the macroeconomic learning literature, is to assume that agents themselves act like statisticians when forming their inflation expectations. In this literature, agents usually know the functional form of the RE equilibrium but not its parameters, which they estimate, hence the term \emph{bounded rationality} (see, e.g., \citealp{evans:01} and \citealp{evan:20} for overviews).

Here, we deviate from complete rationality a bit further: following \citet{hz14} and \cite{HommesMavromatisOzdenZhu2023},  we assume that agents even lack knowledge of the functional form. Instead, expectations of tomorrow's inflation and marginal costs are formed according to a perceived law of motion (PLM) that incorrectly stipulates componentwise AR(1) dynamics disregarding the cross-equation structure of the system; i.e. for $j\in\{y,\pi\}$,
\begin{equation}
    x_{j,t}=\alpha_j+\beta_j(x_{j,t-1}-\alpha_j)+e_{j,t},
\label{eq:PLM}
\end{equation}
with $e_{j,t}$ an error, while $(\alpha_j,\beta_j)$ are parameters unknown to the agent, who has to estimate them.  

 Although an AR(1) forecasting rule may appear restrictive \citep{crumpetal2023,
carvalhoetal2023, eusepi2025short}, a growing body of evidence supports such simple heuristics (that is, expectation formation under model misspecification as in \citealp[e.g.,][Ch.~13]{evans:01}) as an empirically relevant description of inflation expectations: see
\citet{markiewicz2014adaptive} and \citet{mn:2016} for evidence from aggregate survey forecasts and micro data, respectively; \citet{adam2007experimental} and \citet{hommes2021behavioral} for experimental evidence; \citet{hajdini2026} for additional evidence on SAC-type updating. Section~\ref{sec:emp} further shows that the model-implied forecasts closely track the consensus forecasts reported in the Survey of Professional Forecasters (SPF).

 Two questions arise: (1) \emph{How do agents forecast?} and (2) \emph{What do agents actually forecast?} To address the first question, assume that expectations are formed at the end of period $t-1$ using a mean-squared-error-optimal two-period-ahead rule based only on information available at $t-1$. 
 This mechanism, where forecasts are based on lagged rather than current information, is standard in the learning literature (see, e.g., \citet[Section 4.2]{chrismass:18} and the references therein).  This yields, 
\begin{equation}
    x^e_{t+1} = \alpha + B(x_{t-1}-\alpha), \quad B\coloneqq {\sf diag}(\beta_\pi^2, \beta_y^2),
\quad  \alpha \coloneqq (\alpha_\pi,\alpha_y)^\Tr,
\label{eq:iALM}
\end{equation} \bk
where squaring the coefficients $\beta \coloneqq (\beta_\pi,\beta_y)^\Tr$ reflects the two-steps-ahead timing. 

The implied actual law of motion (ALM) is then given by Eqs.~\eqref{eq:pc}-\eqref{eq:var-shock} and \eqref{eq:iALM}, which would obtain {\it if} agents knew the parameters $(\alpha,\beta)$ governing their forecasting rule in Eq.~\eqref{eq:PLM} and which, if $\Delta \coloneqq 1+\varphi(1-\rho_i)(\phi_y+\psi\phi_\pi) \neq 0$, can be summarized as
\begin{align}\label{eq:ALMmat}    
x_t = H(Bx_{t-1}+(I_2-B)\alpha)+g i_{t-1} + M^{-1}Lu_t, \quad L \coloneqq \begin{bmatrix}
    1 & 0 & 0 \\ 0 & 1& -\varphi
\end{bmatrix}, 
\end{align}
where $g \coloneqq -\varphi\rho_i (\psi,1)^\Tr/\Delta$ and    
\[
    H
    \coloneqq
    \frac{1}{\Delta}
    \begin{bmatrix}
        \delta(1+\varphi(1-\rho_i)\phi_y)+\psi\varphi
        &
        \psi\mu
        \\
        \varphi-\varphi(1-\rho_i)\phi_\pi\delta
        &
        \mu
    \end{bmatrix}, \quad M \coloneqq \begin{bmatrix} 1&-\psi \\ \varphi(1-\rho_i)\phi_\pi & 1+  \varphi(1-\rho_i)\phi_y\end{bmatrix}.
\]
Note that ${\sf det}(M) = \Delta$, where $\Delta \geq 1$ on the economically relevant parameter space where $\varphi \geq 0$, $\phi_y \geq 0$, $\phi_\pi > 0$, and $\psi \geq 0$.

To answer the second question, \citet{hz14} and \citet{HommesMavromatisOzdenZhu2023} introduce the notion of {\it behavioural learning equilibria} (BLE). With agents misspecifying the functional form, the idea is to discipline behaviour from becoming overly irrational by imposing that ($a$) the unconditional mean and ($b$) the first-order autocorrelation of $x_t = (\pi_t,y_t)^\Tr$ are correctly perceived. These two restrictions pin down the equilibrium values of $(\alpha,\beta)$. To illustrate, consider the single-equation NKPC of Eq.~\eqref{eq:seNKPC}: Fix agents' beliefs $\ell \coloneqq (\alpha,\omega,r)^\Tr$ about mean ($\alpha$), first auto-covariance ($\omega$) and variance ($r > 0$) so that $\beta = \omega/r$, and let $\pi_t$ denote the stationary process generated by the ALM  $\pi_t   =    \delta(\alpha+\beta^2(\pi_{t-1}-\alpha))    +\psi y_t+u_{t}$. A BLE $\ell^\st$ can then be defined as a solution to the $3 \times 1$ vector of unconditional moment conditions $\frak{h}(\ell^\st) = 0$, where $\frak{h}(\ell) = (\frak{h}^{(\alpha)}(\ell),\frak{h}^{(\omega)}(\ell), \frak{h}^{(r)}(\ell))^\Tr \coloneqq \Ex[\frak{H}(\ell, X_t)]$, with 
\begin{align}\label{eq:psimap}
\frak{H}(\ell,X_t)
    \coloneqq
    \begin{bmatrix}
        \pi_t-\alpha\\
        (\pi_t-\alpha)(\pi_{t-1}-\alpha)-\omega\\
        (\pi_t-\alpha)^2-r
    \end{bmatrix}, \quad X_t \coloneqq (\pi_t,\pi_{t-1})^\Tr.
\end{align}
Solving the moment conditions yields for the mean $
\frak{h}^{(\alpha)}(\ell^\st) = 0 \Leftrightarrow \alpha^\st= \frac{\psi a}{(1-\rho)(1-\delta)},$ 
whereas for the first-order autocorrelation define $\frak{h}^{(\beta)}(\ell) \coloneqq \frac1{r}(\frak{h}^{(\omega)}(\ell)-\beta \frak{h}^{(r)}(\ell))$, which, under $\frak{h}(\ell) = 0$, equals  $F(\beta;\lambda) - \beta$, so that $\beta^\st$ solves the following quartic equation $ \frak{h}^{(\beta)}(\ell^\st) = 0 \Leftrightarrow \beta^\st = F(\beta^\st;\lambda)$, with  
\begin{equation}\label{eq:consAC1}
F(\beta;\lambda) \coloneqq  \delta\beta^2+\frac{\psi^2\rho(1-\delta^2\beta^4)}{\psi^2(\delta\beta^2\rho+1)+(1-\rho^2)(1-\delta\beta^2\rho)(\sigma_u/\sigma_\eps)^2},
\end{equation} 
where $\lambda \coloneqq (\delta,\psi,\rho,\sigma_u^2,\sigma_\eps^2)^\Tr$ (see \citealp{hz14}).  A BLE therefore requires agents' perceived mean and first-order autocorrelation to coincide with those generated by the ALM. The equilibrium mean equals its RE counterpart, whereas persistence generally differs. Eq. (\ref{eq:consAC1}) may admit multiple solutions, corresponding to distinct persistence regimes. Following \citet{blanchard:16} and \citet{JorgensenLansing2025}, low-$\beta$ equilibria can be associated with more firmly anchored expectations, whereas high-$\beta$ imply more persistent inflation dynamics.\footnote{Note also that the low-$\beta$ 
scenario aligns with the original specification of the Phillips curve, namely, the one defined
in terms of inflation levels and the output gap, whereas high-$\beta$ is more consistent with an accelerationist formulation of the curve.}

Agents typically do not know the parameters $(\alpha,\beta)$ of their forecasting model Eq.~\eqref{eq:PLM}. Instead, and in line with the adaptive learning literature (\citealp{evans:01}), we assume agents employ a recursive updating scheme to estimate them; formally, a stochastic approximation algorithm (\citealp{ben:90}). In particular, agents are presumed to use a constant gain version of the sample autocorrelation learning (SAC) algorithm proposed by \cite{hz14} (see also \citealp{HommesMavromatisOzdenZhu2023}) to estimate the subjective parameters $(\alpha_j,\beta_j)$, $j \in \{\pi,y\}$:
\begin{equation}
    \alpha_{j,t}=(1-\gamma)\alpha_{j,t-1}+\gamma x_{j,t},\;
    \beta_{j,t}=
    \frac{\sum_{i=1}^{t-1}(1-\gamma)^{t-1-i}(x_{j,i+1}-\alpha_{j,i})(x_{j,i}-\alpha_{j,i})}
         {\sum_{i=1}^{t}(1-\gamma)^{t-i}(x_{j,i}-\alpha_{j,i-1})^2},
\label{eq:learning}
\end{equation}
for some so-called `gain' or `learning' parameter $\gamma \in (0,1)$ and initial values specified below.\footnote{As discussed in Remark~\ref{rem:initial}, Eq.~\eqref{eq:learning} corresponds to the initialization $\alpha_{j,0}=x_{j,0}$ and $r_{j,0}=\beta_{j,0}=0$, which is also used in our
empirical implementation.} Their expectations are consequently given by
\begin{align}\label{eq:SACexp}
x^e_{j,t+1}=\alpha_{j,t-1}+\beta_{j,t-1}^2(x_{j,t-1}-\alpha_{j,t-1}), \quad j \in \{\pi,y\}.
\end{align}
A larger gain attaches more weight to recent data. With constant gain, the recursive estimates remain time varying and converge in distribution rather than in probability to fixed constants (see also Remark~\ref{rem:smallgam}). Therefore it allows for persistent deviations from
rationality and is often used in empirical applications for its ability in tracking nonstationary  environments \citetext{see, e.g. \citealp{milani:07} or \citealp{chev:10}} including \cite{HommesMavromatisOzdenZhu2023}.

To sum up, under learning the resulting data generating process is
given by Eqs.~\eqref{eq:pc}, \eqref{eq:is}, \eqref{eq:taylor}, and
\eqref{eq:var-shock}, together with agents' expectations formed according
to Eqs.~\eqref{eq:learning} and~\eqref{eq:SACexp}. The unknown parameters
are $\theta\coloneqq(\gamma,\vartheta^\Tr)^\Tr$, $\vartheta \coloneqq (\mu,\varphi,\delta,\psi,\phi_\pi,\phi_y,
\rho_i)^\Tr$, together with the shock parameters governing
\eqref{eq:var-shock}, which we aim to estimate; these parameters in turn
determine the potentially multiple equilibria $(\alpha^\st,\beta^\st)$. In the leading case of the single-equation NKPC~\eqref{eq:seNKPC}, the dgp reduces to
\begin{equation}\label{eq:ALM}
\pi_t=\delta[\alpha_{t-1}+\beta_{t-1}^2(\pi_{t-1}-\alpha_{t-1})]
+\psi y_t+u_t,
\end{equation}
which is governed by the learning gain $\gamma$,
the NKPC coefficients $\vartheta=(\delta,\psi)^\Tr$, collected in
$\theta\coloneqq(\gamma,\vartheta^\Tr)^\Tr$, the parameters $(a,\rho)$ of the
exogenous AR(1) process for $y_t$, and the innovation variances
$(\sigma_u^2,\sigma_\epsilon^2)$. In view of Eq.~\eqref{eq:consAC1}, for the equilibrium analysis, estimates of  $\lambda = (\vartheta^\Tr,\rho,\sigma_u^2,\sigma_\epsilon^2)^\Tr$ are needed.
Crucially, multiplicity below refers to multiple, possibly repeated,
equilibrium roots $\beta^\st = F(\beta^\st;\lambda)$ implied by $\lambda$, rather than to multiple
structural parameter values that generate the same dgp: Section~4 establishes identification and consistency of the structural parameters, while
Section~5.1 studies the number, multiplicity, and sampling behaviour of the
implied equilibria.

The
estimation task is nontrivial because the dgp, which can be viewed as an `{\it observation driven}' model (\citealp{cox:81}), is already in this leading
case highly nonlinear. Nevertheless, as we show in Section~\ref{sec:prop}, the general model allows a Markovian representation and its probabilistic properties can be studied using  tools of Markov chain theory.

\section{Probabilistic properties of the dgp}\label{sec:prop}

To that end, we cast the model in nonlinear state-space form $S_t=\mathcal G(S_{t-1},\eps_t)$, for a continuous map
$\mathcal G$ given in Appendix~\ref{sec:A1}, and demonstrate that the chain is $V$-uniformly ergodic in the sense of \citet[Ch.~16]{meyn:12}. This implies that the law of the chain converges geometrically fast to a unique invariant distribution from any admissible initial state. Moreover, under stationary initialization, the process is strictly stationary and absolutely regular (\citealp{liebscher2005towards}). Together with suitable moment conditions, this provides the LLN and CLT results for sample averages and smooth functionals needed in the subsequent analysis (for discussions of these concepts in the context of nonlinear time series models see, e.g., \citealp{carrasco2002mixing} or \citealp{kristensen2005geometric}).
Ergodicity of the dgp is also important for establishing that the
effect of the initialization of the filtered recursions, based on
the observed path of $x_t$, vanishes fast enough (commonly referred to
as \emph{invertibility}; see, e.g. \citealp{blasques:22}) which is crucial for estimation and inference.
 Once standard Markov-chain regularity conditions are established (see Lemma~\ref{lem:markovprop} below), a convenient characterization of $V$-uniform ergodicity (\citealp[Theorem~15.0.1 and Theorem~16.0.1]{meyn:12}) is the existence of a function $V\geq1$ and constants $b<\infty$ and $\lambda<1$ such that the so-called Foster-Lyapunov drift condition $\Ex[V(S_t)\mid S_{t-1}=S]\leq b+\lambda V(S)$ holds for every admissible state $S$. Below we provide a discussion of primitive sufficient and necessary conditions, based on the joint spectral radius governing the
dynamics of the chain.

To set the stage, note that we can represent $\beta_{j,t}$ also recursively i.e.
\[
\beta_{j,t} = \beta_{j,t-1}+ \frac{\gamma}{r_{j,t}} ((x_{j,t}-\alpha_{j,t-1})(x_{j,t-1}-\alpha_{j,t-1})-\beta_{j,t-1}(x_{j,t}-\alpha_{j,t-1})^2),
\]
and $r_{j,t} =  r_{j,t-1}+\gamma ((x_{j,t}-\alpha_{j,t-1})^2-r_{j,t-1})$, where $j \in \{\pi,y\}$. Define the $8\times1$ vector $s_t\coloneqq(x_t^\Tr,\alpha_t^\Tr,\beta_t^\Tr,r_t^\Tr)^\Tr$ and $S_t\coloneqq(s_t^\Tr,u_t^\Tr,i_t)^\Tr\in\mathbb R^{12}$, taking values in the admissible space $\mathsf S\coloneqq\mathbb R^2\times\mathbb R^2\times[-1,1]^2\times\mathbb R_{++}^2$ for $s_t$ and $\mathbb R^4$ for $(u_t^\Tr,i_t)^\Tr$. Although $[-1,1]$ is the natural range of a sample autocorrelation, an arbitrary numerical initialization with $\beta_{j,0}\in[-1,1]$ might result in $|\beta_{j,t}|>1$ for some $t \geq 1$.\footnote{To illustrate, suppress $j$, and let \(\gamma=1-\gamma \eqqcolon q =1/2\). If $x_0=3,$ $\alpha_0=0,$ $r_0=1,$ $\beta_0=0,$ then $s_0 = {\sf S}$ is admissible. However, choosing \(x_1=1\) gives
$\nu_1\coloneqq x_1-\alpha_0=1,$ $r_1=qr_0+\gamma\nu_1^2=1,$
and, since \(\omega_0=\beta_0r_0=0\), $\omega_1
=
q\omega_0+\gamma\nu_1(x_0-\alpha_0)
=
3/2.$
Consequently, \(\beta_1=\omega_1/r_1=3/2>1\). More generally, the reason is that, under arbitrary initialization $s_0 \in {\sf S}$, \(r_t\) and
\(\omega_t\), although interpreted as a recursive variance and
autocovariance, need not initially satisfy the Cauchy-Schwarz inequality because, due to initialization, they do not fully share the same history. This is analogous to multivariate GARCH models, where
separately updating variances and covariances does not automatically
produce a valid covariance matrix. In the present setting, the condition \(\mathfrak s_j(s)>0\) mimics the restrictions resulting from the Cauchy-Schwarz inequality if recursions were based on a shared data set.} Hence {\sf S} is not, by itself, forward invariant (i.e. $s_0 \in {\sf S}$ does not guarantee $s_t \in {\sf S}$ for $t \geq 1$). The following open set isolates the forward-invariant part ${\sf S}$: $\mathsf S_0\coloneqq \{s \in {\sf S}: \mathfrak{s}_j(s) > 0, \, j \in \{y,\pi\}\},$
\begin{equation}
\mathfrak{s}_j(s) \coloneqq r_j-\frac{|\omega_j|}{\sqrt{1-\gamma}}-\frac{\gamma}{2(1-\gamma)^2}(x_j-\alpha_j)^2,  \qquad \omega_j \coloneqq \beta_jr_j, \quad s \in {\sf S}.
\label{eq:S0}
\end{equation}
In particular, $s \in {\sf S}_0$ implies $\beta_{j}^2<1-\gamma$. If $\alpha_0=x_0$, then $s_0\in{\sf S}_0$ reduces to the single initial restriction $B_0 = {\sf diag}(\beta_{\pi,0}^2,\beta_{y,0}^2) \in [0,1-\gamma]^2.$ The following lemma shows that \({\sf S}_0\) is not a restriction on the dynamics because any deviation from \({\sf S}_0\) caused by initialization is only temporary: in particular, the probability of not having entered ${\sf S}_0$ by time $t$ decays at least geometrically, with decay $\sqrt{1-\gamma}$ per period, and the recursion enters \({\sf S}_0\) almost surely in finite time.

\begin{lemma}\label{lem:invariant}
Suppose $\Delta \neq 0$ and $\gamma \in (0,1)$. \emph{($i$)} If $s_0\in\mathsf S_0$, then $s_t\in\mathsf S_0$ for all $t\ge0$, for every shock path; in particular $\beta_{j,t}^2\le1-\gamma$ for all $t$. \emph{($ii$)} If, in addition, conditionally on ${\cal F}_{t-1}$ the components of $\eps_t$ are  independent and have conditional densities uniformly bounded by $\bar  f < \infty$, then for any admissible $s_0\in\mathsf S$, there exists an almost surely finite random time $N(\omega)$ such that $s_t\in\mathsf S_0$ for all $t\ge N(\omega)$. Moreover, 
$$
\mathbb{P}\{N > t\} \leq  \sqrt{\frac{(1-\gamma)^{t}}{\gamma}} 4\sqrt{6}\bar f\frac{ \mmax\{-\mathfrak s_\pi(s_0),0\}^{\frac1{2}}+\mmax\{-\mathfrak s_y(s_0),0\}^{\frac1{2}}}{\lambda_{\min}(M^{-1}LL^\Tr(M^{-1})^\Tr)^{\frac1{2}}} .
$$
\end{lemma}

\begin{remark}\label{rem:initial}
Similar to \textup{\cite{hz14}}, the initialization used below in the finite sample exercise is $    \alpha_0=x_0=\fra \in {\mathbb R}^2,$ $r_{j,0}= \beta_{j,0}=0,$   $j\in\{y,\pi\}.$
This initial value does not belong to ${\sf S}$, because ${\sf S}$ requires $r_j>0$. However, after one update, $r_{j,1}=\gamma(x_{j,1}-\alpha_{j,0})^2,$ which has the intuitive consequence $
    \beta_{j,1}=0$. Under the condition of Lemma~\textup{\ref{lem:invariant}  ($ii$)}, $r_{j,1}>0$ almost surely. Hence $s_1\in{\sf S}_0$ almost surely, and by Lemma~\textup{\ref{lem:invariant}  ($i$)}, $s_t\in{\sf S}_0$ for all $t\ge1$ almost surely. In particular, it is admissible for the initialization-forgetting result below, even though it is not an element of the natural state space ${\sf S}$ at time zero.
\end{remark}

The chain $S_t = \mathcal{G}(S_{t-1},\eps_t)$ belongs to the class of nonlinear state space models defined in \citet[p. 33]{meyn:12}. Moreover, under Assumption \ref{ass:density} below, $S_t$ is weak Feller (\citealp[Prop. 6.1.2]{meyn:12}). We cannot, however, directly rely on off-the-shelf results typically used in the nonlinear time series literature to verify geometric ergodicity of $S_t$. For example, the otherwise very general results in \citet[Section~A.1]{tong1990non} do not apply because, among other reasons, the underlying skeleton difference equation fails to be globally Lipschitz. For a similar reason, we cannot invoke \citet[Thm 2.8]{strau:06} or the ergodicity result from \citet[Lem 3]{ds:93} used, for instance, by  \citet[Appendix D]{adam:16}. Instead, we derive the stochastic properties of the model from first principles, following the exposition in \citet{chot:19}. In doing so, we impose the following assumption:
\renewcommand{\theassumption}{A}

\begin{assumption}\label{ass:density}
The innovations $\eps_t \coloneqq (\eps_{\pi,t},\eps_{y,t},\eps_{i,t})^\Tr$ are independently and identically distributed \textnormal{({\sf IID})}, independent across components, and have finite second moments. For each \(j\in\{\pi,y,i\}\), the component
\(\eps_{j,t}\) admits a bounded, lower semicontinuous density
 $\forall w\in \mathbb R:\; f_j(w) > 0.$
\end{assumption}

While some mild deviations from the {\sf IID} assumption might in principle be possible, existence of lower semicontinous error-densities with full support is essential for both our ergodicity proof as well as parameter identification. Under Assumption~\ref{ass:density}, we can show, by applying \citet[Theorem~4.4]{chot:19}, that $S_t$ explores all meaningful regions of the state space ($\varphi$-irreducibility) and avoids periodic behaviour (aperiodicity).

\begin{lemma}\label{lem:markovprop}
Suppose Assumption~\textup{\ref{ass:density}} holds, $\gamma \in (0,1)$, $\Delta \neq 0$, and let $\mathsf X_0 \coloneqq {\sf S_0} \times \mathbb R^4.$ If \(S_0 \in \mathsf X_0\), then \((S_t)_{t \geq 0}\) is a \(\varphi\)-irreducible and aperiodic \(T\)-chain on \(\mathsf X_0\).
\end{lemma}

To obtain ergodicity we also need to preclude divergence to infinity. 
This can be ensured by the drift condition referenced earlier, for which we require the following parameter constraints. 

\renewcommand{\theassumption}{B1}
\begin{assumption}\label{ass:para}
$\Delta \neq 0$, with $\Delta$ defined in Eq.~\eqref{eq:ALMmat}, $\rho(R)<1$, and $\mu\in[0,1]$, $\varphi\ge0$, $\psi\in {\mathbb R}$, $|\delta|<1$, $\phi_\pi>0$, $\phi_y\ge0$, $\gamma\in(0,1)$, $|\rho_i|<1$, and
$
   \Upsilon \coloneqq (1-\mu)(1-\delta)+\varphi(\phi_y(1-\delta)+\psi(\phi_\pi-1))>0. $
\end{assumption}

The condition $\Upsilon>0$ is a generalized Taylor principle that reduces at $\mu=1$, $\varphi > 0$, and $\psi \geq 0$ to the textbook case. In view of Eq.~\eqref{eq:ALMmat}, $\Delta \neq 0$ allows the model to be uniquely solved for $x_t$. Here, $\psi \geq 0$ implies directly $\Delta \geq 1$. Discounting relaxes the parameter restrictions implied by $\Upsilon > 0$, and smoothing does not tighten it. In view of Eq.~\eqref{eq:ALMmat}, we note that the coefficient matrix $H$ becomes upper triangular at $\varphi = 0$, so that $\Upsilon>0$ requires $\mu \in [0,1)$ if $\varphi = 0$ to avoid unit-root type behaviour. The gain $\gamma\in(0,1)$ is left unrestricted, covering the range of estimates typically found in empirical work \citep[cf.][Fig.~3]{berardi2017empirical}. 

When $\varphi=0$ (i.e. the policy rate does not feed back into output through the IS curve), Assumption~\ref{ass:para} is sufficient for the stability arguments used below. When $\varphi>0$, however, past interest rates affect current output through the IS curve, while current output and inflation feed back into the future policy rate through the Taylor rule. The condition $\Upsilon > 0$ is then no longer enough to control the dynamic feedback generated by recursively updated beliefs.

To see this, let $ z_t \coloneqq (x_t^\Tr,\alpha_t^\Tr,i_t)^\Tr$ and consider $z_t=A(B_{t-1})z_{t-1}+D u_t,$ where $D$ is a $5 \times 3$ matrix
\begin{align}
    \label{eq:ABmat}
   A(B)=
    \begin{bmatrix}
        HB & H(I-B) & g\\
        \gamma HB& (1-\gamma)I+\gamma H(I-B) & \gamma g\\
        (1-\rho_i)\phi^\Tr HB &
        (1-\rho_i)\phi^\Tr H(I-B) &
        \rho_i+(1-\rho_i)\phi^\Tr g
    \end{bmatrix},
\end{align}
see Eq.~\eqref{eq:ALMmat} for the definitions of $g$ and $H$. Thus the model admits a representation similar to stochastic recurrence equations (SRE) (see, e.g. \citealp{strau:06}). Because $B_t$ is a time-varying coefficient matrix, the relevant stability is whether the products $A(B_{t-1})A(B_{t-2})\cdots A(B_0)$ remain uniformly contracting along all admissible paths $B_t$ (see, e.g. \citealp{straumann2005estimation}). \emph{If} the process \(\beta_t\) were already known to be stationary and ergodic, a natural primitive stability condition would be a stochastic
Lyapunov-exponent condition 
$
    \lim_{m\rightarrow \infty}
    \frac1m
    \Ex[\operatorname{\sf log}
    \|
        A(B_{m-1})\cdots A(B_0)
    \|] <0 ,
$
under the stationary law of \(\beta_t\); see \citet{bougerol:93} and
\citet{strau:06} for related SRE arguments. 
In the present model, however, stationarity and ergodicity of \(B_t\) are themselves part of what has to
be established. Instead, on $\mathsf S_0$, and eventually almost surely from every admissible
initialization under Lemma~\ref{lem:invariant} ($ii$), the SAC recursion
satisfies $B_t\in[0,1-\gamma]^2.$ We therefore assume that the joint spectral radius (JSR) (\citealp{jungers2009joint}) of the set of matrices ${\cal A} \coloneqq \{A(B): B = {\sf diag}(a,b),\ (a,b) \in [0,1-\gamma]^2\}$ is strictly bounded below one.

\renewcommand{\theassumption}{B2}
\begin{assumption}\label{ass:paraJSR}
Assume that $\rho_{\sf JSR}({\cal A}) \coloneqq \iinf\limits_{\|\cdot\|}\ssup\limits_{A \in {\cal A}} \|A\| < 1$, where the infimum is over all matrix induced  norms $\|\cdot\|$.
\end{assumption}

 Because for any $B \in [0,1-\gamma]^2$, $A(B)$ is a convex combination of the four boundary cases
    $A_1=A(0,0),$  $A_2=A(1-\gamma,0),$ $A_3=A(0,1-\gamma),$ $A_4=A(1-\gamma,1-\gamma),$ we have (\citealp[Proposition 1.8]{jungers2009joint}) $\rho_{\sf JSR}({\cal A}) = \rho_{\sf JSR}(A_1,\dots,A_4)$, which, in turn, is equivalent to $\lim_{m\to\infty}
    \mmax_{c\in\{1,\ldots,4\}^m}
    \frac1m
    \operatorname{\sf log}
    \|
        A_{c_m}\cdots A_{c_1}
    \|
    <0,$
for any matrix induced norm $\|\cdot\|$. Thus Assumption~\ref{ass:paraJSR} can be viewed as a \emph{worst-case} counterpart of the expected negative Lyapunov-exponent conditions commonly used for SREs. 

Analogous conditions have been used to establish stationarity and ergodicity of nonlinear time series processes; see, e.g.
\cite{Kheifets20042020} and the references therein. As this literature illustrates, finding interpretable sufficient conditions is not always easy. A useful starting point is that
$
L(m) \coloneqq \mmax_{c\in\{1,\ldots,4\}^m}\rho(A_{c_m}\cdots A_{c_1})^{\frac1{m}} \leq \rho_{\sf JSR}(\mathcal A)\le \mmax_{c\in\{1,\ldots,4\}^m}
    \|
        A_{c_m}\cdots A_{c_1}
    \|^{\frac1{m}} \eqqcolon U(m),$ for every matrix-induced norm $\|\cdot\|$, integer $m\geq1$, and spectral radius $\rho(\cdot)$. In particular, for $m = 1$, $\rho(A_j) < 1$ for each $j \in \{1,\dots,4\}$ is necessary for Assumption~\ref{ass:paraJSR} and $\mmax_{1\leq j\leq 4}\|A_j\| < 1$, for some matrix-induced norm $\|\cdot \|$, is sufficient. 
A bit more can be said in certain
special cases, as the following lemma illustrates:

\begin{lemma}\label{lem:sufJSR} Suppose Assumption~\textup{\ref{ass:para}} holds. \emph{($i$)} If $\varphi = 0$, then $\rho_{\sf JSR}({\cal A}) \leq \mmax\{1-\gamma(1-|\delta|),1-\gamma(1-\mu),|\rho_i|\} < 1$. \emph{($ii$)}  If $\rho_i = 0$, $0\leq\delta<1$, and $\psi\geq0$, then $L(1) < 1$. Moreover, $\rho_{\sf JSR}({\cal A}) \leq \mmax\{\rho(|H|), 1-\gamma+\gamma \rho(|H|)\}$, with $|H|$ the entry-wise absolute value of $H$. In this case,  $\rho(|H|)<1 \Leftrightarrow \Delta \Upsilon > 2\psi\mu\varphi(\phi_\pi\delta-1)_+$. Consequently, whenever this inequality holds, $\rho_{\sf JSR}({\cal A})\leq1-\gamma(1-\rho(|H|))<1$. In
particular, \textup{\ref{ass:paraJSR}} holds if $a=0$ for some
$a\in\{\psi,\mu,\varphi\}$, or if $\phi_\pi\delta\leq1$.  
\end{lemma}

Part ($i$) shows that Assumption~\ref{ass:para} implies Assumption~\ref{ass:paraJSR} if $\varphi = 0$. Part ($ii$), the case of no interest-rate smoothing, shows the necessary condition $L(1) < 1$ is implied by \ref{ass:para} on the economically relevant region $\delta \in [0,1)$, $\psi \geq 0$ for any $\gamma \in (0,1)$. To illustrate the sufficient condition, fix standard benchmark values $\varphi = \mu = 1$, $\delta = 0.99$, $\phi_\pi = 1.5$, and $\phi_y = 0.5$. Here, although $\phi_\pi\delta > 1$, the conditions of part ($ii$) are met in regions of flat Phillips curves $\psi \in (0,\psi_-)$ and steeper curves $\psi \in (\psi_+,\infty)$, $\psi_\pm = (17\pm\sqrt{145})/120$, for \emph{any} $\gamma \in (0,1)$. This does not, however, rule out Assumption~\ref{ass:paraJSR} for intermediate values  $\psi \in [\psi_-,\psi_+]$: the sufficient condition fails there, but by part ($ii$) the necessary condition $L(1)<1$ continues to hold. 
Moreover, note that, ceteris paribus, more discounting in the IS curve (lower $\mu$) or a stronger interest-rate channel (higher $\varphi$) both monotonically shrink the unresolved region $[\psi_-,\psi_+]$: In particular, holding
$\varphi=1$ fixed, the interval $[\psi_-,\psi_+]$ disappears below about $\mu^\st = 0.945$; holding $\mu=1$ fixed, it disappears above about $\varphi^\st = 1.175$.

The following proposition yields the probabilistic properties of the dgp:

\begin{proposition}\label{prop:ergod}
Let Assumptions~\textup{\ref{ass:density}}, \textup{\ref{ass:para}}, and
\textup{\ref{ass:paraJSR}} hold. \emph{($i$)} There exists a unique invariant distribution $\pi$, supported
on $\mathsf S_0\times\mathbb R^4$, such that for $S_0\sim\pi$, the process is strictly stationary, ergodic, and geometrically $\beta$-mixing. \emph{($ii$)} The chain is $V$-uniformly geometrically ergodic. \emph{($iii$)} $\Ex[\|r_1\|]<\infty$ and $\Ex[\|\nu_1\|^2]<\infty$, $\nu_t\coloneqq(x_t^\Tr,\alpha_t^\Tr,u_t^\Tr,i_t)^\Tr$. If, in addition, $\Ex[\|\eps_1\|^{2q}]<\infty$ for some $q\geq1$, then $\Ex[\|\nu_1\|^{2q}]<\infty$ and
$\Ex[\|r_1\|^{q}]<\infty$.
\end{proposition}

\begin{remark}\label{rem:smallgam}
As a consequence of the preceding proposition the recursive
estimators of the agent $($viz.\ $\alpha_t$ and $\beta_t)$ converge in distribution but not in
probability for any $\gamma>0$ as $t\rightarrow\infty$. The
limiting law $($as $t\rightarrow\infty)$ of $\alpha_{j,t}$ has mean
$\Ex[x_{j,t}]$ and variance
$\frac{\gamma}{2-\gamma}\sum_{i=-\infty}^{\infty}
\cov(x_{j,t},x_{j,t-i})(1-\gamma)^{|i|}$, $j\in\{y,\pi\}$. A similar
closed-form characterization of the first two moments of the stationary law of
the recursive sample autocorrelation $\beta_t$ is not feasible because it is a
nonlinear ratio; for fixed $\gamma$, we only know from
Lemma~\textup{\ref{lem:invariant}} that its distribution is confined to
$[-\sqrt{1-\gamma},\sqrt{1-\gamma}]$. Following \textup{\citet[Ch.~4]{ben:90}},
more could be said using a `small-$\gamma$' theory of stochastic
approximation $($see also \textup{\citealp[Section 7.4]{evans:01}}$)$. To illustrate, consider the single equation NKPC of Eq.~\eqref{eq:seNKPC}, and collect
the recursions $\ell_t=(\alpha_t,\omega_t,r_t)^\Tr$, with
$\omega_t\coloneqq r_t\beta_t$, so that
$\ell_t=\ell_{t-1}+\gamma \mathfrak{H}(\ell_{t-1},X_t)$; see Eq.~\eqref{eq:psimap}. The recursion can be seen as a noisy Euler discretisation, with step $\gamma$,
of the ordinary differential equation \textup{(ODE)} $\dot\ell(s)=\mathfrak{h}(\ell(s))$ on the
re-scaled time $s=t\gamma$, where $\mathfrak{h}(\ell)\coloneqq\Ex_\ell[\mathfrak{H}(\ell,Y_t)]$ denotes the expectation under the stationary law induced by beliefs
held fixed at $\ell$ \textup{(\citealp[Sec.~1.3]{ben:90})}. Recall that the condition $\mathfrak{h}(\ell)=0$
imposes a BLE, with $\mathfrak{h}^{(\beta)}(\beta)=F(\beta;\lambda)-\beta$. Suppose $\ell^\st$ is a simple zero
of $\mathfrak{h}$, with $r^\st>0$ and $\partial_\beta F(\beta^\st;\lambda)<1$, the latter is equivalent to the common requirement that the eigenvalues of the $3 \times 3$ Jacobian of $\mathfrak{h}(\cdot)$ have strictly negative real parts $($\textup{\citealp[Assumption A ($ii$) on p. 334]{ben:90}}$)$. Then, under regularity conditions,
$\gamma^{-1/2}(\ell_t-\ell^\st)\rightarrow_d\mathcal N(0,V)$ as
$\gamma\rightarrow0_+$ with $\gamma t\rightarrow\infty$, where $V$ is positive definite. Since $\beta$ is a smooth function of $(\omega,r)$ at $r^\st>0$, the delta method gives the corresponding local Gaussian approximation for $\beta_t$. We will, however, maintain throughout the assumption of a constant gain, viewing
$\gamma>0$ as a parameter to be estimated. For robust procedures covering the
case $\gamma\rightarrow0+$, see \textup{\cite{chev:10}}.
\end{remark}

Before turning to parameter estimation, note that this requires the agent's forecast implied by the SAC recursion in Eq.~\eqref{eq:learning}. This recursion, in turn, depends on the gain
\(\gamma\) and on an infinite history $(x_t)_{t \in \mathbb Z}$ of $x_t = (\pi_t,y_t)^\Tr$ that is not observed. We therefore work with a feasible so-called {\it filter}. That is, for a
candidate \(\gamma  \in \Gamma \subseteq (0,1)\) and a fixed initialization \(\fra \coloneqq (\fra_x^\Tr,\fra_\alpha^\Tr,\fra_r^\Tr,\fra_\beta^\Tr)^\Tr\in\mathsf S\), the recursion started at \(s_0 = \fra\), say, yields, for
\(j\in\{\pi,y\}\) and any $t \geq 1$,
\begin{equation}
x^e_{j,t+1}(\gamma,\fra)
\coloneqq
\alpha_{j,t-1}(\gamma,\fra)
+
\beta_{j,t-1}^2(\gamma,\fra)
(x_{j,t-1}
-
\alpha_{j,t-1}(\gamma,\fra)).
\label{eq:feasible-forecast}
\end{equation}
This forecast is computed from the observed path of $(x_t)_{t=1}^n$, the candidate gain
\(\gamma \in \Gamma \subseteq (0,1)\), and \(\fra \in {\sf S}\) through the
recursions~\eqref{eq:learning}.  Because the feasible recursion is initialized at a fixed value
\(\fra\), we distinguish between
\(x^e_{t+1}(\gamma,\fra) \coloneqq (x^e_{\pi,t+1}(\gamma,\fra),x^e_{y,t+1}(\gamma,\fra))^\Tr\) and 
\(x^e_{t+1}(\gamma) \coloneqq (x^e_{\pi,t+1}(\gamma),x^e_{y,t+1}(\gamma))^\Tr\), its stationary counterpart, both constructed from the same strictly stationary process
\((x_t)_{t\in\mathbb Z}\) provided by
Proposition~\ref{prop:ergod}. This distinction is standard in the
literature on stochastic recursions and observation-driven filters;
see, e.g., \cite{bougerol:93}, \cite{strau:06}, and
\cite{blasques:22}. In particular, for every
\(\gamma\in\Gamma\), define $x_{t+1}^e(\gamma)$ as the counterpart of Eq.~\eqref{eq:feasible-forecast} through the stationary candidate recursions of mean
\(\alpha_{j,t}(\gamma)\), variance \(r_{j,t}(\gamma)\), and first auto-covariance \(\omega_{j,t}(\gamma)\). Because $\beta_{j,t}(\gamma)=\omega_{j,t}(\gamma)/r_{j,t}(\gamma)$, the stationary recursion $x_{j,t+1}^e(\gamma)$ is well defined uniformly over \(\Gamma\) whenever
\(
\ubar r_t
\coloneqq 
\iinf_{\gamma\in\Gamma}
\mmin_{j\in\{\pi,y\}}
r_{j,t}(\gamma) > 0,
\) a.s., which is ensured by Assumption~\ref{ass:density}.
The following result establishes uniform geometric
forgetting of the initialization.

\begin{lemma}\label{lem:filter}
Suppose that
\(\Ex[\|x_1\|^2]<\infty\), \(\Gamma=[\ubar\gamma,\bar\gamma]\subset(0,1)
\), and $\Ex[\operatorname{\sf log}^+(1/\ubar r_1)]<\infty,$
then, for every fixed \(d\in\mathbb N_0\) and \(\fra\in\mathsf S\), there exist
\(\varrho_d\in(0,1)\) and an a.s.-finite random variable
\(C_d(\omega,\fra)\) such that, for all \(t\ge1\),
\(
\ssup_{\gamma\in\Gamma}
\|
\partial_\gamma^m
(x_{t+1}^e(\gamma,\fra)
-
x_{t+1}^e(\gamma)
)
\|
\le
C_d(\omega,\fra)\varrho_d^t,\) \(m=0,\ldots,d.\)
\end{lemma}

Note that, apart from the properties of the stationary process $(x_t)_t$ supplied by Proposition~\ref{prop:ergod},
Lemma~\ref{lem:filter} requires only
\(\Gamma\subset(0,1)\) to be compact, an initialization in {\sf S},\footnote{The
initialization of Remark~\ref{rem:initial} does not belong
to \(\mathsf S\). Still, under the maintained conditions, the
corresponding recursion enters \(\mathsf S_0\) after one update almost surely. Hence, the conclusion of Lemma~\ref{lem:filter} also holds for this initialization.} and the mild uniform lower-tail condition $\Ex[\operatorname{\sf log}^+(1/\ubar r_1)]<\infty$
on the stationary variance recursion. The required strict
stationarity and finite second moments of \((x_t)_{t\in\mathbb Z}\)
follow from Proposition~\ref{prop:ergod}. Moreover, under Assumption~\ref{ass:density} and $\Delta \neq 0$, $\Ex[\ubar r_1^{-\tau}]<\infty$, for every $\tau>0,$ so the logarithmic lower-tail condition is automatically satisfied; see Lemma~\ref{lem:negR}.

\section{QML, identification, and consistency}\label{sec:cons}

Observing a sample $(Y_t)_{t=1}^n$, $Y_t\coloneqq(x_t^\Tr,i_t)^\Tr$, of length \(n\) from the stationary process $(S_t)_{t \in \mathbb Z}$ provided by Proposition~\ref{prop:ergod}, we proceed now to
estimating \(\theta\coloneqq(\gamma,\vartheta^\Tr)^\Tr\), 
$\vartheta\coloneqq
(\mu,\varphi,\delta,\psi,\phi_\pi,\phi_y,\rho_i)^\Tr$. 
Following the full-information maximum likelihood approach of
\citet{linde2005}, we estimate $\theta$ using the three structural equations
jointly, profiling out the shock parameters $R \in \mathbb R^{3\times 3}$ and
$\Sigma = {\sf diag}(\sigma_\pi^2,\sigma_y^2,\sigma_i^2)$. As a consequence of Lemma~\ref{lem:filter}, the sample objective below, its score, and its Hessian are uniformly (a.s.) close to their stationary counterparts when suitably scaled. When we now introduce the quasi maximum likelihood estimator (QMLE), we thus suppress the dependence on $\fra$. 

For each $\theta\in\Theta$, the
generalized structural errors $u_t(\theta) \coloneqq (u_{\pi,t}(\theta),u_{y,t}(\theta),u_{i,t}(\theta))^\Tr$ implied by Eqs.~\eqref{eq:pc}-\eqref{eq:taylor} can be written as
\begin{align}
u_t(\theta)
=
B_1(\vartheta)Y_t
-
B_2(\vartheta)W_t(\gamma)
\label{eq:fiml-residuals}
\end{align}
where $W_t(\gamma) \coloneqq (x_{t+1}^e(\gamma)^\Tr,i_{t-1})^\Tr$ and
\begin{align}
B_1(\vartheta)
\coloneqq
\begin{bmatrix}
1 & -\psi & 0\\
0 & 1 & \varphi\\
-(1-\rho_i)\phi_\pi
&
-(1-\rho_i)\phi_y
&
1
\end{bmatrix},
\qquad
B_2(\vartheta)
\coloneqq
\begin{bmatrix}
\delta & 0 & 0\\
\varphi & \mu & 0\\
0 & 0 & \rho_i
\end{bmatrix},
\label{eq:fiml-CP}
\end{align}
so that $u_t = u_t(\theta_0)$. For a $(\theta,R)\in\Theta\times\mathbb R^{3\times3}$, define the corresponding
generalized error $\eps_t(\theta,R) \coloneqq u_t(\theta)-Ru_{t-1}(\theta),$ so that $\eps_t=\eps_t(\theta_0,R_0)$. Because \(\partial\eps_t(\theta,R)/\partial Y_t^\Tr = B_1(\vartheta)\) and $\operatorname{\sf det} B_1(\vartheta) =\operatorname{\sf det} M(\vartheta)
= \Delta(\vartheta),$ with $\Delta(\vartheta)$ and $M(\vartheta)$ defined in Eq.~\eqref{eq:ALMmat}, the conditional (quasi-)density of
$Y_t$ is \(f_Y(Y_t\mid {\cal F}_{t-1};\theta,R,\Sigma)
=
f_\eps(\eps_t(\theta,R);\Sigma)
  |\Delta(\vartheta)|.\)   
  
  Upon specifying $f_\eps(\cdot)$ to be Gaussian, we can now introduce a profiled QMLE: Let $R_n(\theta)
\coloneqq \sum_{t=2}^n
u_t(\theta)u_{t-1}(\theta)^\Tr
(\sum_{t=2}^n u_{t-1}(\theta)u_{t-1}(\theta)^\Tr)^{-1}$ be the OLS
estimator, which under the maintained assumptions is well-defined. Denoting the profiled generalized error $\widehat\eps_t(\theta)
\coloneqq \eps_t(\theta,R_n(\theta))$ and corresponding empirical variances
$\widehat\sigma_{j,n}^2(\theta)
\coloneqq
\frac{1}{n-1}
\sum_{t=2}^n
\widehat \eps^2_{j,t}(\theta),$ $j\in\{\pi,y,i\},$
the profiled quasi-likelihood criterion, up to constants that do not depend on $\theta$, can be defined as
\begin{align}
Q_n(\theta)
\coloneqq
\sum_{j\in\{\pi,y,i\}}
\operatorname{\sf log}\widehat\sigma_{j,n}^2(\theta)
-
2\operatorname{\sf log}|\Delta(\vartheta)|.
\label{eq:fiml-objective}
\end{align}
This criterion coincides with the conditional FIML criterion when
$\eps_t$ is Gaussian; see \citet{linde2005} and
\citet[Section~18]{davidson1993estimation}. A QMLE is then
\begin{align}
\theta_n
\in
\argmin_{\theta\in\Theta}
Q_n(\theta).
\label{eq:fiml-estimator}
\end{align}
In the scalar NKPC of Eq.~(\ref{eq:seNKPC}), where $\operatorname{\sf log}|\Delta(\vartheta)|=0$, this reduces to the nonlinear least squares estimator (NLSE) discussed in more detail in Section~\ref{sec:distnlse}.

\subsection{Identification and consistency}

We study conditions under which the pair
$(\theta_0,R_0)$ uniquely minimizes the population counterpart of the objective~\eqref{eq:fiml-objective}. To this end, note that the structural error in
Eq.~\eqref{eq:fiml-residuals} obeys
$u_t(\theta)=C(\vartheta)u_t+c_t(\theta)$, where
\begin{align}
C(\vartheta)
\coloneqq
B_1(\vartheta)B_1(\vartheta_0)^{-1},\quad 
c_t(\theta)
\coloneqq
C(\vartheta)B_2(\vartheta_0)W_t(\gamma_0)
-B_2(\vartheta)W_t(\gamma),
\label{eq:fiml-Cc}
\end{align}
and $c_t(\theta_0)=0$. Consequently, for any
$(\theta,R)\in\Theta\times\mathbb R^{3\times3}$,
\begin{align}
\eps_t(\theta,R)
=
C(\vartheta)\eps_t+\xi_t(\theta,R),
\label{eq:fiml-eps-decomposition}
\end{align}
where, using $u_t=R_0u_{t-1}+\eps_t$,
\begin{align}
\xi_t(\theta,R)
\coloneqq
(C(\vartheta)R_0-RC(\vartheta))u_{t-1}
+c_t(\theta)-Rc_{t-1}(\theta).
\label{eq:fiml-d}
\end{align}
Using $\Ex[\eps_t\mid{\cal F}_{t-1}]=0$, the conditional mean of the generalized error in~\eqref{eq:fiml-eps-decomposition} is $\xi_t(\theta,R)$, while its second moment matrix is
\begin{align}
\Omega(\theta,R)
\coloneqq
\Ex[\eps_1(\theta,R)\eps_1(\theta,R)^\Tr]
=
C(\vartheta)\Sigma_0C(\vartheta)^\Tr
+
\Ex[\xi_1(\theta,R)\xi_1(\theta,R)^\Tr],
\label{eq:fiml-Omega}
\end{align}
respectively, where $\Sigma_0\coloneqq \Ex[\eps_1\eps_1^\Tr]={\sf diag}(\sigma_\pi^2,\sigma_y^2,\sigma_i^2)$ is, by Assumption~\ref{ass:density}, diagonal. The (unprofiled) population objective is 
$$Q(\theta,R)\coloneqq
\sum_{j\in\{\pi,y,i\}}\operatorname{\sf log}\Omega_{jj}(\theta,R)
-2\operatorname{\sf log}|\Delta(\vartheta)|,$$
which, as shown in
Appendix~\ref{sec:A2}, satisfies
$\ssup\limits_{\theta\in\Theta}|Q_n(\theta)-\mmin\limits_{R \in \mathbb R^{3 \times 3}} Q(\theta,R)| \rightarrow_{a.s.} 0$.

To see that $(\theta_0,R_0)$ is \emph{a} minimizer consider
\begin{align}
\prod_{j\in\{\pi,y,i\}}
\Omega_{jj}(\theta,R)
\stackrel{(a)}{\geq}
\operatorname{\sf det}\Omega(\theta,R)
\stackrel{(b)}{\geq}
\operatorname{\sf det}
C(\vartheta)\Sigma_0C(\vartheta)^\Tr
=
\left[
\frac{\Delta(\vartheta)}
{\Delta(\vartheta_0)}
\right]^2
\operatorname{\sf det}\Sigma_0,
\label{eq:fiml-pop-lower-bound}
\end{align}
that is, $Q(\theta,R)\geq Q(\theta_0,R_0)$ for every
$(\theta,R)\in\Theta\times\mathbb R^{3\times3}$. To establish uniqueness of \emph{the} minimizer, it remains to characterize equality in
Eq.~\eqref{eq:fiml-pop-lower-bound}. Bound $(a)$ (Hadamard's inequality) holds with equality
if and only if $\Omega(\theta,R)$ is diagonal, whereas, since
$\Ex[\xi_1(\theta,R)\xi_1(\theta,R)^\Tr]\succeq0$ and
$C(\vartheta)\Sigma_0C(\vartheta)^\Tr\succ0$, bound $(b)$ holds
with equality if and only if $\xi_t(\theta,R)=0$ a.s. Since the latter
implies
$\Omega(\theta,R)=C(\vartheta)\Sigma_0C(\vartheta)^\Tr$, both bounds
bind precisely when $\xi_t(\theta,R)=0$ a.s. and
$C(\vartheta)\Sigma_0C(\vartheta)^\Tr$ is diagonal; that is, for any $(\theta,R)\in\Theta\times\mathbb R^{3\times3}$, $Q(\theta,R)=Q(\theta_0,R_0)$ if and only if
$\xi_t(\theta,R)=0$ a.s.\ and
$C(\vartheta)\Sigma_0C(\vartheta)^\Tr$ is diagonal.

\begin{proposition}\label{prop:separation}
Suppose Assumptions~\textup{\ref{ass:density}}, \textup{\ref{ass:para}}, and \textup{\ref{ass:paraJSR}} are satisfied, and $\delta>0$, $\psi\geq0$ holds throughout the
parameter space. \textup{($i$)} If, for some
$(\theta,R)\in\Theta\times\mathbb R^{3\times3}$,
$\xi_t(\theta,R)=0$ a.s., then $\gamma=\gamma_0$,
$B_2(\vartheta)=C(\vartheta)B_2(\vartheta_0)$, and
$C(\vartheta)R_0=RC(\vartheta)$. In particular,
$c_t(\theta)=0$ a.s. \textup{($ii$)} If, in addition,
$C(\vartheta)\Sigma_0C(\vartheta)^\Tr$ is diagonal, then
$\vartheta=\vartheta_0$ and $C(\vartheta)=I_3$, so that
$(\theta,R)=(\theta_0,R_0)$.
\end{proposition}

As Proposition~\ref{prop:separation} summarizes, the two restrictions play distinct roles. The condition
$\xi_t(\theta,R)=0$ identifies the gain and the coefficient matrices up to the
transformation $C(\vartheta)$, but not $C(\vartheta)$ itself. Importantly, on that set,  a criterion based on an unrestricted innovation covariance, and not its diagonal elements as in $Q(\theta,R)$, is flat in $\vartheta$. It is therefore the
diagonality restriction that pins $C(\vartheta)$ down, through an orthogonality condition akin to identification through
covariance restrictions in simultaneous-equations models, where the error from
one structural equation acts as an internally generated instrument for another
\citep{hausman1983identification,hausman1987efficient}.

Additional restrictions on the parameter space are required for these two conditions to identify the structural parameters. If
$\delta=\varphi=\mu=0$, agents' forecasts do not enter the
structural equations and the gain $\gamma$ is not identified. We thus require $\delta>0$, which rules this case out while retaining the leading
single-equation NKPC. Moreover, changes in $c_t(\theta)$ may in
principle be offset by changes in $R$. The additional restrictions $\varphi\geq0$ and $\psi\geq0$ exclude such a problematic case that arises via the cancellation $\delta+\psi\varphi=0$. These sign restrictions are all in line with economic theory (\citealp{Woodford2003}).

The same identification strategy, applied directly to the single-equation NKPC in~\eqref{eq:seNKPC}, yields the following result:

\begin{corollary}\label{cor:NLSEident} Consider the single-equation NKPC~\eqref{eq:seNKPC}. Suppose Assumption~\textup{\ref{ass:density}} holds and that $|\rho|<1$, $\gamma \in (0,1)$, and $|\delta|<1$ for  $\delta \neq 0$ throughout the parameter space. If $\delta h_t(\gamma)=\delta_0h_t(\gamma_0)$ a.s., then $(\delta,\gamma)=(\delta_0,\gamma_0),$ with $h_t(\gamma)\coloneqq\alpha_t(\gamma)
+\beta_t(\gamma)^2(\pi_t-\alpha_t(\gamma))$.
\end{corollary}

\begin{remark}\label{rem:tail-sep}
The key assumption is that the conditional law of $\pi_t$ given
$\mathcal F_{t-1}$ has full support. Akin to related identification arguments for
observation-driven models $($see, e.g. \textup{\citealp{francq:04}},
\textup{\citealp[Section~5]{strau:06}}, \textup{\citealp[Proof of Lemma~TA.8]{blasques:22}}, or  \textup{\citealp[Lemma~1]{blasques:24}}$)$, this transforms the stochastic equality into an algebraic identity. Conditional on $\mathcal F_{t-1}$, full
support and continuity imply
\(\delta h_t(\gamma;z)=\delta_0h_t(\gamma_0;z)\), for every $z\in\mathbb R$, where $h_t(\gamma)=h_t(\gamma;\pi_t)$. 
Since both sides are rational functions, the identity yields a polynomial of degree at most nine that vanishes for every real $z$. By a standard consequence of the fundamental theorem of algebra, it must be the zero polynomial; hence the identity also holds over $\mathbb C$. Indeed, the strictly positive denominator $r_t(\gamma;z)^2$, with $r_t(\gamma) \coloneqq r_t(\gamma;\pi_t)$, has two conjugate complex roots,
\(
z_\pm(\gamma)
=
\alpha_{t-1}(\gamma)
\pm
i\tau_{t-1}(\gamma),
\)
$\tau_{t-1}^2(\gamma) \coloneqq (1-\gamma)r_{t-1}(\gamma)/\gamma$, which are fixed conditional on $\mathcal F_{t-1}$. Let
$z_+=z_+(\gamma_0)$. The proof shows that $z_+$ is a
genuine double pole of the true forecast, that is,
\(
a_0
\coloneqq
\delta_0\llim_{z\to z_+}(z-z_+)^2h_t(\gamma_0;z)
\neq0.
\)
However, the identity implies
\(
\delta(z-z_+)^2h_t(\gamma;z)
=
\delta_0(z-z_+)^2h_t(\gamma_0;z).
\)
Taking the limit as $z\to z_+$, the right-hand side converges to $a_0\neq0$.
Hence $h_t(\gamma;z)$ must also have a double pole at $z_+$, since otherwise
the left-hand side would converge to zero. As $z_+(\gamma)$ is the unique root
with positive imaginary part, $z_+(\gamma)=z_+(\gamma_0)$. Comparing real
parts gives
$\alpha_{t-1}(\gamma)=\alpha_{t-1}(\gamma_0)$ a.s.. The recursive definition of
$\alpha_{t-1}(\gamma)$, together with the nonatomic conditional law of
$\pi_{t-1}$, then implies $\gamma=\gamma_0$. Finally, the original identity
and the fact that $h_t(\gamma_0) \neq 0$ a.s. imply
$\delta=\delta_0$.
\end{remark}

Strong consistency of the QMLE follows from Proposition~\ref{prop:separation},
together with a uniform law of large numbers for stationary ergodic processes.

\begin{proposition}\label{prop:const}
Suppose that the conditions of Proposition~\textup{\ref{prop:separation}} are satisfied
and that $\Theta$ is compact. Then, as $n\rightarrow\infty$,
$\theta_n\rightarrow_{a.s.}\theta_0$ and
$R_n(\theta_n)\rightarrow_{a.s.}R_0$.
\end{proposition}

\section{Distributional theory of the NLSE}\label{sec:distnlse}

For reasons of analytical tractability, we develop the distributional theory for the scalar NKPC~\eqref{eq:seNKPC}. Under the
corresponding restrictions, the Gaussian quasi-likelihood factorizes, and maximizing the part that depends on
$\theta\coloneqq(\gamma,\delta,\psi)^\Tr$ is equivalent to minimizing a
nonlinear least-squares criterion. In particular, with a slight abuse of
notation, let $\theta_n\coloneqq(\theta_{\gamma,n},\theta_{\delta,n},
\theta_{\psi,n})^\Tr$ be a minimizer of $Q_n(\theta)
\coloneqq
\sum_{t=1}^n
(\pi_t-f_t(\theta))^2,$ where $f_t(\theta)
\coloneqq
\delta h_{t-1}(\gamma)
+
\psi y_t.$ 
Importantly, compared to the general model, identification of
$\theta$ and strong consistency of the NLSE require only
$\delta_0\neq0$ and leave $\psi$ unrestricted:

\renewcommand{\theassumption}{B'}
\begin{assumption}\label{ass:para1}
  $|\rho|<1$ and $\theta \in \Theta \coloneqq \Gamma \times \mathcal{D} \times \Psi,$ where $\Gamma \coloneqq [\ubar\gamma,\bar\gamma],$ $\mathcal{D} \coloneqq [-\bar\delta,-\ubar\delta] \cup [\ubar\delta,\bar\delta], $ and $\Psi\coloneqq [\ubar\psi,\bar\psi]$ for $0<\ubar\gamma<\bar\gamma<1$, $0<\ubar\delta<\bar\delta<1$,  $-\infty < \ubar\psi < \bar\psi < \infty$.
\end{assumption}

Note that for the scalar NKPC, Assumption~\ref{ass:para1} implies Assumptions~\ref{ass:para} and \ref{ass:paraJSR}. As a consequence of Corollary~\ref{cor:NLSEident}, the NLSE is strongly consistent if Assumptions~\ref{ass:density} and~\ref{ass:para1} hold. 

Having established consistency, we turn to the limiting distribution of
$\theta_n$. The mean value theorem applied componentwise to the first-order condition of the NLSE ensures 
\begin{align}
\sqrt{n}(\theta_n-\theta_0)
=
-\left[\frac1n\nabla_\theta^2Q_n(\bar\theta_n)\right]^{-1}
\frac1{\sqrt n}\nabla_\theta Q_n(\theta_0),
\label{eq:mvt}
\end{align}
where $\bar\theta_n$ is a mean value lying on the line segment connecting $\theta_n$ and $\theta_0$ and that might vary from row to row. Given the consistency of $\theta_n$, standard arguments (see, e.g.
\citealp{amemiya1985advanced}) yield limiting normality if the following holds true: (a) $A\coloneqq\Ex[\nabla_\theta f_1(\theta_0) \nabla_\theta f_1(\theta_0)^\Tr]$ is a finite positive definite $3\times3$ matrix, (b) $\frac1n\nabla_\theta^2Q_n(\bar\theta_n)\rightarrow_p 2A$ for
    any $\bar\theta_n$ such that
    $|\bar\theta_n-\theta_0|\rightarrow_{a.s.}0$, and (c) $\frac1{\sqrt n}\nabla_\theta Q_n(\theta_0)\rightarrow_d
    {\cal N}(0,4A\sigma_u^2)$.
 Due to the highly nonlinear nature of $\nabla_\theta f_t(\theta)$, deriving the matrix $A$ in
closed form is not feasible. Instead, we verify ($a$) in
Lemma~\ref{lem:pd}~($a$) of the appendix by an argument in the spirit of
Remark~\ref{rem:tail-sep}. Part
($c$) is then an immediate consequence of ($a$) and the CLT for homoskedastic
martingale difference sequences, and part ($b$) follows, as argued in
\citet[Lemma 1]{andrews:92}, from a stochastic Lipschitz bound on
$\nabla_\theta^2Q_n$, established in Lemma~\ref{lem:pd}~($b$).
 
These arguments rest on moment bounds for the derivatives of the agents'
recursive estimates $\{\alpha_t(\gamma),r_t(\gamma),\beta_t(\gamma)\}$ uniform
over $\gamma \in \Gamma$, collected in Lemma~\ref{lem:mom} of the appendix.
Here, the derivatives of $r_t$ and $\beta_t$ require more stringent moment conditions on the errors
$v_t=(u_{t},\eps_{t})^\Tr$ than bounding those of $\alpha_t$,
because the former already involve squared data: for example, for the third derivatives that enter the Lipschitz bound, $\Ex[\ssup_{\gamma\in\Gamma}|\partial_\gamma^3\alpha_1(\gamma)|^k]<\infty$
requires only $k$ error moments, whereas the corresponding bound for the third derivative of
$\beta_t$ requires, for some $\zeta > 0$, $6k+\zeta$ so that the $k=\frac4{3}$ moment needed there calls for slightly more than eight error
moments. 

\renewcommand{\theassumption}{C}
\begin{assumption}\label{ass:eight} There exists $\zeta>0$ such that $\Ex[\|v_1\|^{8+\zeta}]<\infty$.
\end{assumption}
 
We are now ready to derive the limiting distribution of the estimator.
 
\begin{proposition}\label{prop:norm} If Assumptions~\textup{\ref{ass:density}},
\textup{\ref{ass:para1}}, and \textup{\ref{ass:eight}} hold and
$\theta_0=(\gamma_0,\delta_0,\psi_0)^\Tr\in{\sf int}(\Theta)$, then
${\sqrt n}(\theta_n-\theta_0)\rightarrow_d{\mathcal N}(0,\Sigma),$ $\Sigma\coloneqq\sigma_u^2A^{-1}\succ0.$
\end{proposition}
 
As is common for extremum estimators (see, e.g. \citealp{newmc:94}), we
restrict the true value $\theta_0$ to fall within the interior of $\Theta$.
This excludes the boundary case $\delta_0=0$, at which $\gamma$ is no longer
{\it jointly} identified with $(\delta,\psi)$. However, using arguments from \cite{saikkonen:95} and
\cite{seo:11},  we can still show that
the remaining components of the NLS estimator remain (weakly)
$\sqrt n$-consistent. 

\begin{corollary}\label{cor:rate}
Suppose Assumptions~\textup{\ref{ass:density}} and~\textup{\ref{ass:eight}} hold,
and that Assumption~\textup{\ref{ass:para1}} holds with $\mathcal D$ replaced by
$[-\bar\delta,\bar\delta]$. If, in addition, $\delta_0=0$, then
$\sqrt{n}(\theta_{n,\delta},\theta_{n,\psi}-\psi_0)^\Tr=O_p(1)$.
\end{corollary}

\subsection{Asymptotic distribution of the roots and the number of roots}\label{sec:roots}

Recall from Eq.~\eqref{eq:consAC1} that an equilibrium is defined via
$\beta^\st_n = F(\beta^\st_n;\lambda_n)$ for
$\lambda_n = (\theta_{\delta,n},\theta_{\psi,n},\rho_n,\hat\sigma_u^2,\hat\sigma_\eps^2)^\Tr$,
where $\theta_n =(\theta_{\gamma,n},\theta_{\delta,n}, \theta_{\psi,n})^\Tr$,
while $\rho_n$ and $\hat\sigma_\eps^2$ denote the OLS estimator of $\rho$ and the
associated error variance $\sigma_\eps^2$, respectively. By Proposition~\ref{prop:norm} and standard arguments, $\sqrt{n}(\lambda_n-\lambda_0) \rightarrow_d {\mathcal N}(0,\Omega)$ with $\Omega\succ0$.  Moreover, let $r(\lambda)$ denote the number of distinct roots of
$G(\beta;\lambda)\coloneqq \beta - F(\beta;\lambda)$ for a given
$\lambda$ and define
$$r_0 \coloneqq r(\lambda_0),\quad r_n \coloneqq r(\lambda_n),\quad
G(\beta) \coloneqq G(\beta;\lambda_0),\quad G_n(\beta) \coloneqq G(\beta;\lambda_n).$$ 
Throughout this section we
impose the following strengthening of Assumption~\ref{ass:para1}:
\renewcommand{\theassumption}{B''}
\begin{assumption}\label{ass:para2} Assumption~\textup{\ref{ass:para1}} holds
with $\rho \in (0,1)$ and $\Theta \coloneqq \Gamma \times {\cal D} \times
\Psi_{\neq 0}$, where $\Psi_{\neq 0} \coloneqq [-\bar\psi,-\ubar\psi] \cup
[\ubar\psi,\bar\psi],$ $0<\ubar\psi<\bar\psi<\infty.$
\end{assumption}

Assumption~\ref{ass:para2} introduces the two restrictions $\psi\neq0$ and
$\rho>0$. The former ensures that $F_\lambda(\beta;\lambda)$ is uniformly
bounded away from zero at $\lambda=\lambda_0$, a crucial condition for the
analysis below, since $\|F_\lambda(\beta;\lambda)\|=\beta^2$ if the true
$\psi_0=0$. Because, by construction, $F(-\beta;\lambda_0)=F(\beta;\lambda_0)$ for any
$\beta\in[-1,1]$ and, by Assumption~\ref{ass:para2},
$\beta+F(\beta;\lambda_0)>0$ for every $\beta\in[0,1]$, $G$ has no root in $[-1,0]$ because there $G(\beta) = -(|\beta|+F(|\beta|;\lambda_0))
<0$. We can therefore restrict the analysis to
$[0,1]$, where Assumption~\ref{ass:para2} ensures
$G(0)<0<G(1)$, and hence the existence of at least one root. As shown by Lemma~\ref{lem:Flambda} of the appendix, this forces $G$ to change sign an odd
number of times on $(0,1)$. Since a root of odd multiplicity produces exactly
one sign change, while a root of even multiplicity produces
none,\footnote{A root $\beta^\st$ of $G(\cdot)$ is said to have
\emph{multiplicity} $m$ if $\partial^j_\beta G(\beta^\st)=0$ for
$j=0,\ldots,m-1$ and $\partial^m_\beta G(\beta^\st)\neq0$. A root is
\emph{simple}, \emph{double}, or \emph{triple}, if $m=1$, $m=2$, or $m=3$,
respectively. Note that $G$ changes sign at $\beta^\st$ if and only if $m$ is
odd.} and since $G$ has at most four roots counted with multiplicity, only three
configurations can arise: (a) all roots are simple, $r_0\in\{1,3\}$; (b) one simple and one double root, $r_0=2$; or (c) one triple root, $r_0=1$. 

\bk As summarized by Corollary~\ref{cor:roots} for each of the three cases (a), (b), and (c), the multiplicity of a root governs how precisely it can be estimated. Indeed, if
$\beta_{i,0}$ has multiplicity $m_i$, then any root $\beta^\st_n$ of $G_n$ near $\beta_{i,0}$ satisfies $\beta^\st_n-\beta_{i,0}=O_p(n^{-1/(2m_i)}).$

\begin{corollary}\label{cor:roots}
Suppose  Assumptions~\textup{\ref{ass:density}}, \textup{\ref{ass:para2}}, and \textup{\ref{ass:eight}} hold, and let $\mathbb{Z}\sim \mathcal{N}(0,\Omega)$. 
\begin{itemize}
\item[\textnormal{(a)}] $r_n\to_p r_0$ and $\sqrt{n}(\beta^\st_{i,n}-\beta_{i,0})
\to_d J_i^\Tr\mathbb Z,$ $J_i \coloneqq J(\beta_{0,i})$, $i=1,\ldots,r_0,$ with
$J(\beta)\coloneqq
\frac{F_\lambda(\beta;\lambda_0)}
{1-F_\beta(\beta;\lambda_0)}.$
\item[\textnormal{(b)}]
Let $\beta_0$ and $\beta_0^\dagger$ denote the simple and the double root,
respectively, where
$G_\beta(\beta_0^\dagger)=0,$
$G_{\beta\beta}(\beta_0^\dagger)\eqqcolon2a,$ and
$G_\lambda(\beta_0^\dagger)\eqqcolon B.$ Then
$\PP\{r_n=1\}\to\tfrac12$ and $\PP\{r_n=3\}\to\tfrac12$. There exist a neighbourhood ${\cal N}_0$ of $\beta_0$
and a sequence of roots $\beta^\st_n$ of $G_n$ such that, with probability
approaching one, $\beta^\st_n$ is the unique root of $G_n$ in ${\cal N}_0$
and, with $J=J(\beta_0)$, $\sqrt n(\beta^\st_n-\beta_0)\to_d J^\Tr\mathbb Z.$

With probability approaching one, on the event $\{r_n=3\}$, $G_n$ has,
besides $\beta^\st_n$, exactly two further roots
$\beta^{\st-}_n<\beta^{\st+}_n$, both converging in probability to
$\beta_0^\dagger$, and, conditionally on $r_n=3$,
\[
(\sqrt n(\beta^\st_n-\beta_0),\,
n^{1/4}(\beta^{\st\pm}_n-\beta_0^\dagger))
\to_d
(J^\Tr\mathbb Z,\,
\pm\sqrt{-a^{-1}B^\Tr\mathbb Z})
\,\big|\,
\{a^{-1}B^\Tr\mathbb Z<0\}.
\]
Moreover, if $cB\neq aD$, with
$G_{\beta\beta\beta}(\beta_0^\dagger)\eqqcolon6c$ and
$G_{\beta\lambda}(\beta_0^\dagger)\eqqcolon D$, then, jointly with the above,
\[
\sqrt n
\left(
\tfrac12(\beta^{\st+}_n+\beta^{\st-}_n)-\beta_0^\dagger
\right)
\to_d
\tfrac{1}{2a}
\left(
\tfrac{c}{a}B-D
\right)^\Tr\mathbb Z
\,\big|\,
\{a^{-1}B^\Tr\mathbb Z<0\},
\]
while the left-hand side is $o_p(1)$ if $cB=aD$.
\item[\textnormal{(c)}] Let $\beta_0$ be the triple root where
$G_\beta(\beta_0)=G_{\beta\beta}(\beta_0)=0$ and define $G_{\beta\beta\beta}(\beta_0)\eqqcolon6c_3$ and $G_\lambda(\beta_0)\eqqcolon B.$
Then $r_n\to_p1$ and
$n^{1/6}(\beta^\st_n-\beta_0) \to_d -(c_3^{-1}B^\Tr\mathbb Z)^{1/3},$
where $w^{1/3}\coloneqq\operatorname{\sf sgn}(w)|w|^{1/3}$.
\end{itemize}
\end{corollary}

\begin{figure}[tb]
\centering
\begin{minipage}[c]{.7\textwidth}
\begin{tikzpicture}
\begin{axis}[
  width=\linewidth,
  height=6cm,
  xlabel={$\beta$},
  ylabel={$G(\beta)$},
  legend pos=south east,
  legend cell align=right,
  legend style={draw=none, fill=none,font=\footnotesize},
  table/col sep=space, 
  grid style={line width=.1pt, draw=gray!20},
  major grid style={line width=.2pt, draw=gray!40},
  scaled y ticks=false,
  yticklabel style={/pgf/number format/fixed, /pgf/number format/precision=4},
  enlarge x limits=false,
  xticklabels={},ytick={0},
  yticklabels={0},
  xlabel near ticks,
    ylabel near ticks,
  label style={font=\footnotesize},      
    tick label style={font=\scriptsize},   
    xlabel style={yshift=2pt},             
    ylabel style={xshift=2pt}
]

\draw[red, line width=.5pt]
  (axis cs:\pgfkeysvalueof{/pgfplots/xmin}, 0)
  -- (axis cs:\pgfkeysvalueof{/pgfplots/xmax}, 0);

\draw[gray, line width=.5pt]
  (axis cs:0.7599, \pgfkeysvalueof{/pgfplots/ymin})
  -- (axis cs:0.7599, \pgfkeysvalueof{/pgfplots/ymax});
\draw[gray, line width=.5pt]
  (axis cs:0.6726, \pgfkeysvalueof{/pgfplots/ymin})
  -- (axis cs:0.6726, \pgfkeysvalueof{/pgfplots/ymax});

\addplot+[smooth, dashed, no marks, thick, black] table[x=ggrid, y=g1] {Figure1_dat.txt}; \addlegendentry{$r_n=1$}
\addplot+[smooth, no marks, ultra thick, black] table[x=ggrid, y=g]  {Figure1_dat.txt}; \addlegendentry{$r_0=2$}
\addplot+[smooth, dotted, no marks, thick, black] table[x=ggrid, y=g2] {Figure1_dat.txt}; \addlegendentry{$r_n=3$}

\end{axis}
\end{tikzpicture}
\end{minipage}%
  \begin{minipage}[c]{.275\textwidth}\vspace*{-.35cm}
\caption{The solid black line is the true $G(\beta)$ at $r_0 = 2$ with two roots indicated by the two grey vertical lines. The dashed and the dotted lines indicate a realized $G_n(\beta)$ conditioning on $r_n = 1$ and $r_n = 3$, respectively.}\label{fig:repeated}
\end{minipage}
\end{figure} 

While case (a) is regular, let us briefly comment part (b) and (c): 
For the case $r_0 = 2$, where $G(\beta)$ merely touches the horizontal axis, an argument similar to
\citet[Proof of Thm 1]{dovonon:18}, yields uniformly over $\beta$ in a neighbourhood around the double root:
\[
\frac1{a}G(\beta;\lambda_n) = (\beta - \beta_0^\dagger)^2+{\mathbb Z}_n+ o_p(n^{-1/2}),
\quad {\mathbb Z}_n \coloneqq \frac1{a}B^\Tr(\lambda_n-\lambda_0),
\]
with $\sqrt{n}\mathbb{Z}_n \rightarrow_d \frac1{a}B^\Tr{\mathbb Z}$. That is, we
have a parabola $\beta \mapsto (\beta-\beta_0^\dagger)^2$ shifted by a random
intercept $\mathbb{Z}_n = O_p(n^{-1/2})$. As shown in
Figure~\ref{fig:repeated}, a random downward shift (${\mathbb Z}_n < 0$) yields,
in addition to the simple root on the left, the dotted line ($r_n = 3$) with two
more crossings at $\beta_n^{\st^\pm}$, while (${\mathbb Z}_n > 0$) shifts the
parabola above the zero line so that the dashed line with no additional crossing
emerges ($r_n = 1$). Because $a^{-1}B^\Tr\mathbb Z$ is centred Gaussian, the events $\{a^{-1}B^\Tr\mathbb Z<0\}$ and
$\{a^{-1}B^\Tr\mathbb Z>0\}$ each have probability $\frac1{2}$. The former yields
two additional roots near $\beta_0^\dagger$, while the latter yields none. Importantly, although the
roots are consistent, the number of roots is not, i.e. $r_n\not\to_p r_0$. Intuitively, consistency of
the repeated roots $\beta_n^{\st^\pm}$ depends on the {\it magnitude} of
$\sqrt{|\mathbb{Z}_n|}$, which is $O_p(n^{-1/4})$. On the other hand, the root
count $r_n$ is determined by the {\it sign} of $\mathbb{Z}_n$, which, by
limiting Gaussianity of $\sqrt{n}\mathbb{Z}_n$, remains a fair coin
asymptotically. Put differently, $\lambda \mapsto r_n(\lambda)$ is discontinuous
at $\lambda_0$ under multiple roots, so $r(\lambda_n)$ need not converge to
$r_0 = 2$ even though $\beta_n^{\st^\pm}$ are consistent on the $\{r_n = 3\}$
event. An interesting by-product is that the double-roots $\beta_n^{\st^\pm}$
have, with opposite sign, the same leading term of order $n^{-1/4}$. For large
$n$, they therefore split symmetrically around the repeated root
$\beta_0^\dagger$. Akin to a two-sample Jackknife, averaging thus eliminates
this leading {\it bias} term. At a root of multiplicity three, finally,
the same logic produces a cubic rather than quadratic local approximation,
yielding the even slower rate $n^{1/6}$ with a
symmetric, non-Gaussian limit. In contrast to
the double root, the local cubic has, with probability approaching one, exactly one real solution,
so that $r_n \rightarrow_p 1$.

 \begin{remark}\label{rem:smallgam1} In view of Remark~\textup{\ref{rem:smallgam}}, recall that
\(F_\beta(\beta^\st;\lambda)<1\), equivalently
\(G_\beta(\beta^\st;\lambda)>0\), is exactly the local stability condition for
a scalar BLE in the limiting case \(\gamma\to0_+\); see also \textup{\citet[Proposition~4]{hz14}}. Indeed,
since $\mathfrak{h}^{(\beta)}(\cdot) = -G(\cdot)$, the condition \(G_\beta(\beta^\st;\lambda)>0\) is the
Hurwitz condition for the Jacobian of the averaged ODE \(\dot\ell=\mathfrak{h}(\ell)\) at
the corresponding equilibrium. Now, for $r_0 = 3$ with $\beta_1^\st < \beta_2^\st < \beta_3^\st$, the sign of \(\mathfrak{h}^{(\beta)}=-G\) alternates across the three simple roots. Only the first and third root can thus be stable, which means that, for $\gamma \rightarrow 0_+$, beliefs started
sufficiently close to $\beta_1^\st$ or $\beta_3^\st$ are attracted to its neighbourhood with high probability, while $\beta_2^\st$ repels and marks the
boundary between their two domains of attraction. Consequently, for $\gamma \rightarrow 0_+$, the limiting Gaussian distribution of $\beta_t$ discussed in Remark~\ref{rem:smallgam} concentrates on whichever stable root lies in the domain of attraction of the initial
belief $($see \textup{\cite{evans1993adaptive}} for a similar discussion$)$.
 \end{remark}

The taxonomy above relies crucially on $G$ being scalar, and breaks down in the general model where our arguments do not extend in a straightforward way. To illustrate, in the multivariate model, where
$G=(G_\pi,G_y)^\Tr$ is two dimensional, even if both components were merely quartic with no common component (which they are not in general; see
\citealp{HommesMavromatisOzdenZhu2023}), B\'ezout's theorem would allow up to
$16$ intersections, already giving rise to
more than 900 multiplicity cases. Moreover, at an intersection, multiplicity alone does not determine possible limiting behaviour as is the case here. 

 \subsection{Inference}\label{sec:inf}

The variance-covariance matrix $\Sigma = \sigma_u^2A^{-1}$, of the limiting distribution of $\theta_n$ is not known in closed form. We therefore estimate $A$ by numerical differentiation of the criterion: with $e_i$ the $i$-th unit vector of $\mathbb R^3$ and a step size $\ell_n$, let
\begin{equation}
\begin{split}\label{eq:numA}
[A_{n}(\theta)]_{i,j} \coloneqq \,& \frac1{2 n}(Q_n(\theta+e_i\ell_n+e_j\ell_n)-Q_n(\theta-e_i\ell_n+e_j\ell_n) \\
&\quad\qquad-Q_n(\theta+e_i\ell_n-e_j\ell_n)+Q_n(\theta-e_i\ell_n-e_j\ell_n))/(2\ell_n)^2,
\end{split}
\end{equation}
$1\leq i,j\leq3$, and set $A_n\coloneqq A_n(\theta_n)$. Recalling $\hat\sigma_u^2 = \frac1{n} Q_n(\theta_n)$, the following result is obtained.
 
\begin{corollary}\label{cor:SEs}
 If the conditions of Proposition~\textup{\ref{prop:norm}} hold and $\ell_n \rightarrow 0$ as $\sqrt{n}\ell_n\rightarrow \infty$, we have $\Sigma_n \coloneqq \hat\sigma_u^2A_n^{-1}  \rightarrow_p \Sigma$, where convergence in probability holds elementwise.
\end{corollary}
 
The rate imposed on the step size is common in the literature; see, e.g. \citet[Theorem 7.4]{newmc:94}, \citet[Section 2.4]{oh2013simulated} or \citet[Corollary 2]{mm:25}.
 
We can use Corollary~\ref{cor:SEs} directly to test hypotheses about $\theta_0 \in \Theta$ using Wald-type tests. An important exception is the boundary case $\delta = 0$, where $\gamma$ is a non-identified nuisance parameter; this is similar to \cite{andpol:1994} and \cite{hansen:1996}. As a solution, we propose the following procedure similar to \cite{hansen:17} (see also \citealp{mm:25}): Under the null $H_0$: $\delta = 0$, the ALM reduces to $\pi_t = \psi y_t+u_t$, where $\psi$ can be estimated by the OLS estimator $\tilde\psi_n$, say, with residual variance $\tilde\sigma_n^2 = \frac1{n}\sum_{t=1}^n(\pi_t-\tilde\psi_n y_t)^2$. Define $\hat\sigma_n^2(\gamma) \coloneqq \frac1{n}Q_n^\st(\gamma)$ and
$\hat\sigma_n^2 \coloneqq \hat\sigma_n^2(\gamma_n)$, where $Q_n^\st(\cdot)$ is
the objective upon concentrating out $(\delta,\psi)$ and $\gamma_n$ its minimizer (see
Remark~\ref{rem:prof}). The proposed test statistic is the following
{\it sup}{\sf F} statistic:
${sup}{\sf F}\coloneqq\ssup_{\gamma \in \Gamma} F_n(\gamma)$,
$F_n(\gamma) \coloneqq n(\tilde\sigma_n^2-\hat\sigma_n^2(\gamma))/\hat\sigma_n^2(\gamma)$.
Note that $F_n(\gamma)$ is decreasing in $\hat\sigma_n^2(\gamma)$, which by
construction is minimised at $\gamma_n$; the statistic therefore takes the
value ${\it sup}{\sf F} = n(\tilde\sigma_n^2-\hat\sigma_n^2)/\hat\sigma_n^2$. Moreover, by Frisch-Waugh-Lovell, $F_n(\gamma)$ equals the squared $t$-statistic for the exclusion of $h_{t-1}(\gamma)$ in a regression of $\pi_t$ on $(h_{t-1}(\gamma),y_t)^\Tr$.
 
\begin{corollary}\label{cor:sup} Suppose the conditions of Corollary~\textup{\ref{cor:rate}} hold and $H_0:$ $\delta  = 0$, then
\(
{sup}{\sf F} \rightarrow_d \ssup\limits_{\gamma \in \Gamma} {\mathbb U}(\gamma)^2,
\)
where $ {\mathbb U}(\gamma)$ is a Gaussian process with $\cov[{\mathbb U}(\gamma_1),{\mathbb U}(\gamma_2)] = \Ex[h^\st_{1}(\gamma_1)h^\st_{1}(\gamma_2)]/\sqrt{\Ex[h^\st_{1}(\gamma_1)^2]\Ex[h^\st_{1}(\gamma_2)^2]}$,
where $h_t^\star(\gamma)$ denotes the population residual from projecting
$h_t(\gamma)$ on $y_t^\Tr$.
\end{corollary}
 
Another interesting question is how to conduct inference with respect to the equilibria $\beta$. Here, we propose uniform confidence bands for the unknown curve $\beta \mapsto G(\beta)$. To fix ideas, note that, by the functional Delta method, we have in $\ell^\infty([0,1])$, $\sqrt{n}(G(\beta,\lambda_n) - G(\beta,\lambda_0)) \rightsquigarrow {\mathbb G}(\beta),$
where ${\mathbb G}(\beta)$ is a Gaussian process with kernel $\cov[{\mathbb G}(\beta_1),{\mathbb G}(\beta_2)] = F_\lambda(\beta_1)^\Tr\Omega F_\lambda(\beta_2)$. We can therefore compute uniform $1-\alpha$ confidence bands as summarized by the following Corollary:
 
\begin{corollary}\label{cor:CI} For a consistent estimator $s_n^2(\beta)$ of $s^2(\beta) \coloneqq F_\lambda(\beta)^\Tr\Omega F_\lambda(\beta)$ such that $\ssup_\beta|s_n(\beta)-s(\beta)| = o_p(1)$, fix a significance level $\alpha \in (0,1)$ and suppose Assumptions \textup{\ref{ass:density}}, \textup{\ref{ass:para2}}, and \textup{\ref{ass:eight}} hold, then $\PP\{\forall \beta \in [0,1]: G(\beta) \in [G_n(\beta) \pm c_\alpha n^{-1/2}s_n(\beta)]\} \rightarrow 1-\alpha,$ where $c_\alpha$ is such that $\PP\{\ssup\limits_{\beta \in [0,1]} \frac{|{\mathbb G}(\beta)|}{s(\beta)} \leq c_\alpha\} = 1-\alpha$.
\end{corollary}
 
If preferred, a non-studentized, {\it percentile} version of the confidence bands could be obtained by letting $[G_n(\beta) \pm \tilde c_\alpha n^{-1/2}]$, with $\tilde c_{\alpha}$ such that $\PP\{\ssup_{\beta} |{\mathbb G}(\beta)| \leq \tilde c_\alpha\} = 1-\alpha$. Note that, contrary to pointwise confidence intervals at selected roots $\beta$, our confidence bands are constructed for the function $\beta \mapsto G_n(\beta;\lambda_0)$ itself, exploiting smoothness in $\lambda$, and are thus unaffected by potentially repeated roots where $G_\beta$ vanishes.  Since the limiting processes involved in Corollaries \ref{cor:sup} and \ref{cor:CI} depend on unknown nuisance parameters,  Appendix~\ref{app:inf} shows that in both cases, inference can be based on a Gaussian multiplier bootstrap to obtain consistent bootstrap $p$-values and consistent bootstrap critical values, respectively.

\section{Monte Carlo simulation}\label{sec:mc} 

In this section, we present finite-sample evidence based on simulated data from the NKPC~\eqref{eq:seNKPC}.\footnote{Appendix~\ref{sec:addMC} contains additional finite sample evidence in support of the QMLE also for the multivariate model.} We consider two scenarios: Scenario~(A) with three behavioural equilibria ($r_0 = 3$) and Scenario~(B) with two equilibria ($r_0 = 2$). The Monte Carlo dgp is defined by Eqs.~\eqref{eq:seNKPC}, \eqref{eq:learning}, and \eqref{eq:ALM}, with {\sf IID} Gaussian innovations $v_t = (u_t,\varepsilon_t)^\Tr \sim \mathcal N(0,\textup{\sf diag}(\sigma_u^2,\sigma_\varepsilon^2))$. For Scenario~(A), the parameter values are inspired by our empirical analysis. In particular, the parameter vector estimated by NLS is set to $\theta = (0.076,0.998,0.090)^\Tr$, the AR(1) dynamics~\eqref{eq:seNKPC} are parametrised via $(a,\rho) = (-0.02,0.93)$, and, for the innovations, we set $(\sigma_u,\sigma_\varepsilon) = (0.44,0.76)$. This configuration gives rise to three behavioural equilibria $\beta = (0.174,0.856,0.999)^\Tr$. The implied values of $\lambda = (\delta,\psi,\rho,\sigma_u^2,\sigma_\varepsilon^2)^\Tr$ are broadly in line with those in \citet[Sec~4.2]{hz14}. For Scenario~(B), we keep the value of the gain and the AR(1) dynamics, and choose
$\lambda = (0.95,0.2734,0.9,1,1)^\Tr$, which gives rise to a simple root at $\beta = 0.9766$ and a repeated root at $\beta^\dagger = 0.5551$.

\begin{figure}[!h]
\centering
\begin{minipage}[c]{.7\textwidth}
\begin{tikzpicture}
\begin{axis}[
  width=\linewidth,
  height=5cm,
  xlabel={$\beta$},
  ylabel={$G(\beta)$},
  legend pos=south east,
  legend cell align=right,
  legend style={draw=none, at={(rel axis cs:0.9,.05)}, fill=none,font=\footnotesize},
  table/col sep=space, 
  grid style={line width=.1pt, draw=gray!20},
  major grid style={line width=.2pt, draw=gray!40},
  scaled y ticks=false,
  yticklabel style={/pgf/number format/fixed, /pgf/number format/precision=4},
  enlarge x limits=false,
  xticklabels={},ytick={0},
  yticklabels={0},
  xlabel near ticks,
    ylabel near ticks,
  label style={font=\footnotesize},      
    tick label style={font=\scriptsize},   
    xlabel style={yshift=2pt},             
    ylabel style={xshift=2pt},  
    xmin = 0.1, xmax = 1.01,
  ymin = -.11, ymax = .13,
]

\draw[red, line width=.5pt]
  (axis cs:\pgfkeysvalueof{/pgfplots/xmin}, 0)
  -- (axis cs:\pgfkeysvalueof{/pgfplots/xmax}, 0);

Thicker grey dotted vertical lines
\draw[gray, line width=.5pt]
  (axis cs:0.9986868, \pgfkeysvalueof{/pgfplots/ymin})
  -- (axis cs:0.9986868, \pgfkeysvalueof{/pgfplots/ymax});
\draw[gray, line width=.5pt]
  (axis cs:0.9206429, \pgfkeysvalueof{/pgfplots/ymin})
  -- (axis cs:0.9206429, \pgfkeysvalueof{/pgfplots/ymax});
  \draw[gray, line width=.5pt]
  (axis cs:0.1345104, \pgfkeysvalueof{/pgfplots/ymin})
  -- (axis cs:0.1345104, \pgfkeysvalueof{/pgfplots/ymax});
\draw[gray, dashed, line width=.5pt]
  (axis cs:0.5551, \pgfkeysvalueof{/pgfplots/ymin})
  -- (axis cs:0.5551, \pgfkeysvalueof{/pgfplots/ymax});
  \draw[gray, dashed,line width=.5pt]
  (axis cs:.9766 , \pgfkeysvalueof{/pgfplots/ymin})
  -- (axis cs:.9766 , \pgfkeysvalueof{/pgfplots/ymax});
  
\addplot+[smooth, no marks, no marks, thick, black] table[x=betas, y=G1] {Figure2_dat.txt}; \addlegendentry{(A)}
\addplot+[smooth, dashed, no marks, black] table[x=betas, y=G2]  {Figure2_dat.txt}; \addlegendentry{(B)}
\end{axis}
\end{tikzpicture}
\end{minipage}%
  \begin{minipage}[c]{.275\textwidth}\vspace*{-.5cm}
\caption{The solid and dashed lines indicate the true $G(\beta)$ for scenario (A) and (B), respectively. The roots corresponding to $r_0 = 3$ and $r_0=2$ are indicated by grey vertical lines.}\label{fig:mcfun} 
\end{minipage}
\vspace*{-.5cm}
\end{figure} 

\begin{table}[!hp]
\centering
\scriptsize 
\setlength{\tabcolsep}{3pt}
\renewcommand{\arraystretch}{.7}
\resizebox{\linewidth}{!}{
\begin{tabular}{lrcccrrccc}
\midrule
 (A)	&		&	$\gamma =   0.076$	&	$\delta = 0.998$	&	$\psi = 0.090$	&	&		&	$\beta_1 = 0.174$	&	$\beta_2 = 0.856$	&	$\beta_3 = 0.999$	 \\
\midrule																
250	&	mean	&	0.0754	&	0.9696	&	0.0929	&	&	$\PP\{r=1\}$	&		&	{\bf 0.4085}	&	\\
	&	bias	&	-0.0006	&	-0.0285	&	0.0029	&	&	mean	&	&	0.2250	&		\\
	&	sd	&	0.0171	&	0.0551	&	0.0160	&	&	sd	&		&	0.2911	&	\\[-2pt]
\cmidrule(r){8-10}	\\[-8.4pt]																	
	&	size	&	0.0595	&	0.0415	&	0.0525	&	&	$\PP\{r=3\}$	&		&	{\bf 0.5915 }	&		\\
	&	cover	&	0.9335	&		&		&	&	mean	&	0.1924	&	0.8516	&	0.9961	\\
	&	cover-s	&	0.9085	&		&		&	&	sd	&	0.0832	&	0.0929	&	0.0086	\\[-2pt]
\cmidrule(r){1-5}	\cmidrule(r){7-10}																	
500	&	mean	&	0.0760	&	0.9860	&	0.0911	&	&	$\PP\{r=1\}$	&	&	 {\bf 0.2310}	&		\\
	&	bias	&	0.0000	&	-0.0121	&	0.0011	&	&	mean	&	&	0.1967	&		\\
	&	sd	&	0.0123	&	0.0240	&	0.0110	&	&	sd	&		&	0.2434	&	\\[-2pt]
\cmidrule(r){8-10}	\\[-8.4pt]																	
	&	size	&	0.0565	&	0.0440	&	0.0500	&	&	$\PP\{r=3\}$	&		&	{\bf 0.7690}	&		\\
	&	cover	&	0.9290	&		&		&	&	mean	&	0.1863	&	0.8547	&	0.9963	\\
	&	cover-s	&	0.9280	&		&		&	&	sd	&	0.0692	&	0.0802	&	0.0072	\\[-2pt]
\cmidrule(r){1-5}	\cmidrule(r){7-10}																	
1,000	&	mean	&	0.0764	&	0.9935	&	0.0905	&	&	$\PP\{r=1\}$	&		&	{\bf 0.0695}	&		\\
	&	bias	&	0.0004	&	-0.0046	&	0.0005	&	&	mean	&		&	0.1229	&		\\
	&	sd	&	0.0086	&	0.0104	&	0.0077	&	&	sd	&		&	0.0292	&		\\[-2pt]
\cmidrule(r){8-10}	\\[-8.4pt]																	
	&	size	&	0.0455	&	0.0285	&	0.0495	&	&	$\PP\{r=3\}$	&		&	{\bf 0.9305}	&		\\
	&	cover	&	0.9350	&		&		&	&	mean	&	0.1778	&	0.8598	&	0.9972	\\
	&	cover-s	&	0.9490	&		&		&	&	sd	&	0.0494	&	0.0600	&	0.0050	\\[-2pt]
\cmidrule(r){1-5}	\cmidrule(r){7-10}																	
2,000	&	mean	&	0.0758	&	0.9962	&	0.0904	&	&	$\PP\{r=1\}$	&	&	{\bf 0.0120}	&		\\
	&	bias	&	-0.0002	&	-0.0018	&	0.0004	&	&	mean	&		&	0.1298	&		\\
	&	sd	&	0.0060	&	0.0048	&	0.0054	&	&	sd	&		&	0.0224	&		\\[-2pt]
\cmidrule(r){8-10}																		
	&	size	&	0.0520	&	0.0345	&	0.0480	&	&	$\PP\{r=3\}$	&		&	\textbf{0.9880}	&		\\
	&	cover	&	0.9330	&		&		&	&	mean	&	0.1759	&	0.8582	&	0.9983	\\
	&	cover-s	&	0.9505	&		&		&	&	sd	&	0.0357	&	0.0441	&	0.0030	\\
\midrule
(B)	&		&	$\gamma = 0.076$	&	$\delta = 0.95 $	&	$\psi = 0.2734 $	&	&		&	$\beta^\dagger$	&	$\beta^\dagger=0.55$	&	$\beta=0.977$	 \\
\midrule				
250	&	mean	&	0.0759	&	0.9361	&	0.2772	&	&	$\PP\{r=1\}$	&		&	{\bf 0.7550}	&	\\
	&	bias	&	-0.0001	&	-0.0139	&	0.0039	&	&	mean	&		&	0.6382	&		\\
	&	sd	&	0.0156	&	0.0507	&	0.0310	&	&	sd	&		&	0.3632	&		\\[-2pt]
	\cmidrule(r){8-10}	\\[-8.4pt]																
	&	size	&	0.0540	&	0.0450	&	0.0460	&	&	$\PP\{r=3\}$	&		&	{\bf 0.2450}	&		\\
	&	cover	&	0.9270	&		&		&	&	mean	&	0.3429	&	0.7671	&	0.9755	\\
	&	cover-s	&	0.9125	&		&		&	&	sd	&	0.0954	&	0.0906	&	0.0242	\\
   	&	 	&		&		&		&	&		&	\multicolumn{2}{c}{$\textstyle{\frac{1}{2}}(\beta_1+\beta_2)$}		&		 \\[-1pt]    \cmidrule(r){8-9} 		
	&		&		&		&		&	&	mean	&	\multicolumn{2}{c}{0.5550}		&		\\
	&		&		&		&		&	&	sd	&	\multicolumn{2}{c}{0.0255}		&		\\[-2pt]
	\cmidrule(r){1-5}	\cmidrule(r){7-10}																
500	&	mean	&	0.0762	&	0.9438	&	0.2746	&	&	$\PP\{r=1\}$	&		&	{\bf0.6525}	&		\\
	&	bias	&	0.0002	&	-0.0062	&	0.0012	&	&	mean	&		&	0.7166	&		\\
	&	sd	&	0.0113	&	0.0261	&	0.0216	&	&	sd	&		&	0.3428	&		\\[-2pt]
	\cmidrule(r){8-10}																	
	&	size	&	0.0520	&	0.0490	&	0.0515	&	&	$\PP\{r=3\}$	&	&	{\bf0.3475}	&		\\
	&	cover	&	0.9255	&		&		&	&	mean	&	0.3764	&	0.7572	&	0.9641	\\
	&	cover-s	&	0.9205	&		&		&	&	sd	&	0.0771	&	0.0804	&	0.0239	\\
   	&	 	&		&		&		&	&		&	\multicolumn{2}{c}{$\textstyle{\frac{1}{2}}(\beta_1+\beta_2)$}		&		 \\[-1pt]    \cmidrule(r){8-9} 		
	&		&		&		&		&	&	mean	&	\multicolumn{2}{c}{0.5668}		&		\\
	&		&		&		&		&	&	sd	&	\multicolumn{2}{c}{0.0236}		&		\\[-2pt]
	\cmidrule(r){1-5}	\cmidrule(r){7-10}																
1,000	&	mean	&	0.0766	&	0.9477	&	0.2737	&	&	$\PP\{r=1\}$	&		&	{\bf0.5615}	&		\\
	&	bias	&	0.0006	&	-0.0023	&	0.0004	&	&	mean	&		&	0.8238	&		\\
	&	sd	&	0.0077	&	0.0138	&	0.0152	&	&	sd	&		&	0.2913	&		\\[-2pt]
	\cmidrule(r){8-10}																	
	&	size	&	0.0470	&	0.0400	&	0.0445	&	&	$\PP\{r=3\}$	&		&	{\bf0.4385}	&		\\
	&	cover	&	0.9385	&		&		&	&	mean	&	0.3893	&	0.7468	&	0.9637	\\
	&	cover-s	&	0.9430	&		&		&	&	sd	&	0.0680	&	0.0793	&	0.0168	\\
   	&	 	&		&		&		&	&		&	\multicolumn{2}{c}{$\textstyle{\frac{1}{2}}(\beta_1+\beta_2)$}		&		 \\[-1pt]    \cmidrule(r){8-9} 		
	&		&		&		&		&	&	mean	&	\multicolumn{2}{c}{0.5680}		&		\\
	&		&		&		&		&	&	sd	&	\multicolumn{2}{c}{0.0166}		&		\\[-2pt]
	\cmidrule(r){1-5}	\cmidrule(r){7-10}																
2,000	&	mean	&	0.0760	&	0.9487	&	0.2739	&	&	$\PP\{r=1\}$	&		&	{\bf0.5015}	&		\\
	&	bias	&	0.0000	&	-0.0013	&	0.0006	&	&	mean	&		&	0.9253	&		\\
	&	sd	&	0.0054	&	0.0087	&	0.0105	&	&	sd	&		&	0.1863	&		\\[-2pt]
	\cmidrule(r){8-10}																	
	&	size	&	0.0585	&	0.0445	&	0.0480	&	&	$\PP\{r=3\}$	&		&	{\bf0.4985}	&		\\
	&	cover	&	0.9415	&		&		&	&	mean	&	0.4066	&	0.7263	&	0.9659	\\
	&	cover-s	&	0.9425	&		&		&	&	sd	&	0.0608	&	0.0752	&	0.0136	\\
   	&	 	&		&		&		&	&		&	\multicolumn{2}{c}{$\textstyle{\frac{1}{2}}(\beta_1+\beta_2)$}		&		 \\[-1pt]    \cmidrule(r){8-9} 		
	&		&		&		&		&	&	mean	&	\multicolumn{2}{c}{0.5664}		&		\\
	&		&		&		&		&	&	sd	&	\multicolumn{2}{c}{0.0133}		&		\\		
    \bottomrule
\end{tabular}
}\caption{Simulation results for scenario (A) in the upper panel with $r_0 =3$ roots and for scenario (B) in the lower panel with $r_0=2$ roots  based on 2,000 Monte Carlo repetitions. }\label{tab:MC}
\end{table}
We report, for the NLS estimator $\theta_n$ of $\theta = (\gamma,\delta,\psi)^{\Tr}$, the mean, bias, standard deviation, and the empirical size of two-sided $t$-tests (using, in accordance with Corollary~\ref{cor:SEs}, numerical standard errors) at a nominal 5\% level. We also present the empirical coverage of the percentile ({\it cover}) and studentized ({\it cover-s}) confidence bands (see  Corollary~\ref{cor:CI}) for the curves $\beta \mapsto G(\beta)$ plotted in Figure~\ref{fig:mcfun}, based on $B = 4{,}999$ bootstrap replications at a 95\% coverage level. Finally, we report the mean and standard deviation of the estimated equilibria, together with the observed frequency at which the true number of roots is recovered, i.e.\ $\mathbb{P}\{r = r_0\}$.

The findings are summarized in Table~\ref{tab:MC}. In both scenarios, the NLS estimator appears consistent and the normal approximation for the $t$-statistics is fairly accurate, while the empirical coverage of the studentized confidence bands is slightly below the nominal 95\% level for smaller $n$. Moreover, in line with Corollary~\ref{cor:roots}, we observe that the estimated number of roots $r_n$ either converges to the true value $r_0$ or, in Scenario~(B), oscillates between $r = 1$ and $r = 3$ with approximately equal probability. In the latter case, the convergence of the estimates associated with the repeated root appears to be an order of magnitude slower than the $\sqrt{n}$-rate observed for simple roots, while averaging the two nearby estimates corresponding to the repeated root improves performance.

\section{Empirical application}\label{sec:emp}

To illustrate the empirical relevance of our theoretical results, we estimate the scalar
NKPC using quarterly US data from 1960:Q1 to
2019:Q4 ($n=240$). Inflation $\pi_t$
is measured by the CPI growth rate. Since marginal costs
are proportional to the output gap or the real labour share \citetext{see, e.g., \citealp{Woodford2003} or \citealp{MavroeidisPlagborgMollerStock2014}, Section 2.1}, we consider two standard slack
measures: {\sf A} the output gap measure from the Congressional Budget Office and
{\sf B} the percentage change in real unit labour costs \citetext{see also \citealp{gali1999inflation} or \citealp{sbordone2002prices}}.\footnote{Figure \ref{fig:empraw} in Appendix \ref{app:empirical} displays these data.} Table \ref{tab:empAB} reports full-sample estimates, together with 95\% confidence intervals, for the NKPC (upper panel) and the AR(1) dynamics (lower panel) of the slack variable. 

\begin{table}[ht]
\centering\small
\begin{tabular}{l cc cc}
\toprule
 & \multicolumn{2}{c}{\sf A} & \multicolumn{2}{c}{\sf B} \\ 
& Baseline & Hybrid &Baseline & Hybrid \\ \midrule
$\gamma$ & 0.0761 & 0.07780 & 0.0463  & 0.0552 \\[-5pt]
& {\scriptsize [0.073,\ 0.079]} & {\scriptsize [0.075,\ 0.081]} & {\scriptsize [0.045,\ 0.048]} & {\scriptsize [0.053,\ 0.057]} \\
$\delta$ & 0.9986 & 0.7858 & 0.9017 & 0.6183 \\[-5pt]
& {\scriptsize [0.998,\ 0.999]} & {\scriptsize [0.785,\ 0.786]} & {\scriptsize [0.897,\ 0.906]} & {\scriptsize [0.607,\ 0.629]} \\
$\psi$ & 0.0890 & 0.0761 & 0.1259 &  0.0924 \\[-5pt]
&  {\scriptsize [0.087,\ 0.091]} & {\scriptsize [0.074,\ 0.078]} & {\scriptsize [0.122,\ 0.130]} & {\scriptsize [0.088,\ 0.096]}  \\
$\varphi$ & -- & 0.2134 & -- & 0.2957  \\[-5pt]    
& {\scriptsize } & {\scriptsize [0.212,\ 0.214]} &{\scriptsize } & {\scriptsize [0.285,\ 0.306]} \\
$\sigma_u$ & 0.4439 & 0.4373 & 0.4673 & 0.4559 \\[-2pt]
\cmidrule(r){2-5}
$a$ & \multicolumn{2}{c}{ -0.0259}  & \multicolumn{2}{c}{0.5661}  \\[-5pt] & \multicolumn{2}{c}{\scriptsize [-0.125,\ 0.073]}  &\multicolumn{2}{c}{\scriptsize [0.401,\ 0.731]} \\
$\rho$ & \multicolumn{2}{c}{0.9376}  & \multicolumn{2}{c}{0.1575} \\[-5pt]
& \multicolumn{2}{c}{\scriptsize [0.893,\ 0.982]} & \multicolumn{2}{c}{\scriptsize [0.031 ,\ 0.283]} \\
$\sigma_\eps$ & \multicolumn{2}{c}{0.7614} & \multicolumn{2}{c}{1.1101}  \\
\bottomrule
\end{tabular} 
\caption{Point estimates and 95\% confidence intervals based on numerical standard errors. Columns 3 and 5 refer to the hybrid version of the model, with lagged inflation.}\label{tab:empAB}
\end{table}

 The NLS estimates are broadly consistent with the empirical macroeconomic literature \citetext{see, among others, \citealp{milani:07}, \citealp{chev:10}, \citealp{Lansing2009}, and \citealp{HommesMavromatisOzdenZhu2023}}. The coefficient on expected inflation is large, while the Phillips-curve slope is positive, significant, and relatively small. Adding lagged inflation ($\varphi$) leaves the estimated gain and slope broadly unchanged but lowers the coefficient on expected inflation. This is unsurprising because the AR(1) forecasting rule already incorporates lagged inflation dynamics, so the two terms partly capture the same persistence.
 
\begin{figure}[!h]
\centering
\begin{minipage}[c]{.675\textwidth}
\begin{tikzpicture}
  \begin{axis}[
    width=10cm,
    height=5.25cm,
    xmin=0, xmax=1,
    axis lines=box,
    xlabel={},
    ylabel={},
    grid=none,
    table/col sep=tab,
    tick label style={font=\scriptsize},
    legend style={
      font=\scriptsize,
      draw=none,
      fill=none,
      at={(0.97,0.03)},
      anchor=south east
    },
    legend cell align=left
  ]

    \addplot[gray, dashed, forget plot]
      coordinates {(0,0) (1,0)};

    \addplot[name path=G1up, draw=none, forget plot]
      table[x=grid, y=G1up]{Figure3_dat.txt};

    \addplot[name path=G1low, draw=none, forget plot]
      table[x=grid, y=G1low]{Figure3_dat.txt};

    \addplot[ 
      fill=pink,
      fill opacity=0.25,
      draw=none,
      forget plot
    ]
      fill between[of=G1low and G1up];

    \addplot[name path=G2up, draw=none, forget plot]
      table[x=grid, y=G2up]{Figure3_dat.txt};

    \addplot[name path=G2low, draw=none, forget plot]
      table[x=grid, y=G2low]{Figure3_dat.txt};

    \addplot[
      fill=blue,
      fill opacity=0.1,
      draw=none,
      forget plot
    ]
      fill between[of=G2low and G2up];

    \addplot[pink, thick]
      table[x=grid, y=G1]{Figure3_dat.txt};

    \addplot[blue,  thick]
      table[x=grid, y=G2]{Figure3_dat.txt};

    \legend{{unit labour costs}, { output gap}}

  \end{axis}
\end{tikzpicture}
\end{minipage}%
  \begin{minipage}[c]{.3\textwidth}\vspace*{-.1cm}
\caption{Plot of empirical fixed point functions $\beta \mapsto G_n(\beta)$ together with 90\% confidence bands based on $B=$ 4,999 bootstrap replications.}\label{fig:empci}
\end{minipage}
\end{figure} 
Figure~\ref{fig:empci} displays the corresponding $\beta \mapsto G_n(\beta)$ map delivering the behavioural equilibria. In the baseline specification {\sf A}, the estimated parameters imply three behavioural equilibria $\beta_1 = 0.1914$, $\beta_2 = 0.8316$, $\beta_3=	0.9995$; in specification {\sf B}  we find a unique equilibrium $\beta = 0.01341$. The difference between the two specifications is primarily driven by the persistence of the slack variable. The estimated autoregressive coefficient $\rho$ is close to one for the output gap but considerably smaller for unit labour costs. This is consistent with the theoretical property that, holding the remaining parameters fixed, $\rho\rightarrow0$ implies a unique equilibrium with $\beta \rightarrow 0$. The lower persistence of unit labour costs therefore weakens the feedback between expectations and realized inflation and rules out a high-persistence belief regime. Under Specification~{\sf A}, the intermediate solution $\widehat{\beta}_2$ is unstable according to the learning criterion of \citet[Proposition~4]{hz14}: an agent who asymptotically uses the entire history of the data, corresponding to $\gamma \rightarrow 0$, cannot learn this equilibrium (see also Remark~\ref{rem:smallgam1}). The estimates thus identify two stable belief regimes, characterized by low and high perceived inflation persistence.

Akin to \citet{JorgensenLansing2025}, $\beta$ plays a similar role to their Kalman gain as an anchoring-measure of expectations.
This interpretation is also consistent with the broader literature on information rigidities, which studies expectation formation through the responsiveness of forecast
revisions to incoming information and forecast errors \citetext{see, e.g., \citealp{CoibionGorodnichenko2015} and \citealp{CoibionGorodnichenkoKamdar2018}}. In our model, $\beta$ is not the Kalman gain but measures the persistence agents attribute to deviations of inflation from its estimated mean: such deviations enter the forecasting rule Eq.~\eqref{eq:iALM} with coefficient $\beta^2$. Hence, low values of $|\beta|$ imply only limited effects on subsequent forecasts,
whereas high values imply persistent effects. In this sense, following \citet{JorgensenLansing2025b} and \citet{kosh:forth}, we interpret $|\beta|$ as an inverse measure of short-run anchoring around a fixed long-run inflation mean. The recursively estimated path $\beta_t$ therefore provides a
time-varying indicator of anchoring. Our evidence suggests relatively de-anchored expectations up to around 2000 and stronger anchoring thereafter. This pattern closely resembles that reported by \citet{Lansing2009} and \citet{JorgensenLansing2025b} and is broadly consistent with the survey-based
anchoring measure of \citet{naggert2023anchoring}, as shown in Figure \ref{fig:BetaAnchoring}, Appendix \ref{app:empirical}. In terms of the behavioural equilibria, it is consistent with the high-$\beta$ equilibrium being more relevant prior to the early 2000s and the low-$\beta$ equilibrium thereafter. 

As an external assessment of the learning mechanism, Figure~\ref{fig:InflPred} compares the model implied forecasts with the one-year-ahead SPF consensus forecasts for core inflation, available from 1980Q1 onward \citep{FedPhil2024}. Although survey expectations are not used in estimation, the model generated forecasts closely track the SPF series under both measures of slack. Furthermore, the model-implied forecast errors display an autocorrelation structure similar to that observed in the SPF data. This close correspondence is consistent with the view advocated by \citet{CoibionGorodnichenkoKamdar2018} that survey expectations constitute an informative benchmark for evaluating alternative models of expectation formation. As an additional robustness exercise, we replace model-implied expectations with SPF forecasts and re-estimate the Phillips curve. The resulting estimates are broadly similar to those reported in Table~\ref{tab:empAB}. Taken together, these findings suggest that the learning recursion provides a parsimonious approximation to observed expectations and that the structural estimates are not driven by idiosyncratic features of the assumed updating mechanism.

\begin{figure}[tb]
\centering
\begin{minipage}[c]{.7\textwidth}
\begin{tikzpicture}
\begin{axis}[
  width=10cm,
  height=6cm,
  xmin=1960,
  xmax=2020,
  axis lines=box,
  xtick={1960,1970,1980,1990,2000,2010,2020},
  xticklabel style={
    /pgf/number format/fixed,
    /pgf/number format/precision=0,
    /pgf/number format/set thousands separator={}
  },
  xlabel={},
  ylabel={$\pi$},
      tick label style={font=\scriptsize},
  grid=none,
  table/col sep=space,
  legend columns=1,
  legend image post style={scale=0.8},
  legend cell align={left},
  legend style={
    at={(0.02,0.02)},
    anchor=south west,
    draw=none,
    fill=white,
    fill opacity=0.75,
    text opacity=1,
    font=\tiny,
    column sep=0.2cm,
    row sep=0.02cm
  },
]

\addplot+[no marks, dotted, black] table[
  x expr=1960 + \coordindex/4,
  y=pi
] {Figure4_dat.txt};
\addlegendentry{{ Actual inflation}}

\addplot+[no marks, thick, green] table[
  x expr=1960 + \coordindex/4,
  y=fore_ulc
] {Figure4_dat.txt};
\addlegendentry{{ Model based forecasts (ulc)}}

\addplot+[no marks, thick, blue] table[
  x expr=1960 + \coordindex/4,
  y=fore_og
] {Figure4_dat.txt};
\addlegendentry{{ Model based forecasts (output gap)}}

\addplot+[no marks, thick, red] table[
  x expr=1960 + \coordindex/4,
  y=SPF
] {Figure4_dat.txt};
\addlegendentry{{ Survey of Professional Forecasters (Core Inflation)}}

\end{axis}
\end{tikzpicture}
\end{minipage}%
  \begin{minipage}[c]{.275\textwidth}
\caption{Comparison between observed inflation (dotted black), model-based predictions based on different measures of slack (blue and green) and consensus forecast obtained from the SPF on core inflation (red).}
\label{fig:InflPred}
\end{minipage}
\end{figure} 

Finally, to assess parameter stability, we follow \citet{JorgensenLansing2025} and split the sample at the first quarter in 1999, when  \citet{JorgensenLansing2025} argue that U.S. inflation expectations had become anchored. For both measures of slack, the estimated gain falls sharply to around 0.01 in the second subsample, consistent with substantially slower belief updating. The output-gap slope remains fairly stable, declining from 0.094 to 0.085, whereas the unit labor cost slope falls from about 0.13 to 0.035. The latter decline is therefore much more pronounced, although the relatively small subsamples warrant caution.

\section{Conclusion}\label{sec:conc}
 This paper  develops estimation and inference methods for a New Keynesian model with constant-gain learning and potentially multiple behavioural equilibria. Under mild conditions, the model is geometrically ergodic, while the structural parameters can be consistently estimated. In the scalar case we also derived distributional theory. These techniques are applied to U.S.\ inflation data, revealing multiple belief regimes in the data.
 
 Several extensions are left for future work: A natural next step is to extend the distributional theory also to the general model. It would also be useful to 
relax some of our regularity conditions, to exploit panel data to increase
effective sample size, and to adapt the analysis to decreasing-gain learning
rules using results from earlier work on the econometrics of models with adaptive learning like \citealp{chev:10}, \cite{chev:17}, \cite{chrismass:18}, \cite{mayer:22}, or \cite{mm:25}. Finally, as emphasized by \citet{preston2005}, replacing expectations by
forecasts from agents' PLM in the one-period equilibrium conditions is
equivalent to their multiperiod decision problem only under rational
expectations, so that allowing decisions to depend on long-horizon forecasts is
a further natural extension.

\addcontentsline{toc}{section}{References}
{\singlespacing 
\bibliography{bibl}

@article{naggert2023anchoring,
  title={The anchoring of US inflation expectations since 2012},
  author={Naggert, Kristoph and Rich, Robert W and Tracy, Joseph},
  journal={Federal Reserve Bank of Cleveland, Economic Commentary 2023-11},
  number={2023-11},
  year={2023},
  publisher={Federal Reserve Bank of Cleveland}
}

@article{JorgensenLansing2025b,
  title={A simple measure of anchoring for short-run expected inflation in FIRE models},
  author={J{\o}rgensen, Peter Lihn and Lansing, Kevin J},
  journal={Economics Letters},
  volume={246},
  pages={112050},
  year={2025}
}

@article{linde2005,
	author = {Jesper Lind{\'e}},
	journal = {Journal of Monetary Economics},
	pages = {1135-1149},
	title = {Estimating New-Keynesian Phillips curves: A full information maximum likelihood approach},
	volume = {52},
	year = {2005}
}

@article{markiewicz2014adaptive,
  title={Adaptive learning and survey data},
  author={Markiewicz, Agnieszka and Pick, Andreas},
  journal={Journal of Economic Behavior \& Organization},
  volume={107},
  pages={685--707},
  year={2014}
}

@article{cox:81,
 author = {D. R. Cox and Gudmundur Gudmundsson and Georg Lindgren and Lennart Bondesson and Erik Harsaae and Petter Laake and Katarina Juselius and Steffen L. Lauritzen},
 journal = {Scandinavian Journal of Statistics},
 number = {2},
 pages = {93--115},
 title = {Statistical Analysis of Time Series: Some Recent Developments [with Discussion and Reply]},
 urldate = {2025-12-02},
 volume = {8},
 year = {1981}
}

@article{kasy:15,
  title={Non-parametric inference on the number of equilibria},
  author={Kasy, Maximilian},
  journal={The Econometrics Journal},
  volume={18},
  number={1},
  pages={1--39},
  year={2015},
  publisher={Oxford University Press Oxford, UK}
}

@article{blanchard:16,
  title={{The Phillips Curve: Back to the '60s?}},
  author={O. Blanchard},
  journal={American Economic Review},
  volume={106},
  pages={31-34},
  year={2016}
}

@article{francq:04,
  title={Maximum likelihood estimation of pure GARCH and ARMA-GARCH processes},
  author={Francq, Christian and Zakoïan, Jean-Michel},
  journal={Bernoulli},
  volume={10},
  pages={605--637},
  year={2004}
}

@article{dovonon:18,
  title={The asymptotic properties of GMM and indirect inference under second-order identification},
  author={Dovonon, Prosper and Hall, Alastair R},
  journal={Journal of Econometrics},
  volume={205},
  number={1},
  pages={76--111},
  year={2018},
  publisher={Elsevier}
}

@article{bougerol:93,
author = {Bougerol, Philippe},
title = {Kalman Filtering with Random Coefficients and Contractions},
journal = {SIAM Journal on Control and Optimization},
volume = {31},
pages = {942-959},
year = {1993}
}

@article{sbordone2002prices,
  title={Prices and unit labor costs: a new test of price stickiness},
  author={Sbordone, Argia M},
  journal={Journal of Monetary Economics},
  volume={49},
  number={2},
  pages={265--292},
  year={2002},
  publisher={Elsevier}
}

@article{gali1999inflation,
  title={Inflation dynamics: A structural econometric analysis},
  author={Galí, Jordi and Gertler, Mark},
  journal={Journal of Monetary Economics},
  volume={44},
  number={2},
  pages={195--222},
  year={1999}
}

@book{jungers2009joint,
  title={The joint spectral radius: theory and applications},
  author={Jungers, Rapha{\"e}l},
  volume={385},
  year={2009},
  publisher={Springer Science \& Business Media}
}

@article{Kheifets20042020,
author = {Igor L. Kheifets and Pentti J. Saikkonen},
title = {Stationarity and ergodicity of vector STAR models},
journal = {Econometric Reviews},
volume = {39},
pages = {407--414},
year = {2020}
}

@article{adam2007experimental,
  title={Experimental evidence on the persistence of output and inflation},
  author={Adam, Klaus},
  journal={The Economic Journal},
  volume={117},
  number={520},
  pages={603--636},
  year={2007}
}

@article{clarida2000monetary,
  title={Monetary policy rules and macroeconomic stability: evidence and some theory},
  author={Clarida, Richard and Gali, Jordi and Gertler, Mark},
  journal={The Quarterly Journal of Economics},
  volume={115},
  pages={147--180},
  year={2000}
}

@article{evans1993adaptive,
  title={Adaptive forecasts, hysteresis, and endogenous fluctuations},
  author={Evans, George W and Honkapohja, Seppo},
  journal={Economic Review-Federal Reserve Bank of San Francisco},
  volume={3},
  pages={3-13},
  year={1993}
}

@book{davidson1993estimation,
  title={Estimation and Inference in Econometrics},
  author={Davidson, Russell and MacKinnon, James G},
  year={1993},
  publisher={Oxford University Press}
}

@book{straumann2005estimation,
  title={Estimation in Conditionally Heteroscedastic Time Series Models},
  author={Straumann, Daniel},
  year={2005},
  publisher={Springer}
}

@article{hausman1983identification,
  title={Identification in linear simultaneous equations models with covariance restrictions: An instrumental variables interpretation},
  author={Hausman, Jerry A and Taylor, William E},
  journal={Econometrica: Journal of the Econometric Society},
  pages={1527--1549},
  year={1983},
}

@article{hausman1987efficient,
  title={Efficient estimation and identification of simultaneous equation models with covariance restrictions},
  author={Hausman, Jerry A and Newey, Whitney K and Taylor, William E},
  journal={Econometrica: Journal of the Econometric Society},
  pages={849--874},
  year={1987}
}

@article{liebscher2005towards,
  title={Towards a unified approach for proving geometric ergodicity and mixing properties of nonlinear autoregressive processes},
  author={Liebscher, Eckhard},
  journal={Journal of Time Series Analysis},
  volume={26},
  pages={669--689},
  year={2005}
}

@article{gabaix20,
  author  = {Gabaix, Xavier},
  title   = {A Behavioral New {K}eynesian Model},
  journal = {American Economic Review},
  year    = {2020},
  volume  = {110},
  pages   = {2271--2327}
}

@article{mns17,
  title={The power of forward guidance revisited},
  author={McKay, Alisdair and Nakamura, Emi and Steinsson, J{\'o}n},
  journal={American Economic Review},
  volume={106},
  pages={3133--3158},
  year={2016}
}

@book{meyn:12,
  title={Markov Chains and Stochastic Stability},
  author={Meyn, Sean P and Tweedie, Richard L},
  year={2012},
  publisher={Springer}
}

@article{hajdini2026,
  title={Mis-specified forecasts and myopia in an estimated new Keynesian model},
  author={Hajdini, Ina},
  journal={American Economic Journal: Macroeconomics},
  volume={18},
  number={2},
  pages={188--226},
  year={2026}
}

@article{hommes2021behavioral,
  title={Behavioral and experimental macroeconomics and policy analysis: A complex systems approach},
  author={Hommes, Cars},
  journal={Journal of Economic Literature},
  volume={59},
  number={1},
  pages={149--219},
  year={2021}
}

@article{hz14,
title = {Behavioral learning equilibria},
journal = {Journal of Economic Theory},
volume = {150},
pages = {778-814},
year = {2014},
author = {Cars Hommes and Mei Zhu}
}

@article{oh2013simulated,
  title={Simulated method of moments estimation for copula-based multivariate models},
  author={Oh, Dong Hwan and Patton, Andrew J},
  journal={Journal of the American Statistical Association},
  volume={108},
  pages={689--700},
  year={2013},
  publisher={Taylor \& Francis}
}

@article{blasques:24,
title = {Autoregressive conditional betas},
journal = {Journal of Econometrics},
volume = {238},
number = {2},
pages = {105630},
year = {2024},
author = {F. Blasques and Christian Francq and Sébastien Laurent}
}

@article{christiano:24,
  title={Slow Learning},
  author={Christiano, Lawrence and Eichenbaum, Martin S and Johannsen, Benjamin K},
  year={2024},
  journal={Working Paper, National Bureau of Economic Research, 32358}
}

@article{seo:11,
  title={Estimation of nonlinear error correction models},
  author={Seo, Myung Hwan},
  journal={Econometric Theory},
  volume={27},
  pages={201--234},
  year={2011}
}

@article{saikkonen:95,
  title={Problems with the asymptotic theory of maximum likelihood estimation in integrated and cointegrated systems},
  author={Saikkonen, Pentti},
  journal={Econometric Theory},
  volume={11},
  pages={888--911},
  year={1995}
}

@article{strau:06,
author = {Daniel Straumann and Thomas Mikosch},
title = {Quasi-maximum-likelihood estimation in conditionally heteroscedastic time series: A stochastic recurrence equations approach},
volume = {34},
journal = {The Annals of Statistics},
pages = {2449--2495},
year = {2006}
}

@article{blasques:22,
  title={Maximum likelihood estimation for score-driven models},
  author={Blasques, Francisco and van Brummelen, Janneke and Koopman, Siem Jan and Lucas, Andre},
  journal={Journal of Econometrics},
  volume={227},
  pages={325--346},
  year={2022}
}

@book{amemiya1985advanced,
  title={Advanced Econometrics},
  author={Amemiya, Takeshi},
  year={1985},
  publisher={Harvard University Press}
}

@article{chev:17,
title = {Learning can generate long memory},
journal = {Journal of Econometrics},
volume = {198},
pages = {1-9},
year = {2017},
author = {Guillaume Chevillon and Sophocles Mavroeidis}
}

@book{bachmann:22,
  title={Handbook of Economic Expectations},
  author={Bachmann, Ruediger and Topa, Giorgio and van der Klaauw, Wilbert},
  year={2023},
  publisher={Elsevier}
}

@article{davide:25,
  title={Multiple Equilibria and the Phillips Curve: Do Agents Always Underreact?},
  author={Casarin, Roberto and Peruzzi, Antonio and Raggi, Davide},
  journal={Ca'Foscari University of Venice, Department of Economics Research Paper Series, N. 10/WP/2025},
  year={2025}
}

@article{mm:25,
  title={Least squares estimation in nonstationary nonlinear cohort panels with learning from experience},
  author={Mayer, Alexander and Massmann, Michael},
  journal={Journal of Business \& Economics Statistics},
  year={2025}
}

@article{kosh:forth,
author = {Kostyshyna, Olena and Salle, Isabelle and Truong, Hung},
title = {Anchored or Adrift? A Note on Measuring Inflation Expectations Anchoring},
journal = {Oxford Bulletin of Economics and Statistics},
year={2026},
publisher={Wiley Online Library}
}

@article{milani:07,
title = {Expectations, learning and macroeconomic persistence},
journal = {Journal of Monetary Economics},
volume = {54},
pages = {2065-2082},
year = {2007},
author = {Fabio Milani}
}

@article{adam:16,
author = {Adam, Klaus and Marcet, Albert and Nicolini, Juan Pablo},
title = {Stock market volatility and learning},
journal = {The Journal of Finance},
volume = {71},
pages = {33-82},
year = {2016}
}

@incollection{newmc:94,
title = {Large sample estimation and hypothesis testing},
booktitle = {Handbook of Econometrics},
chapter = 36,
publisher = {Elsevier},
volume = {4},
pages = {2111-2245},
year = {1994},
author = {Whitney K. Newey and Daniel McFadden}
}

@article{mayer:23,
title = {Two-step estimation in linear regressions with adaptive learning},
journal = {Statistics \& Probability Letters},
volume = {195},
year = {2023},
author = {Alexander Mayer}
}

@book{ben:90,
  title={Adaptive Algorithms and Stochastic Approximations},
  author={Benveniste, Albert and M{\'e}tivier, Michel and Priouret, Pierre},
  year={1990},
  publisher={Springer}
}

@article{chev:10,
  title={Inference in models with adaptive learning},
  author={Chevillon, Guillaume and Massmann, Michael and Mavroeidis, Sophocles},
  journal={Journal of Monetary Economics},
  volume={57},
  pages={341--351},
  year={2010}
}

@article{chot:19,
  title={Verifiable conditions for the irreducibility
and aperiodicity of {M}arkov chains by
analyzing underlying deterministic models},
  author={Chotard, Alexandre and Auger, Anne},
  journal={Bernoulli},
  volume={25},
  pages={112--147},
  year={2019}
}

@article{hansen:1996b,
  title={Stochastic equicontinuity for unbounded dependent heterogeneous arrays},
  author={Hansen, Bruce E},
  journal={Econometric Theory},
  volume={12},
  pages={347--359},
  year={1996}
}

@article{mayer:22,
  title={Estimation and inference in adaptive learning models with slowly decreasing gains},
  author={Mayer, Alexander},
  journal={Journal of Time Series Analysis},
  volume={43},
   pages={720--749},
  year={2022}
}

@article{chrismass:18,
  title={Estimating structural parameters in regression models with adaptive learning},
  author={Christopeit, Norbert and Massmann, Michael},
  journal={Econometric Theory},
  volume={34},
  pages={68--111},
  year={2018}
}

@article{chrismass:19,
  title={Strong consistency of the least squares estimator in regression models with adaptive learning},
  author={Christopeit, Norbert and Massmann, Michael},
  journal={Electronic Journal of Statistics},
  volume={13},
  pages={1646--1693},
  year={2019}
}

@article{hansen:17,
  title={Regression kink with an unknown threshold},
  author={Hansen, Bruce E},
  journal={Journal of Business \& Economic Statistics},
  volume={35},
  pages={228--240},
  year={2017}
}

@book{evans:01,
  title={Learning and Expectations in Macroeconomics},
  author={Evans, George W and Honkapohja, Seppo},
  year={2001},
  publisher={Princeton University Press}
}

@article{andrews:92,
  title={Generic uniform convergence},
  author={Andrews, Donald W. K.},
  journal={Econometric Theory},
  volume={8},
  pages={241--257},
  year={1992}
}

@article{marsar:89,
author = {Albert Marcet and Thomas J Sargent},
title = {Convergence of least squares learning mechanisms in self-referential linear stochastic models},
journal = {Journal of Economic Theory},
volume = {48},
pages = {337-368},
year = {1989}
}

@article{mn:2016,
    author = {Malmendier, Ulrike and Nagel, Stefan},
    title = { Learning from inflation experiences},
    journal = {The Quarterly Journal of Economics},
    volume = {131},
     pages = {53--87},
    year = {2016}
    }

@article{preston2005,
 author = {Bruce Preston},
 journal = {International Journal of Central Banking},
 pages = {81--126},
  title = {Learning about Monetary Policy Rules When Long-Horizon Expectations Matter},
 volume = {1},
 year = {2005}
}

@incollection{eusepi2025short,
  title={A short history in defense of adaptive learning},
  author={Eusepi, Stefano and Preston, Bruce},
  booktitle={The Routledge Handbook of Economic Expectations in Historical Perspective},
  pages={39--66},
  year={2025},
  publisher={Routledge}
}

@article{hansen:1996,
 author = {Bruce E. Hansen},
 journal = {Econometrica},
 pages = {413--430},
  title = {Inference when a nuisance parameter is not identified under the null hypothesis},
 volume = {64},
 year = {1996}
}

@article{andpol:1994,
 author = {Donald W. K. Andrews and Werner Ploberger},
 journal = {Econometrica},
 pages = {1383--1414},
 title = {Optimal tests when a nuisance parameter is present only under the alternative},
 volume = {62},
 year = {1994}
}

@article{CoibionGorodnichenko2015,
	author = {O. Coibion and Y. Gorodnichenko},
	journal = {American Economic Journal: Macroeconomics},
	pages = {197--232},
	title = {{Is the Phillips Curve Alive and Well after all? Inflation Expectations and the Missing Disinflation}},
	volume = {7},
	year = {2015}}

@article{CoibionGorodnichenkoKamdar2018,
	author = {O. Coibion and Y. Gorodnichenko and R. Kamdar},
	journal = {Journal of Economic Literature},
	pages = {1447--1491},
	title = {{The Formation of
Expectations, Inflation, and the Phillips Curve}},
	volume = {56},
	year = {2018}}

@techreport{FedPhil2024,
	author = {{Federal Reserve of Philadelphia}},
	institution = {Federal Reserve Bank of Philadelphia},
	title = {Survey of Professional Forecasters: Documentation},
	year = {2024}}

@article{ds:93,
 author = {Darrell Duffie and Kenneth J. Singleton},
 journal = {Econometrica},
 number = {4},
 pages = {929--952},
 publisher = {[Wiley, Econometric Society]},
 title = {Simulated Moments Estimation of Markov Models of Asset Prices},
 urldate = {2025-11-12},
 volume = {61},
 year = {1993}
}

@article{HommesMavromatisOzdenZhu2023,
	author = {C. Hommes and K. Mavromatis and T. Özden and M. Zhu},
	journal = {Quantitative Economics},
	title = {{Behavioral Learning Equilibria in New Keynesian Models}},
	volume = {14},
	pages = {1401--1445},
        year = {2023}}

@article{kristensen2005geometric,
  title={Geometric ergodicity of a class of Markov chains with applications to time series models},
  author={Kristensen, Dennis},
  journal={mimeo},
  year={2005}
}

@incollection{evan:20,
  title={Adaptive learning in macroeconomics},
  author={Evans, George W and McGough, Bruce},
  booktitle={Oxford Research Encyclopedia of Economics and Finance},
  publisher= {Oxford University Press},
  pages={--},
  year={2020}
}

@article{carrasco2002mixing,
  title={Mixing and moment properties of various GARCH and stochastic volatility models},
  author={Carrasco, Marine and Chen, Xiaohong},
  journal={Econometric Theory},
  volume={18},
  number={1},
  pages={17--39},
  year={2002},
  publisher={Cambridge University Press}
}

@article{berardi2017empirical,
  title={Empirical calibration of adaptive learning},
  author={Berardi, Michele and Galimberti, Jaqueson K},
  journal={Journal of Economic Behavior \& Organization},
  volume={144},
  pages={219--237},
  year={2017},
  publisher={Elsevier}
}

@article{Lansing2009,
	author = {K.J. Lansing},
	journal = {Review of Economic Dynamics},
	title = {{Time-Varying U.S. Inflation Dynamics and the New Keynesian Phillips
Curve}},
	volume = {12},
	pages = {304--326},
        year = {2009}}

@article{JorgensenLansing2025,
	author = {P.L. J{\o}rgensen and K.J. Lansing},
	journal = {European Economic Review},
	title = {{Anchored inflation expectations and the slope of the Phillips curve}},
	volume = {178},
	pages = {1--38},
        year = {2025}}

@article{Lucas1972,
	author = {R.E. Lucas},
	journal = {Journal of Economic
Theory},
	title = {{Expectations and the Neutrality of Money}},
	volume = {4},
	pages = {103--124},
        year = {1972}}

@article{MavroeidisPlagborgMollerStock2014,
	author = {S. Mavroeidis and M. Plagborg-Møller and J.H. Stock},
	journal = {Journal of Economic Perspectives},
	title = {{Empirical Evidence of Inflation Expectations in the New Keynesian Phillips Curve}},
	volume = {52},
	pages = {124--188},
        year = {2014}}

@article{Muth1961,
	author = {J.F. Muth},
	journal = {Econometrica},
	title = {{Rational Expectations and the Theory of Price Movements}},
	volume = {29},
	pages = {315--335},
        year = {1961}}

@book{tong1990non,
  title={Non-linear Time Series: A Dynamical System Approach},
  author={Tong, Howell},
  year={1990},
  publisher={Oxford University Press}
}

@book{Woodford2003,
	author = {M. Woodford},
	publisher = {Princeton University Press},
	title = {{Interest and Prices: Foundations of a Theory of Monetary Policy}},
	year = {2003}}

@article{CoibionGorodnichenko2012,
	author = {O. Coibion and Y. Gorodnichenko},
	journal = {Journal of Political Economy},
	pages = {116-159},
	title = {What can survey forecasts tell us about informational rigidities?},
	volume = {120},
	year = {2012}}

@article{carvalhoetal2023,
Author = {Carvalho, Carlos and Eusepi, Stefano and Moench, Emanuel and Preston, Bruce},
Title = {Anchored Inflation Expectations},
Journal = {American Economic Journal: Macroeconomics},
Volume = {15},
Number = {1},
Year = {2023},
Pages = {1–47}
}

@incollection{crumpetal2023,
	author = {Richard K. Crump and Stefano Eusepi and Emanuel Moench and Bruce Preston},
	booktitle = {Handbook of Economic Expectations},
	editor = {R{\"u}diger Bachmann and Giorgio Topa and Wilbert {van der Klaauw}},
	pages = {507-540},
	publisher = {Academic Press},
	title = {The term structure of expectations},
	year = {2023}
    }

@article{HommesSorger1998,
	author = {C. Hommes and G. Sorger},
	journal = {Macroeconomic Dynamics},
	pages = {287-321},
	title = {{Consistent Expectations Equilibria}},
	volume = {2},
	year = {1998}}
}
\newpage\clearpage
\begin{appendices}

\renewcommand{\theequation}{\Alph{section}.\arabic{equation}}
\renewcommand{\thelemma}{\Alph{section}.\arabic{lemma}}
\renewcommand{\theremark}{\Alph{section}.\arabic{remark}}
\setcounter{equation}{0}
\setcounter{lemma}{0}
\setcounter{proposition}{0}
\section{Proofs}

This appendix contains the proofs of the main results. Proofs of the auxiliary results can be found in Appendix~\ref{sec:appaux}.

\subsection{Results of Section \ref{sec:prop}}\label{sec:A1}

{\bf Proof of Lemma~\ref{lem:invariant}.} ($i$) 
Let $q\coloneqq1-\gamma$, so that, by \eqref{eq:S0},
\[
\mathfrak s_j(s_t)
=
r_{j,t}-\frac{|\omega_{j,t}|}{\sqrt q}
-\frac{\gamma}{2q^2}d_{j,t}^2, \quad d_{j,t}\coloneqq x_{j,t}-\alpha_{j,t}, \; \omega_{j,t} \coloneqq r_{j,t}\beta_{j,t},
\qquad j\in\{\pi,y\},
\]
and $s_t \in{\sf S}_0 \Leftrightarrow \mathfrak s_j(s_t) > 0$, $j\in\{\pi,y\}$. Now suppress $j$ and for $t\geq1$, let $\nu_t\coloneqq x_t-\alpha_{t-1}$ and note that $d_t=q\nu_t,$ $r_t=qr_{t-1}+\gamma\nu_t^2,$ and $\omega_t=q\omega_{t-1}+\gamma\nu_t d_{t-1}$. Hence, by the triangle inequality,
\begin{align*}
\mathfrak s(s_t)
= \ & qr_{t-1}+\frac{\gamma}{2}\nu_t^2
-\frac{1}{\sqrt q}
 \left|q\omega_{t-1}+\gamma\nu_t d_{t-1}\right| \geq
q\,\mathfrak s(s_{t-1})
+\frac{\gamma}{2}
(
|\nu_t|-\frac{|d_{t-1}|}{\sqrt q}
)^2 \\
\geq \ & q\,\mathfrak s(s_{t-1}).
\end{align*}
Therefore, $\mathfrak s(s_{t-1})>0 \Rightarrow \mathfrak s(s_t)>0$ and thus $s_{t-1}\in{\sf S}_0 \Rightarrow s_t\in{\sf S}_0$ for every $t\geq1$.

\underline{Part ($ii$).} Fix $s_0\in{\sf S}$. For
$j\in\{\pi,y\}$, define the first entry time $\tau_j\coloneqq\iinf\{t\geq0:\mathfrak s_j(s_t) > 0\}$.
On $\{\tau_j>t-1\} = \{\tau_j \geq t\}$, $\mathfrak s_j(s_k)\leq 0$ for $k=0,\ldots,t-1$; i.e. entry in ${\sf S}_0$ has not happened by $t-1$. By part ($i$), regardless of sign, we get $\mathfrak s_j(s_k)\geq q\,\mathfrak s_j(s_{k-1})$ and hence $|\mathfrak s_j(s_{t-1})|\leq q^{t-1}\mathfrak s_j(s_0)^{-},$ $\mathfrak s_j(s_0)^{-}\coloneqq\mmax\{-\mathfrak s_j(s_0),0\}.$
On $\{\tau_j>t-1\}$, we know $\mathfrak s_j(s_{t-1})\leq 0$, and, conditional on $S_{t-1}=S$, $\mathfrak s_j(s_t)\leq 0$ can occur only if
\[
||\nu_{j,t}|-a_j(S)|
\leq b_j(S), \quad a_j(S)\coloneqq q^{-1/2}|d_j(S)|, \quad b_j(S)^2 \coloneqq \frac{2q|\mathfrak s_j(S)|}{\gamma},
\]
where $d_j(S)\coloneqq x_j-\alpha_j$ denotes the value of $d_{j,t-1}$ implied by $S$. Clearly, $a_j(S)$ and $b_j(S)$ are functions of $S$ alone. So, conditional on $S_{t-1}=S$, $\mathfrak s_j(s_t)\leq 0$ happens if $\nu_{j,t}$ falls into $\{\nu\in\mathbb R: ||\nu|-a_j(S)| \leq b_j(S)\}$, which has length bounded above by $4b_j(S)$. Moreover, we can write, conditional on $S_{t-1}=S$,
\begin{align}\label{eq:reducedform}
\nu_{j,t}=x_{j,t}-\alpha_{j,t-1}=m_j(S)+c_j^\Tr\eps_t,
\end{align}
where $m_j(S)$ is a fixed function of $S$ and, in the notation of
Eq.~\eqref{eq:ALMmat}, $c_j^\Tr\coloneqq e_j^\Tr M^{-1}L$. Since
${\sf det}(M)=\Delta\ne0$, $M$ is invertible and since $L$ has full row rank, it holds $c_j=e_j^\Tr M^{-1}L\ne0$. 

Since $\|c_j\|_2^2=\sum_kc_{j,k}^2\le3\mmax_kc_{j,k}^2$, Assumption~\ref{ass:density} gives, for the conditional density $f_{\nu,j}(\cdot\mid S_{t-1}=S)$ of
$\nu_{j,t}$,
\[
\mmax_{j\in\{\pi,y\}}\ssup_{S \in \mathbb R^{12}: r>0}\ssup_{v\in\mathbb R}
f_{\nu,j}(v\mid S_{t-1}=S)
\ \le\
\kappa
\coloneqq
\frac{\sqrt3\bar f}
{\lambda_{\min}\bigl(M^{-1}LL^\Tr(M^{-1})^\Tr\bigr)^{1/2}}
\ \in(0,\infty).
\]

Because, by part ($i$), once $\mathfrak s_j(s_t)>0$ it remains positive. Hence, for $t \geq 1$, $\{\tau_j>t\} = \{\mathfrak s_j(s_t) \leq 0\} \cap \{\tau_j \geq t\}$, and $\{\tau_j\ge t\}$ is $\sigma(S_{t-1})$- measurable, we thus get
\begin{align*}
 {\mathbb P}\{\tau_j>t\mid S_{t-1}\} = \,& \mathbb{P}\{\mathfrak s_j(s_t) \leq 0 \mid S_{t-1}\}{\bf 1}\{\tau_j \geq t\}\\ 
\leq \,&  \kappa\, 4b_j(S_{t-1}){\bf 1}\{\tau_j \geq t\} \\
\leq \,& \bar f \sqrt{\frac{96\,\mathfrak \mmax\{-s_j(s_0),0\}}{\gamma\lambda_{\min}\bigl(M^{-1}LL^\Tr(M^{-1})^\Tr\bigr)}}
\ q^{t/2}{\bf 1}\{\tau_j \geq t\}.
\end{align*}
So, $\tau_j<\infty$ a.s. as claimed. Finally, setting
$N\coloneqq\max\{\tau_\pi,\tau_y\}$, part~($i$) gives
$s_t\in{\sf S}_0$ for $t\geq N$ a.s., and, taking expectations yields the claim. \hfill $\square$

{\bf Proof of Lemma~\ref{lem:markovprop}.} Recall $q =  1-\gamma$. Because $r_{j,t}>0$, $j \in \{\pi,y\}$, it suffices to establish the result for $S_t \coloneqq (s_t^\Tr,u_t^\Tr,i_t)^\Tr$, where, with a slight abuse of notation, we re-define $s_t \coloneqq (x_t^\Tr,\alpha_t^\Tr, \omega_t^\Tr,r_t^\Tr)^\Tr$, with $\omega_t = (\omega_{\pi,t},\omega_{y,t})^\Tr$, $\omega_{j,t} \coloneqq \beta_{j,t}r_{j,t}$. By Lemma~\ref{lem:invariant}, ${\sf S}_0 \coloneqq \{s \in \mathbb R^6 \times \mathbb R^2_{++}: \frac{|\omega_j|}{\sqrt{q}}+\frac{\gamma}{2q^2}(x_j-\alpha_j)^2 < r_j,\ j\in\{y,\pi\} \}$ is forward-invariant for $s_t$  and thus ${\sf X}_0 \coloneqq {\sf S}_0 \times \mathbb R^4$ is an open forward-invariant set for the chain $S_t$. The proof is now essentially an application of \citet[Theorem~4.4]{chot:19}. First, note that $S_t = {\cal G}(S_{t-1},\eps_t)$ is a non-linear state space chain (as in \citealp[Section~2.2.2]{meyn:12}) on an open forward invariant space. Moreover, $r_{j,t}>0$ ensures that the map ${\cal G}: {\sf X}_0 \times \mathbb R^3 \longrightarrow {\sf X}_0$ is continuously differentiable in both arguments, satisfying \citet[Assumption~A5]{chot:19} and implying that $S_t$ is weak Feller (\citealp[Proposition~6.1.2]{meyn:12}).

It remains to verify the two conditions of
\citet[Theorem~4.4]{chot:19} for the associated deterministic so-called {\it control
model}, obtained by replacing $\eps_t$ with a {\it control} sequence $w_t\in\mathbb R^3$ (\citealp[Section~3.3.1]{chot:19}):
\begin{enumerate}
    \item[\textbf{C1}] The control model possesses a steadily attracting state.
    \item[\textbf{C2}] At that state, for some finite $m$ and some control sequence, the controllability matrix
\[
{\cal C}_m
\coloneqq
[\nabla_{w_1}S_m|\cdots|\nabla_{w_m}S_m],
\]
has full rank equal to the dimension of the state space.
\end{enumerate}

Importantly, these conditions cannot hold directly for $S_t$ on ${\sf X}_0$. To see this, define the $3 \times 1$ vector $\zeta_t
\coloneqq \zeta(S_t)$ for
\[
\zeta(S) \coloneqq 
\begin{bmatrix}
x-M^{-1}Lu\\
i-(1-\rho_i)\phi^\Tr x-u_{i}
\end{bmatrix},
\]
and note that, by Eq.~\eqref{eq:ALMmat}, we get
\begin{equation}
\zeta_t
=
B
\begin{bmatrix} h_{t-1}\\ i_{t-1}\end{bmatrix},
\qquad
B
\coloneqq
\begin{bmatrix} H&g\\ 0_{1\times2}&\rho_i\end{bmatrix} \in \mathbb R^{3\times 3}.
\label{eq:P1-effective}
\end{equation}
Hence, the 3 components of $\zeta_t$ satisfy $3-r$, $r \coloneqq {\sf rank}(B)$, linear restrictions. A direct computation gives
${\sf det}(B)=\rho_i{\sf det}(H)=\rho_i\mu\delta/\Delta$, so
$r<3$ precisely when $\rho_i\mu\delta=0$; this includes the
leading single-equation NKPC ($\mu=\varphi=0$) and the case of no interest-rate smoothing ($\rho_i=0$). If $r=3$ we could work directly with $S_t$ on ${\sf X}_0$. However, to cover all possible cases $r \in \{0,1,2,3\}$, the full-rank condition must therefore be formulated relative to the effective dimension. 

To fix ideas, consider the singular value decomposition $B=U\Sigma V^\Tr,$ where $U,V\in\mathbb R^{3\times r}$ have orthonormal columns and
$\Sigma\in\mathbb R^{r\times r}$ is diagonal and positive
definite; the columns of $U$ form an orthonormal basis for the (column) space of $B$. Define, for $t\ge1$, $\eta_t\coloneqq U^\Tr\zeta_t\in\mathbb R^{r},$ so that $\zeta_t=UU^\Tr\zeta_t=U\eta_t$, and, by
\eqref{eq:P1-effective} and $U^\Tr U=I_{r}$,
\begin{equation}
\eta_t
=\Sigma V^\Tr
\begin{bmatrix} h_{t-1}\\ i_{t-1}\end{bmatrix},
\qquad t\ge1.
\label{eq:P1-eta}
\end{equation}
Writing
$U\eta_t=(a_t^\Tr,b_t)^\Tr$, $a_t\in\mathbb R^2$,
$b_t\in\mathbb R$, the original variables are recovered as
\begin{equation}
x_t=M^{-1}Lu_t+a_t,
\qquad
i_t=b_t+(1-\rho_i)\phi^\Tr x_t+u_{i,t}.
\label{eq:P1-recover}
\end{equation}

Thus, for each $t\geq1$, the full state $S_t$ can equivalently be
represented by
\begin{equation}
Y_t
\coloneqq
(\alpha_t^\Tr,\omega_t^\Tr,r_t^\Tr,u_t^\Tr,\eta_t^\Tr)^\Tr
\in\mathbb R^{9+r},
\label{eq:P1-Y}
\end{equation}
since $S_t$ can be recovered from $Y_t$ using
\eqref{eq:P1-recover}. Conversely, $Y_t$ is uniquely determined by
$S_t$, because $\eta_t=U^\Tr\zeta(S_t).$ More generally, let
$\iota:\mathbb R^{9+r}\longrightarrow\mathbb R^{12}$ 
denote the map that reconstructs $S$ from $Y$ using
\eqref{eq:P1-recover}, and define
\[
\pi:\mathbb R^{12}\longrightarrow\mathbb R^{9+r},
\qquad
\pi(S)
\coloneqq
(
\alpha^\Tr,\omega^\Tr,r^\Tr,u^\Tr,
(U^\Tr\zeta(S))^\Tr
)^\Tr.
\]
Both maps are continuous. Since $U^\Tr U=I_r$, $\pi(\iota(Y))=Y$
for every $Y\in\mathbb R^{9+r}$. Moreover, by
\eqref{eq:P1-effective}, $\zeta(S_t)$ belongs to the column space of
$B$, which is spanned by the columns of $U$. Hence, $UU^\Tr\zeta(S_t)=\zeta(S_t),$ and therefore
$\iota(\pi(S_t))=S_t,$ $t\geq1.$
Define ${\sf Y}_0
\coloneqq
\{
Y\in\mathbb R^{9+r}:\iota(Y)\in{\sf X}_0
\}.$ Since ${\sf X}_0$ is open and $\iota$ is continuous, ${\sf Y}_0$ is
open. It is also forward invariant under
${\cal G}_Y(Y,\eps)
\coloneqq
\pi({\cal G}(\iota(Y),\eps)).$
Indeed, ${\cal G}(\iota(Y),\eps)\in{\sf X}_0$ by
Lemma~\ref{lem:invariant}, and its $\zeta$-value belongs to the column
space of $B$ by \eqref{eq:P1-effective}. Consequently, if
$S_0\in{\sf X}_0$ and $Y_t=\pi(S_t)$ for $t\geq1$, then
$Y_t
=
{\cal G}_Y(Y_{t-1},\eps_t).$
The map ${\cal G}_Y$ has the same differentiability as ${\cal G}$,
and the innovation distribution is unchanged. Hence, the regularity,
weak-Feller, and innovation-density conditions also hold for the chain $Y_t$.

{\textbf{C1} \it Steadily attracting state.}
For $k\ge1$, $Y\in{\sf Y}_0$, and controls
$w_1,\ldots,w_k\in\mathbb R^3$, define the iterated control model
by ${\cal G}_Y^{(1)}(Y;w_1)\coloneqq{\cal G}_Y(Y,w_1)$ and
$${\cal G}_Y^{(k)}(Y;w_1,\ldots,w_k)
\coloneqq{\cal G}_Y({\cal G}_Y^{(k-1)}
(Y;w_1,\ldots,w_{k-1}),w_k).$$ A state
$Y^\star\in{\sf Y}_0$ is \emph{steadily attracting} if, for every
$Y_0\in{\sf Y}_0$ and every open neighbourhood $U$ of $Y^\star$,
there exists $T\in\mathbb Z_{>0}$ such that, for every $k\ge T$,
there is a control sequence $w_1,\ldots,w_k \in {\cal O}\coloneqq\{w\in\mathbb R^3:f_\eps(w)>0\}$ with
\[
{\cal G}_Y^{(k)}(Y_0;w_1,\ldots,w_k)\in U,
\]
see \citet[p.~122]{chot:19}. Note that by Assumption~\ref{ass:density}, $f_\eps = f_{\eps_\pi}f_{\eps_y}f_{\eps_i}$, with $f_j(w)>0$ for any $w \in \mathbb \mathbb R$. Hence ${\cal O}= \mathbb R^3$.

Because the recursion $r_{j,t} = qr_{j,t-1}+\gamma \nu_{j,t}^2$, $\nu_{j,t}\coloneqq x_{j,t}-\alpha_{j,t-1}$
is an (IGARCH type) intercept-free second-moment recursion, the control model has no suitable
fixed point that lies in ${\sf S}_0$; i.e. any path of $(x_t,i_t)$ converging to {\it constants} makes the forecast errors $\nu_{j,t}$ vanish and hence implies \(r_{j,t}\to0\), which lies on the boundary of \({\sf S}_0\). We therefore proceed in two steps. {\it First} (Step 1), we construct a (two-)periodic controlled solution, along which the forecast error \(\nu_{j,t}\) alternates in sign while \(\nu_{j,t}^{2}\) remains strictly positive, and select one of its two state values as the candidate steadily attracting state \(Y^\star\). {\it Second} (Step 2), for each terminal horizon \(k\), we construct a possibly different \(k\)-step control path $w_k$ so that its terminal date corresponds to \(Y^\star\); stability of the remaining recursions then makes the effect of the arbitrary initial state vanish as \(k\to\infty\). Thus the terminal states of these horizon-specific paths converge to \(Y^\star\), as required by the definition of a steadily attracting state. The proof of the following lemmas can be found in Section~\ref{sec:appaux} below:

\begin{lemma}\label{lemA:attract}
    Suppose Assumptions~\textup{\ref{ass:density}} and \textup{\ref{ass:para}} hold.
Then the control model associated with ${\cal G}_Y$ on ${\sf Y}_0$ possesses a
steadily attracting state $Y^\star$. 
\end{lemma}

 \textbf{C2} \emph{Controllability condition at \(Y^\star\).} Recall that \(r=\operatorname{\sf rank}(B)\), so that
\({\sf dim}({\sf Y}_0)=9+r\).  We show in Section~\ref{sec:appaux} below that there exists a five-step
admissible control sequence:

\begin{lemma}\label{lemA:control}
Suppose Assumptions~\textup{\ref{ass:density}} and \textup{\ref{ass:para}} hold, and
let $Y^\star$ be a steadily attracting state as constructed in the proof of
Lemma~\textup{\ref{lemA:attract}}. Then there exist admissible controls
$w_1,\ldots,w_5\in{\cal O}$ such that, along the controlled trajectory started at
$Y^\star$,
\[
\operatorname{\sf rank}({\cal C}_5)=9+r,\qquad
{\cal C}_5=[\nabla_{w_1}{\cal G}_Y^{(5)}\mid\cdots\mid\nabla_{w_5}{\cal G}_Y^{(5)}].
\]
\end{lemma}

{\bf Proof of Lemma~\ref{lem:sufJSR}.} Throughout, it is convenient to work in the reordered coordinates
$[\,y,\alpha_y\mid\pi,\alpha_\pi\mid i\,]$. To keep the notation light, we use the same $H$, $g$, $\phi$, $A(B)$, $\mathcal A$ for the reordered objects; in
particular, the rows and columns of all $2\times2$ blocks are henceforth ordered
as $(y,\pi)$, and $B={\sf diag}(\beta_y^2,\beta_\pi^2)$.
 
\underline{Part~($i$).} If $\varphi=0$, then $g=0$, $\Delta=1$, and, by Eq.~\eqref{eq:ALM},
\[
H=
\begin{bmatrix}
\mu&0\\
\psi\mu&\delta
\end{bmatrix}
\]
is lower triangular. Consequently, every matrix in $\mathcal A$ is block lower
triangular with diagonal blocks
\[
T_h(w)
=
\begin{bmatrix}
hw & h(1-w)\\
\gamma hw & 1-\gamma+\gamma h(1-w)
\end{bmatrix},
\qquad
h\in\{\mu,\delta\},\quad
w\in[0,1-\gamma],
\]
corresponding to the pairs $(y,\alpha_y)$ and $(\pi,\alpha_\pi)$, respectively, and
the scalar block $\rho_i$. Hence, by \citet[Prop.~1.5]{jungers2009joint},
\[
\rho_{\sf JSR}(\mathcal A)
=
\mmax\{
\rho_{\sf JSR}(\mathcal T_\mu),
\rho_{\sf JSR}(\mathcal T_\delta),
|\rho_i|
\},
\]
where $\mathcal T_h\coloneqq\{T_h(w):w\in[0,1-\gamma]\}$. Moreover,
$\|T_h(w)\|_\infty
\leq
\mmax\{|h|,1-\gamma+\gamma|h|\}
=
1-\gamma(1-|h|)$,
since $|h|\leq1$. Therefore,
\[
\rho_{\sf JSR}(\mathcal A)
\leq
\mmax\left\{
1-\gamma(1-\mu),
1-\gamma(1-|\delta|),
|\rho_i|
\right\}.
\]
Assumption~\ref{ass:para} with $\varphi=0$ implies $\mu<1$, while $|\delta|<1$ and
$|\rho_i|<1$ hold by assumption. Hence, $\rho_{\sf JSR}(\mathcal A)<1$.
 
\underline{Part~($ii$).}  If $\rho_i=0$, then $g=0$, and the $i$-coordinate does not affect
$(x,\alpha)$. Hence, again by \citet[Prop.~1.5]{jungers2009joint},
$\rho_{\sf JSR}(\mathcal A)$ equals the joint spectral radius of the
$(x,\alpha)$-block family $\mathcal N\coloneqq\{N(B):B={\sf diag}(a,b),\,
(a,b)\in[0,1-\gamma]^2\}$, where
\[
N(B) \coloneqq
\begin{bmatrix}
HB & H(I-B)\\
\gamma HB & (1-\gamma)I+\gamma H(I-B)
\end{bmatrix}.
\]
Moreover, as discussed in the main text, for any norm $V$ on $\mathbb R^4$ and its
induced matrix norm $\|\cdot\|_V$,
\[
\rho_{\sf JSR}(\mathcal A)
=\rho_{\sf JSR}(\mathcal N)
\leq
\ssup_{N\in\mathcal N}\|N\|_V.
\]
We therefore construct a common norm $V$ for which the right-hand side is strictly
bounded by one.
 
Let $K\coloneqq|H|$, where the absolute value is taken entrywise. Under $\psi\geq0$
and $0\leq\delta<1$ we have $\Delta\geq1$ and, in the coordinates fixed above,
\[
K
=
\begin{bmatrix} K_{yy} & K_{y\pi} \\ K_{\pi y} & K_{\pi\pi}  \end{bmatrix}
=
\frac{1}{\Delta}
\begin{bmatrix}
\mu & \varphi|1-\phi_\pi\delta|\\
\psi\mu & \psi\varphi+(1+\varphi\phi_y)\delta
\end{bmatrix}.
\]
A direct calculation gives
\[
{\sf det}(I-K)
=
\frac{
\Delta\Upsilon
-
2\psi\mu\varphi(\phi_\pi\delta-1)_+
}{\Delta^2},
\]
using that $z+|z|=2z_+$ for any $z\in\mathbb R$. Note also that
$K_{yy}=\mu/\Delta<1$: this is immediate if $\Delta>1$, whereas $\Delta=1$ forces
$\varphi(\phi_y+\psi\phi_\pi)=0$, in which case $\Upsilon=(1-\mu)(1-\delta)>0$ and
hence $\mu<1$. Since $K$ is a nonnegative $2\times2$ matrix with $K_{yy}<1$, we
therefore have $\rho(K)<1\Leftrightarrow{\sf det}(I-K)>0$.
 
Suppose first that $K_{y\pi},K_{\pi y}>0$, so that $K$ is irreducible. By the
Perron-Frobenius theorem, $\rho(K)$ is a simple eigenvalue of $K$; i.e. there is a
(Perron) vector $p$ with positive entries such that $Kp=\rho(K)p$. For
$z\in\mathbb R^2$, define the Perron-weighted maximum norm
\[
\|z\|_p
\coloneqq
\mmax_j\frac{|z_j|}{p_j},
\]
and, for $u\coloneqq(u_1^\Tr,u_2^\Tr)^\Tr$ with $u_1,u_2\in\mathbb R^2$, set
$V(u)\coloneqq\mmax\{\|u_1\|_p,\|u_2\|_p\}$. Writing
\[
N(B)u=
\begin{bmatrix}
\widetilde u_1\\
\widetilde u_2
\end{bmatrix},
\]
we have $\widetilde u_1=H(Bu_1+(I-B)u_2)$ and
$\widetilde u_2=(1-\gamma)u_2+\gamma\widetilde u_1$. Since
$0\leq B\leq(1-\gamma)I$, we get $|Bu_1+(I-B)u_2|\leq V(u)p$ componentwise, and
therefore
$|\widetilde u_1|\leq K|Bu_1+(I-B)u_2|\leq\rho(K)V(u)p$, so that
$\|\widetilde u_1\|_p\leq\rho(K)V(u)$. Similarly,
$\|\widetilde u_2\|_p\leq(1-\gamma+\gamma\rho(K))V(u)$. Hence
\[
V(N(B)u)
\leq
\mmax\{\rho(K),1-\gamma+\gamma\rho(K)\}V(u),
\]
uniformly over $B$. Therefore, by the definition
\[
\|N(B)\|_V
\coloneqq
\ssup_{u\neq0}\frac{V(N(B)u)}{V(u)},
\]
we obtain
\[
\rho_{\sf JSR}(\mathcal A)
=\rho_{\sf JSR}(\mathcal N)
\leq
\ssup_{N\in\mathcal N}\|N\|_V
\leq
\mmax\{\rho(|H|),1-\gamma+\gamma\rho(|H|)\}.
\]
If $K$ is reducible (i.e. $K_{y\pi}K_{\pi y}=0$), then $H$ is triangular and the
same bound follows from the block-triangular argument of part~($i$). If
$\Delta\Upsilon>2\psi\mu\varphi(\phi_\pi\delta-1)_+$, then $\rho(|H|)<1$, and
therefore
$\rho_{\sf JSR}(\mathcal A)\leq1-\gamma+\gamma\rho(|H|)<1$. Finally, if
$\psi\mu\varphi(\phi_\pi\delta-1)_+=0$, the displayed condition reduces to
$\Delta\Upsilon>0$, which follows from Assumption~\ref{ass:para}.
 
It remains to verify the necessary condition. Under Assumption~\ref{ass:para} with
$\rho_i=0$, $\psi\geq0$ and $\delta\in[0,1)$, each of the four corner matrices is
Schur stable. Therefore
$L(1)=\mmax_{1\leq j\leq4}\rho(A_j)<1$. \hfill $\square$.

{\bf Proof of Proposition~\ref{prop:ergod}}. By Lemma~\ref{lem:markovprop}, under
Assumptions~\ref{ass:density} and~\ref{ass:para}, the chain
\((S_t)\) is a \(\varphi\)-irreducible and aperiodic \(T\)-chain on
\(\mathsf X_0\), and, by \citet[Thm.~6.2.5]{meyn:12}, every compact subset of \(\mathsf X_0\) is
petite. By \citet[Lem.~15.2.8]{meyn:12}, it is therefore sufficient
to construct a measurable function $V:\mathsf X_0\to[1,\infty)$
that is unbounded off petite sets, together with constants
\(\lambda_0\in(0,1)\) and \(b_0<\infty\), such that
\[
\Ex[V(S_t)\mid S_{t-1}=S]
\le
\lambda_0V(S)+b_0,
\qquad
S\in\mathsf X_0.
\]
We (1) first establish the drift inequality and (2) then show that the sublevel
sets of \(V\), defined below as ${\sf C}_L
\coloneqq
\{S\in\mathsf X_0:V(S)\le L\}$, $L < \infty$, are compact, and hence petite.

Step (1).  Recall that
\[
S_t=(z_t^\Tr,\beta_t^\Tr,r_t^\Tr,u_t^\Tr)^\Tr,
\qquad
z_t=A(B_{t-1})z_{t-1}+Du_t,
\]
where \(A(B)\) is defined in Eq.~\eqref{eq:ABmat},
\(D\in\mathbb R^{5\times3}\) is a fixed coefficient matrix, and
\(B_{t-1}\in[0,1-\gamma]^2\). Assumption~\ref{ass:paraJSR} and the
common-norm characterization of the joint spectral radius imply that
there exist a norm \(\|\cdot\|_*\) on \(\mathbb R^5\) and a constant
\(\chi\in(0,1)\) such that $ \ssup_{B\in[0,1-\gamma]^2}\|A(B)\|_*
\le \chi.$ Consequently, $\|A(B)z\|_* \le \chi\|z\|_*,$ for any $z\in\mathbb R^5$, $B\in[0,1-\gamma]^2.$ Moreover, since \(\rho(R)<1\), there exist a norm
\(\|\cdot\|_R\) on \(\mathbb R^3\) and a constant
\(\chi_u\in(0,1)\) such that $\|Ru\|_R\le\chi_u\|u\|_R.$
Choose \(\delta_u>0\) sufficiently small that $\varrho_u\coloneqq(1+\delta_u)\chi_u^2<1.$
Since $u_t=Ru_{t-1}+\eps_t,$ we have $\|u_t\|_R^2 \leq (\chi_u\|u_{t-1}\|_R+\|\eps_t\|_R)^2$. Note that by the `Peter-Paul' variant of Young's $2ab \leq \eta a^2+b^2/\eta$ for any $\eta>0$ so that $(a+b)^2 \leq (1+\eta)a^2+(1+\eta^{-1})b^2$. Because \(\Ex[\|\eps_1\|_R^2]<\infty\), letting $\eta = \delta_u$ gives with $a = \chi_u\|u_{t-1}\|_R$ and $b=\|\eps_t\|_R$, 
\begin{equation}
\Ex[\|u_t\|_R^2\mid S_{t-1}=S]
\le
c_3+\varrho_u\|u\|_R^2
\label{eq:drift-u}
\end{equation}
for some \(c_3<\infty\). Since all norms are equivalent on finite-dimensional spaces, there is
a \(K_D<\infty\) such that $\|Dv\|_*\le K_D\|v\|_R,$ $
v\in\mathbb R^3.$
Choose \(\delta_z>0\) sufficiently small that
$\varrho_z\coloneqq(1+\delta_z)\chi^2<1.$
It follows from similar arguments that
$\|z_t\|_*^2
\le
\varrho_z\|z_{t-1}\|_*^2
+
(1+\delta_z^{-1})K_D^2\|u_t\|_R^2.$
Combining this inequality with~\eqref{eq:drift-u}, there exist
\(c_1,c_2<\infty\) such that
\begin{equation}
\Ex[\|z_t\|_*^2\mid S_{t-1}=S]
\le
c_1+c_2\|u\|_R^2+\varrho_z\|z\|_*^2.
\label{eq:drift-z}
\end{equation}
Next, write $\bar r_t\coloneqq r_{y,t}+r_{\pi,t}.$ Since $r_{j,t}
=
(1-\gamma)r_{j,t-1}+\gamma\nu_{j,t}^2,$ $\nu_{j,t}=x_{j,t}-\alpha_{j,t-1},$
we have
$\bar r_t = (1-\gamma)\bar r_{t-1}+\gamma\|\nu_t\|^2.$
Because \(x_t\) is a subvector of \(z_t\) and \(\alpha_{t-1}\) is a
subvector of \(z_{t-1}\), there is a \(c_4<\infty\) such that
$\|\nu_t\|^2
\le
c_4(\|z_t\|_*^2+\|z_{t-1}\|_*^2).$
Combining this with~\eqref{eq:drift-z}, there exist
\(c_5,c_6,c_7<\infty\) such that
\begin{equation}
\Ex\!\left[
\bar r_t\mid S_{t-1}=S
\right]
\le
c_5+c_6\|z\|_*^2+c_7\|u\|_R^2
+(1-\gamma)\bar r.
\label{eq:drift-r}
\end{equation}

Recall that Step (2) we have to show that every ${\sf C}_L$ is compact in ${\sf X}_0$. Because ${\sf X}_0$ is open, it remains thus to control also the boundary where $\mathfrak s_j \downarrow 0$. Recall, for \(j\in\{y,\pi\}\),
\[
\mathfrak s_j(S) =
r_j-\frac{|\omega_j|}{\sqrt q}
-\frac{\gamma}{2q^2}(x_j-\alpha_j)^2.
\]
Thus, $\mathsf X_0
=
\{S = (s^\Tr,u^\Tr,i)^\Tr \in {\sf S}:\mathfrak s_j(s)>0,\ j\in\{y,\pi\}\}.$
The proof of Lemma~\ref{lem:invariant} implies
\[
\mathfrak s_j(S_t)
\ge
\frac{\gamma}{2}
(
|\nu_{j,t}|-\frac{|d_{j,t-1}|}{\sqrt q}
)^2,
\]
and hence $\mathfrak s_j(s_t)^{-1/4}
\le
(\frac{\gamma}{2})^{1/4}
|
|\nu_{j,t}|-q^{-1/2}|d_{j,t-1}|
|^{-1/2}.$ We will now show that $\Ex[\mathfrak s_j(S_t)^{-1/4}\mid S_{t-1}=S]<\infty$ so that $s_j(s_t)^{-1/4}$ can be used for the construction of the drift. In particular, as shown in the proof of Lemma~\ref{lem:invariant}, conditional on \(S_{t-1}=S\), the densities $f_{\nu,j}(v \mid S_{t-1} = S)$ of \(\nu_{j,t}\) is uniformly bounded
\[
\mmax\limits_{j \in \{\pi,y\}} \ssup\limits_{S \in {\sf S}_0}\ssup\limits_{v \in \mathbb R}  f_{\nu,j}(v \mid S_{t-1} = S) \leq \kappa < \infty.
\]
For \(a\ge0\), define $Y_{j,t}(a)
\coloneqq
\lvert |\nu_{j,t}|-a\rvert.$  For every \(r>0\),
\[
\{Y_{j,t}(a)<r\}
\subset
\{\nu_{j,t}\in(a-r,a+r)\}
\cup
\{\nu_{j,t}\in(-a-r,-a+r)\}.
\]
Since the interval on the right has length at most \(4r\), $\mathbb{P}\{Y_{j,t}(a)<r\mid S_{t-1}=S\}
\le
4\kappa r.$
Using the tail-integral representation of a nonnegative random variable,
\begin{align*}
\Ex[Y_{j,t}(a)^{-1/2}\mid S_{t-1}=S]
&=
\int_0^\infty
\mathbb{P}\{Y_{j,t}(a)^{-1/2}>t\mid S_{t-1}=S\}
\,\mathrm dt
\\
&\le
\int_0^1 1\,\mathrm dt
+
\int_1^\infty
\mathbb{P}\{Y_{j,t}(a)<t^{-2}\mid S_{t-1}=S\}
\,\mathrm dt
\\
&\le
1+4\kappa\int_1^\infty t^{-2}\,\mathrm dt=
1+4\kappa.
\end{align*}
Hence, 
\[
\ssup\limits_{S\in\mathsf X_0}
\mmax\limits_{j\in\{\pi,y\}}
\ssup\limits_{a\ge0}
\Ex[
\lvert |\nu_{j,t}|-a\rvert^{-1/2}
\mid S_{t-1}=S
]
\le
1+4\kappa
<
\infty,
\]
which shows that, 
\begin{equation}
\ssup\limits_{S\in\mathsf X_0}
\Ex[
\mathfrak s_j(S_t)^{-1/4}
\mid S_{t-1}=S
]
\le (\frac\gamma{2})^{\frac1{4}}(1+4\kappa) \eqqcolon c_8 <\infty.
\label{eq:drift-slack}
\end{equation}

Choose $\lambda_0
\in
(
\mmax\{\varrho_u,\varrho_z,1-\gamma\},
1
)$
and \(a_r>0\) sufficiently small that $
\varrho_z+a_rc_6<\lambda_0.$ Next, choose \(a_u>0\) sufficiently large that $c_2+a_rc_7+a_u\varrho_u
<
\lambda_0a_u.$
Fix any \(a_s>0\) and define
\[
V(S)
\coloneqq
1+\|z\|_*^2
+a_u\|u\|_R^2
+a_r(r_y+r_\pi)
+a_s\sum_{j\in\{y,\pi\}}\mathfrak s_j(S)^{-1/4}.
\]
Combining
\eqref{eq:drift-u}--\eqref{eq:drift-slack} gives
\begin{align*}
\Ex[V(S_t)\mid S_{t-1}=S]\le b_0+ (\varrho_z+a_rc_6)\|z\|_*^2 + \ &
(c_2+a_rc_7+a_u\varrho_u)\|u\|_R^2
\\
+ \ &
a_r(1-\gamma)(r_y+r_\pi)
\end{align*}
for some \(b_0<\infty\). By the choices of \(a_r\), \(a_u\), and
\(\lambda_0\),
\begin{equation}
\Ex[V(S_t)\mid S_{t-1}=S]
\le
b_0+\lambda_0V(S),
\qquad
S\in\mathsf X_0.
\label{eq:global-drift}
\end{equation}

Step (2). Following \citet[p.~188]{meyn:12}, define, for every \(L<\infty\),
the sublevel set
\[
{\sf C}_L
\coloneqq
\{S\in\mathsf X_0:V(S)\le L\}.
\]
To apply \citet[Lem.~15.2.8]{meyn:12}, it remains to verify that
\(V\) is unbounded off petite sets, that is, that every sublevel set
\({\sf C}_L\) is petite. By Lemma~\ref{lem:markovprop}, every compact
subset of \(\mathsf X_0\) is petite. It is therefore sufficient to
show that \({\sf C}_L\) is compact for every \(L<\infty\).

For every \(S\in {\sf C}_L\),
$\|z\|_*^2\le L,$ $\|u\|_R^2\le\frac{L}{a_u},$ $r_y+r_\pi\le\frac{L}{a_r},$
and
\[
\mathfrak s_j(S)
\ge
\left(\frac{a_s}{L}\right)^4
\eqqcolon
\underline s_L>0.
\]
Since
$0<\mathfrak s_j(S)\le r_j$,
we also have
$\underline s_L
\le
r_j
\le
\frac{L}{a_r}$.
Moreover, \(\mathfrak s_j(S)>0\) implies $|\beta_j|<\sqrt q.$ Hence all coordinates of \(S\) are bounded on \({\sf C}_L\).

Let \((S_n)\) be a sequence in \({\sf C}_L\) such that $S_n\to S
\in \mathbb R^{12}.$ By continuity of \(\mathfrak s_j\),
$\mathfrak s_j(S)
= \llim_{n\to\infty}\mathfrak s_j(S_n) \ge \underline s_L>0,$
and likewise
$r_j
=
\llim_{n\to\infty}r_{j,n}
\ge
\underline s_L>0.$ Hence \(S\in\mathsf X_0\). Since \(V\) is continuous on
\(\mathsf X_0\),
\(
V(S)
=
\llim_{n\to\infty}V(S_n)
\le
L,
\)
so \(S\in{\sf C}_L\). Thus \({\sf C}_L\) is closed in
\(\mathbb R^{12}\). Since it is also bounded, it is compact. By Lemma~\ref{lem:markovprop}, \({\sf C}_L\) is
petite. Consequently, \(V\) is unbounded off petite sets in the sense
of \cite[Lem.~15.2.8]{meyn:12}.

Finally, choose any
$\lambda\in(\lambda_0,1)$ and take
$L\ge\frac{b_0}{\lambda-\lambda_0}.$
If \(S\notin {\sf C}_L\), then
$b_0
\le
(\lambda-\lambda_0)V(S)$,
so that~\eqref{eq:global-drift} implies $\Ex[V(S_t)\mid S_{t-1}=S] \le \lambda V(S).$
If \(S\in {\sf C}_L\), then
$\Ex[V(S_t)\mid S_{t-1}=S]
\le
\lambda V(S)+b_0.$
Therefore,
\[
\Ex[V(S_t)\mid S_{t-1}=S]
\le
\lambda V(S)+b_0\mathbf 1_{{\sf C}_L}(S),
\qquad
S\in\mathsf X_0,
\]
where \({\sf C}_L\) is petite. 

Hence, every sublevel set \({\sf C}_L\) is compact and therefore
petite by Lemma~\ref{lem:markovprop}. Consequently, \(V\) is
unbounded off petite sets in the sense of
\citet[p.~188]{meyn:12}. Accordingly, there exist a petite set \(C\subset\mathsf X_0\),
a constant \(\lambda\in(0,1)\), and a constant \(b<\infty\) such that
\[
\Ex[V(S_t)\mid S_{t-1}=S]
\le
\lambda V(S)+b\mathbf 1_C(S),
\qquad
S\in\mathsf X_0.
\]
Since \((S_t)\) is also \(\varphi\)-irreducible and aperiodic,
\citet[Thm.~15.0.1]{meyn:12} implies that the chain is geometrically
ergodic, while \citet[Thm.~16.0.1]{meyn:12} yields
\(V\)-uniform geometric ergodicity.

By Lemma~\ref{lem:markovprop}, the chain is
\(\varphi\)-irreducible and aperiodic, and the set \(C\) is petite.
The drift condition therefore implies, by
\citet[Thm.~15.0.1]{meyn:12}, that the chain is positive Harris
recurrent and admits a unique invariant distribution \(\pi\) satisfying
\(\pi(V)<\infty\). By \citet[Thm.~16.0.1]{meyn:12}, the chain is
\(V\)-uniformly geometrically ergodic. Consequently, under
\(S_0\sim\pi\), it is strictly stationary, ergodic, and geometrically
\(\beta\)-mixing. Since \(V\) dominates
\(\|\nu\|^2+\|r\|\), the baseline moment assertions in
part~($iii$) follow. Under
\(\Ex[\|\eps_1\|^{2p}]<\infty\), the same argument applied to the
higher-order Lyapunov function \(V_p\) yields
\(\pi(V_p)<\infty\), and hence the remaining assertions in
part~($iii$). This proves the claim. \hfill $\square$

{\bf Proof of Lemma~\ref{lem:filter}.}  Throughout, let $C(\omega)$ denote a positive and a.s.-finite random variable that might depend on fixed constants such as the initial value $\fra$, the order of the partial derivative $q$, or $\eta$, with $\eta > 0$ defined below, but is independent of $t$ and $\gamma$. Suppress the component
index \(j\in\{\pi,y\}\), and prove the result for the corresponding scalar
recursion. Moreover, let $\varrho $ be a constant such that $\varrho \in (1-\ubar\gamma,1)$.Let $q_\gamma \coloneqq 1-\gamma,$ and set $\bar q\coloneqq1-\ubar\gamma<1.$ Moreover, let $\Delta a_t(\gamma)
\coloneqq
a_t(\gamma,\fra)-a_t(\gamma)$, $\nu_t(\gamma,\fra)
\coloneqq x_t-\alpha_{t-1}(\gamma,\fra),$ $
d_t(\gamma,\fra)
\coloneqq
x_t-\alpha_t(\gamma,\fra),$
and define \(\nu_t(\gamma)\coloneqq
x_t-\alpha_{t-1}(\gamma)\) and \(d_t(\gamma)\coloneqq
x_t-\alpha_{t}(\gamma)\) analogously. The
learning recursions can be written as
$
\alpha_t(\gamma,\fra)
=
q_\gamma\alpha_{t-1}(\gamma,\fra)
+
\gamma x_t,
$ $r_t(\gamma,\fra)
=
q_\gamma r_{t-1}(\gamma,\fra)
+
\gamma\nu_t(\gamma,\fra)^2,$
and $\omega_t(\gamma,\fra)
=
q_\gamma\omega_{t-1}(\gamma,\fra)
+
\gamma\nu_t(\gamma,\fra)d_{t-1}(\gamma,\fra).$
Moreover, $d_t(\gamma,\fra)
=q_\gamma\nu_t(\gamma,\fra)$.
The stationary recursions satisfy the same identities after
suppressing \(\fra\).

In particular, $\alpha_t(\gamma)
=
\gamma\sum_{k=0}^{\infty}q_\gamma^k x_{t-k},
$ $r_t(\gamma)
=
\gamma\sum_{k=0}^{\infty}
q_\gamma^k\nu_{t-k}(\gamma)^2,$ and $\omega_t(\gamma)
=
\gamma\sum_{k=0}^{\infty}
q_\gamma^k
\nu_{t-k}(\gamma)d_{t-k-1}(\gamma).$
Differentiation with respect to \(\gamma\) introduces only polynomial
factors in \(k\), whereas the weights remain bounded uniformly over
\(\Gamma\) by summable sequences of the form $C_m(1+k)^m\bar q^k,$ $
m=0,\ldots,d,$ where $C_m$ are constants. Cauchy-Schwarz in conjunction with $\Ex[\|x_1\|^2]<\infty$, therefore yields
\begin{align}
\Ex[
\ssup_{\gamma\in\Gamma}
\{
|\partial_\gamma^m\alpha_t(\gamma)|^2
+
|\partial_\gamma^m r_t(\gamma)|
+
|\partial_\gamma^m\omega_t(\gamma)|
\}
]
&<\infty
\label{eq:filter-proof-level-mom}
\end{align}
for \(m=0,\ldots,M\). We make use of the following result that is used in similar form elsewhere (e.g. \citealp[Lemma 2.1]{strau:06}):

\begin{lemma}\label{lem:subexp}
    If $A_t$ is stationary and $\Ex[\operatorname{\sf log}^+|A_1|]<\infty$, then for every $\eta > 0$, there exists an a.s. finite random variable $C_\eta(\omega)$ such that $|A_t|\le C_\eta(\omega)e^{\eta t}$.
\end{lemma}

{\footnotesize {\bf Proof of Lemma~\ref{lem:subexp}.} By stationarity and Markov's inequality
\[
\sum_{t=1}^{\infty}\PP\{|A_t|>e^{\eta t}\}
=
\sum_{t=1}^{\infty}\PP\{\operatorname{\sf log}^+|A_1|>\eta t\}
\le
\frac{1}{\eta}\Ex[\operatorname{\sf log}^+|A_1|]
<\infty.
\]
Hence, by the first Borel-Cantelli lemma, with probability one only finitely many events $\{|A_t|>e^{\eta t}\}$ occur, or, equivalently, there exists some finite $T(\omega) <\infty$ such that \(|A_t|\le e^{\eta t}\) for every $t \ge T(\omega)$, which proves the result. \hfill $\square$}

Consequently, for every \(\eta>0\), Lemma~\ref{lem:subexp} implies the existence of an a.s.-finite
\(C(\omega)\) such that $|x_t| \le C(\omega)e^{\eta t}$ and
\begin{align}
\ssup_{\gamma\in\Gamma}
\{
|\partial_\gamma^m\alpha_t(\gamma)|
+
|\partial_\gamma^m r_t(\gamma)|
+
|\partial_\gamma^m\omega_t(\gamma)|
\}
&\le
C(\omega)e^{\eta t},
\label{eq:filter-proof-level-subexp}
\end{align}
for every \(t\ge1\) and \(m=0,\ldots,M\). Clearly, the same holds for the derivatives of $\nu_t(\gamma)$ and $d_t(\gamma)$.
Moreover, by the same argument $R_t^{-1}\le C(\omega)e^{\eta t}$,
or, more generally, for every fixed integer \(k\ge1\),
\begin{equation}
R_t^{-k}
\le
C(\omega)e^{k\eta t},
\qquad
t\ge1.
\label{eq:filter-proof-inverse-R}
\end{equation}

Since the initialized and stationary filters are driven by the same
observation \(x_t\),
$\Delta\alpha_t(\gamma)
=
q_\gamma\Delta\alpha_{t-1}(\gamma)
=
q_\gamma^t(\fra_\alpha-\alpha_0(\gamma)).$
For \(m=0,\ldots,M\), Leibniz' rule gives
\[
\partial_\gamma^m\Delta\alpha_t(\gamma)
= {\fra}_{\alpha} \partial_\gamma^m q_\gamma^t-
\sum_{k=0}^m
\binom{m}{k}
(\partial_\gamma^k q_\gamma^t)(\partial_\gamma^{m-k}\alpha_0(\gamma)).
\]
By Eq.~\eqref{eq:filter-proof-level-mom}, $\ssup_{\gamma\in\Gamma}
|\partial_\gamma^m\Delta\alpha_0(\gamma)|<\infty$ a.s., while
from \(q_\gamma=1-\gamma\), we have for any \(\varrho\in(1-\ubar\gamma,1)\) and some $C_{k,\varrho} <\infty$, $\ssup_{\gamma\in\Gamma}
|\partial_\gamma^k q_\gamma^t|
\le
C_{k,\varrho}\varrho^t,$ $t\ge0$. Thus, from $\Delta\nu_t(\gamma)
=
-\Delta\alpha_{t-1}(\gamma),$ $\Delta d_t(\gamma)
=
-\Delta\alpha_t(\gamma),$ we get
\begin{equation}
\ssup_{\gamma\in\Gamma}
\{|\partial_\gamma^m\Delta\alpha_t(\gamma)|+
|\partial_\gamma^m\Delta\nu_t(\gamma)|
+
|\partial_\gamma^m\Delta d_t(\gamma)|\}
\le
C(\omega)\varrho^t,
\label{eq:filter-proof-alpha}
\end{equation}
and, by similar arguments,
\begin{equation}
\ssup_{\gamma\in\Gamma}
\{
|\partial_\gamma^m\Delta r_t(\gamma)|
+
|\partial_\gamma^m\Delta\omega_t(\gamma)|
\}
\le
C(\omega)\varrho^t,
\qquad
m=0,\ldots,M.
\label{eq:filter-proof-r-omega}
\end{equation}
Define
\[
R_t(\fra)
\coloneqq
\iinf_{\gamma\in\Gamma}
\mmin_{j\in\{\pi,y\}}
r_{j,t}(\gamma,\fra).
\]
By~\eqref{eq:filter-proof-r-omega}, $R_t(\fra)
\ge
R_t-C(\omega)\varrho^t.$ Choose \(\eta>0\) sufficiently small that $\varrho e^\eta<1.$

It follows from~\eqref{eq:filter-proof-inverse-R} that
\[
\frac{C(\omega)\varrho^t}{R_t}
\le
C(\omega)(\varrho e^\eta)^t
\longrightarrow0
\quad\text{a.s.}
\]
Consequently,$R_t(\fra)\ge\frac12R_t$ for all sufficiently large \(t\), almost surely.

Suppose first that the initialization has strictly positive variance
coordinates. Since
\[
r_t(\gamma,\fra)
=
q_\gamma r_{t-1}(\gamma,\fra)
+
\gamma\nu_t(\gamma,\fra)^2
\ge
(1-\bar\gamma)^t r_0(\fra),
\]
the initialized variance is uniformly positive over \(\Gamma\) on
every finite initial segment. Absorbing that finite segment into the
random constant gives, for every fixed integer \(k\ge1\),
\begin{equation}
\ssup_{\gamma\in\Gamma}r_t(\gamma)^{-k}
+
\ssup_{\gamma\in\Gamma}r_t(\gamma,\fra)^{-k}
\le
C(\omega)e^{k\eta t},
\qquad
t\ge1.
\label{eq:filter-proof-inverse-both}
\end{equation}

Since
 $\beta_t(\gamma,\fra)
=
\frac{\omega_t(\gamma,\fra)}{r_t(\gamma,\fra)},$
 $\beta_t(\gamma)
=
\frac{\omega_t(\gamma)}{r_t(\gamma)},$
 we have
\[
\begin{aligned}
\Delta\beta_t(\gamma)\coloneqq
\beta_t(\gamma,\fra)-\beta_t(\gamma)=
\Delta\omega_t(\gamma)r_t(\gamma,\fra)^{-1}
+
\omega_t(\gamma)
\{
r_t(\gamma,\fra)^{-1}-r_t(\gamma)^{-1}
\}.
\end{aligned}
\]
Moreover,
\[
r_t(\gamma,\fra)^{-1}-r_t(\gamma)^{-1}
=
-
\frac{\Delta r_t(\gamma)}
{r_t(\gamma,\fra)r_t(\gamma)}.
\]
After differentiating the latter identity \(m\) times, every resulting
term contains at least one derivative of \(\Delta r_t(\gamma)\),
possibly of order zero, multiplied by derivatives of the stationary
or initialized variance levels and finite powers of their inverses.
Therefore,
\eqref{eq:filter-proof-level-subexp},
\eqref{eq:filter-proof-r-omega}, and
\eqref{eq:filter-proof-inverse-both}, with \(\eta>0\) chosen
sufficiently small, imply
\begin{equation}
\ssup_{\gamma\in\Gamma}
\left|
\partial_\gamma^m
\{
r_t(\gamma,\fra)^{-1}-r_t(\gamma)^{-1}
\}
\right|
\le
C(\omega)\varrho^t,
\qquad
m=0,\ldots,d.
\label{eq:filter-proof-inverse-difference}
\end{equation}
Applying Leibniz' rule to the expression for
\(\Delta\beta_t(\gamma)\) gives
\begin{equation}
\ssup_{\gamma\in\Gamma}
|\partial_\gamma^m\Delta\beta_t(\gamma)|
\le
C(\omega)\varrho^t,
\qquad
m=0,\ldots,d.
\label{eq:filter-proof-beta}
\end{equation}

The stationary slope process and its derivatives grow at most
subexponentially by
\eqref{eq:filter-proof-level-subexp} and
\eqref{eq:filter-proof-inverse-both}. Together with
\eqref{eq:filter-proof-beta}, this also gives
\begin{equation}
\ssup_{\gamma\in\Gamma}
|\partial_\gamma^m\beta_t(\gamma,\fra)|
\le
C(\omega)e^{\eta t},
\qquad
m=0,\ldots,d,
\label{eq:filter-proof-beta-level}
\end{equation}
for every \(\eta>0\), after adjusting \(C(\omega)\).

Finally, for \(j\in\{\pi,y\}\), define
\[
h_{j,t}(\gamma,\fra)
\coloneqq
\alpha_{j,t}(\gamma,\fra)
+
\beta_{j,t}(\gamma,\fra)^2
(x_{j,t}-\alpha_{j,t}(\gamma,\fra)),
\]
and define \(h_{j,t}(\gamma)\) analogously. Then
\[
\begin{aligned}
h_{j,t}(\gamma,\fra)-h_{j,t}(\gamma)
={}&
(1-\beta_{j,t}(\gamma,\fra)^2)
\Delta\alpha_{j,t}(\gamma)
\\
&+
(
\beta_{j,t}(\gamma,\fra)^2
-
\beta_{j,t}(\gamma)^2
)(
x_{j,t}-\alpha_{j,t}(\gamma)).
\end{aligned}
\]
The level processes and their derivatives grow at most
subexponentially by
\eqref{eq:filter-proof-level-subexp} and
\eqref{eq:filter-proof-beta-level}, while the differences in
\(\alpha_{j,t}\) and \(\beta_{j,t}\), together with their derivatives
up to order \(d\), decrease geometrically by
\eqref{eq:filter-proof-alpha} and
\eqref{eq:filter-proof-beta}. A final application of Leibniz' rule,
with \(\eta>0\) chosen sufficiently small, therefore gives
\[
\ssup_{\gamma\in\Gamma}
|\partial_\gamma^m
\{
h_{j,t}(\gamma,\fra)-h_{j,t}(\gamma)
\}|
\le
C(\omega)\varrho^t,
\qquad
m=0,\ldots,d.
\]
This proves the claim. \hfill$\square$

Finally, we show that $\Ex[\ubar r_1^{-\tau}] < \infty$ for any $\tau > 0$, which implies $\Ex[\log^+(1/\ubar r_1)]<\infty$, and, in turn,  $\ubar r_t > 0$ a.s..

\begin{lemma}\label{lem:negR}
Suppose Assumption~\ref{ass:density} holds, $\Delta\neq0$,
$\Gamma=[\underline\gamma,\bar\gamma]\subset(0,1)$, and $(x_t)$ is strictly
stationary with $\Ex[\|x_1\|^2]<\infty$. Then, for every $\tau>0$,
$\Ex[\ubar r_1^{-\tau}]<\infty.$
\end{lemma}

\subsection{Results of Section \ref{sec:cons}}\label{sec:A2}

{\bf Proof of Proposition~\ref{prop:separation}.} To set the stage, some notation is introduced or, for convenience, restated.

In view of the discussion preceding Lemma~\ref{lem:filter}, for each
$\gamma\in\Gamma$ and $j\in\{\pi,y\}$, let
$\{\alpha_{j,t}(\gamma),r_{j,t}(\gamma),\omega_{j,t}(\gamma)\}_{t\in\mathbb Z}$
denote the $\mathcal F_t$-measurable stationary filter, driven by the
stationary process $(x_t)_{t\in\mathbb Z}$. Write $M\coloneqq M(\vartheta_0)$,
$L\coloneqq L(\vartheta_0)$ and $m_j^\Tr\coloneqq e_j^\Tr M^{-1}L$ for
$j\in\{\pi,y\}$, where $e_\pi,e_y\in\mathbb R^2$ are the canonical unit
vectors; since $\Delta(\vartheta_0)\neq0$, the matrix $M$ is nonsingular, and,
$L$ having rank two, $m_j\neq0$. Also, let $e_3\in\mathbb R^3$ denote the
third canonical unit vector and define
$s^\Tr\coloneqq(1-\rho_{i,0})\phi_0^\Tr M^{-1}L+e_3^\Tr$; neither $m_j$ nor
$s$ depends on $t$. By the reduced form in Eq.~\eqref{eq:ALMmat}, conditional
on $\mathcal F_{t-1}$ the variables $u_t$, $x_t$ and $i_t$ are affine
functions of $\varepsilon_t$. For $z\in\mathbb R^3$, representing a possible
realization of $\varepsilon_t$, define $u_t(z)\coloneqq R_0u_{t-1}+z$,
$x_{j,t}(z)\coloneqq\overline x_{j,t}+m_j^\Tr z$ and
$i_t(z)\coloneqq\overline\imath_t+s^\Tr z$, where $\overline x_{j,t}$ and
$\overline\imath_t$ are $\mathcal F_{t-1}$-measurable. Thus (a.s.)
$u_t=u_t(\varepsilon_t)$, $x_{j,t}=x_{j,t}(\varepsilon_t)$ and
$i_t=i_t(\varepsilon_t)$.
 
For $\gamma\in\Gamma$, put $q_\gamma\coloneqq1-\gamma$ and define
\[
\nu_{j,t}(\gamma;z)\coloneqq x_{j,t}(z)-\alpha_{j,t-1}(\gamma),
\qquad
\alpha_{j,t}(\gamma;z)\coloneqq q_\gamma\alpha_{j,t-1}(\gamma)+\gamma x_{j,t}(z),
\]
so that $x_{j,t}(z)-\alpha_{j,t}(\gamma;z)=q_\gamma\nu_{j,t}(\gamma;z)$.
Writing $d_{j,t-1}(\gamma)\coloneqq x_{j,t-1}-\alpha_{j,t-1}(\gamma)$, define
further
\[
r_{j,t}(\gamma;z)\coloneqq q_\gamma r_{j,t-1}(\gamma)+\gamma\nu_{j,t}(\gamma;z)^2,
\qquad
\omega_{j,t}(\gamma;z)\coloneqq q_\gamma\omega_{j,t-1}(\gamma)
+\gamma\nu_{j,t}(\gamma;z)d_{j,t-1}(\gamma),
\]
and $\beta_{j,t}(\gamma;z)\coloneqq\omega_{j,t}(\gamma;z)/r_{j,t}(\gamma;z)$
as well as
$\tau_{j,t-1}(\gamma)\coloneqq\{q_\gamma r_{j,t-1}(\gamma)/\gamma\}^{1/2}$.
Conditional on $\mathcal F_{t-1}$, the quantity $d_{j,t-1}(\gamma)$ and the
candidate filter states at time $t-1$ are fixed; the argument $z$ enters the
recursions only through $x_{j,t}(z)$. Accordingly, let
\begin{equation}\label{eq:Wt-gamma-z}
W_{t+1}(\gamma;z)
=
\begin{bmatrix}
x_{\pi,t+2}^e(\gamma;z)\\
x_{y,t+2}^e(\gamma;z)\\
i_t(z)
\end{bmatrix},
\end{equation}
where, for $j\in\{\pi,y\}$,
\begin{equation}\label{eq:xejt-gamma-z}
\begin{aligned}
x_{j,t+2}^e(\gamma;z)
\coloneqq{}&
\alpha_{j,t}(\gamma;z)
+\beta_{j,t}(\gamma;z)^2
(x_{j,t}(z)-\alpha_{j,t}(\gamma;z))\\
={}&
\alpha_{j,t}(\gamma;z)
+q_\gamma\nu_{j,t}(\gamma;z)
\frac{
(q_\gamma\omega_{j,t-1}(\gamma)
+\gamma\nu_{j,t}(\gamma;z)d_{j,t-1}(\gamma))^2
}{
(q_\gamma r_{j,t-1}(\gamma)
+\gamma\nu_{j,t}(\gamma;z)^2)^2
} .
\end{aligned}
\end{equation}
By construction, $W_{t+1}(\gamma;\varepsilon_t)=W_{t+1}(\gamma)$ a.s.;
moreover, since $r_{j,t-1}(\gamma)>0$ a.s. under
Assumption~\ref{ass:density}, the denominator in \eqref{eq:xejt-gamma-z} is
strictly positive for every $z\in\mathbb R^3$, so that
$z\mapsto W_{t+1}(\gamma;z)$ is continuous.
 
\begin{lemma}\label{lem:rank}
Suppose Assumptions~\ref{ass:density} and~\ref{ass:para} hold. Then
$\Ex[W_{1}(\gamma_0)W_{1}(\gamma_0)^\Tr]\succ0$.
\end{lemma}
 
Since $\Ex[W_{1}(\gamma_0)W_{1}(\gamma_0)^\Tr]\succeq 0$, the $3 \times 3$ matrix is singular only if there is
$\iota=(\iota_1,\iota_2,\iota_3)^\Tr\neq0$ such that the corresponding linear
combination of the entries of $W_{t+1}(\gamma_0)$ vanishes, that is
\begin{equation}\label{eq:rank-degenerate}
\iota^\Tr W_{t+1}(\gamma_0) =\iota_1x^e_{\pi,t+2}(\gamma_0)+\iota_2x^e_{y,t+2}(\gamma_0)+\iota_3i_t=0
\qquad a.s.
\end{equation}
We show that \eqref{eq:rank-degenerate} forces $\iota=0$.
 
By Eq.~\eqref{eq:rank-degenerate} and the full support of the conditional law of $\eps_t$, Eq.~\eqref{eq:rank-degenerate} implies, conditional on ${\cal F}_{t-1}$, $\iota^\Tr W_{t+1}(\gamma_0;z) = 0$ for any $z \in \mathbb R^3$. Since $M^{-1}L$ has rank two, $m_\pi$ and $m_y$ are linearly
independent, so that there is $w\in\mathbb R^3$ with $m_y^\Tr w=0$ and
$m_\pi^\Tr w\neq0$. Fix such a
$w$ and restrict $z$ to the line $z=\zeta w$, $\zeta\in\mathbb R$. Along this line $x_{y,t}(\zeta w)=\overline x_{y,t},$ $x_{\pi,t}(\zeta w)=\overline x_{\pi,t}+\zeta\,m_\pi^\Tr w,$ i.e. $x_{y,t}(\zeta w)$ is constant in $\zeta$, whereas inflation varies in $\zeta$. As noted above, $z$ enters $x_{j,t+1}^e$ only via $x_{j,t}(z)$. In particular, $\pi_t$ does not enter $x_{y,t+1}^e$. Consequently $x^e_{y,t+2}(\gamma_0;\zeta w)=x^e_{y,t+2}(\gamma_0;0)$ for every
$\zeta$ and, $i_t(\zeta w)$ being affine in $\zeta$, equation
\eqref{eq:rank-degenerate} reads, conditionally on $\mathcal F_{t-1}$ and as a
function of the scalar $v\coloneqq x_{\pi,t}(\zeta w)$,
\begin{equation}\label{eq:rank-scalar}
\iota_1x^e_{\pi,t+2}(\gamma_0;v)+\kappa_1v+\kappa_0=0
\qquad\text{for every }v\in\mathbb R,
\end{equation}
where $\kappa_1\coloneqq\iota_3s^\Tr w/m_\pi^\Tr w$ and
$\kappa_0\coloneqq\iota_2x^e_{y,t+2}(\gamma_0;0)+\iota_3\overline\imath_t
-\kappa_1\overline x_{\pi,t}$ are $\mathcal F_{t-1}$-measurable and do not
depend on $v$. Akin to Remark~\ref{rem:tail-sep}, Equation \eqref{eq:rank-scalar} is a rational function of a single real
variable plus an affine remainder. Its denominator $r_{\pi,t}(\gamma_0;v)^2$
being strictly positive on $\mathbb R$, clearing denominators turns
\eqref{eq:rank-scalar} into an identity between polynomials in $v$, valid for
every real and hence for every complex $v$. The quadratic
$r_{\pi,t}(\gamma_0;\cdot)$ has the two conjugate roots
$v_\pm=\alpha_{\pi,t-1}(\gamma_0)\pm i\tau_{\pi,t-1}(\gamma_0)$, at which
$x^e_{\pi,t+2}(\gamma_0;\cdot)$ has double poles; these are genuine a.s.,
because at $v_+$ the bracket in the numerator of \eqref{eq:xejt-gamma-z}
equals
$q_{\gamma_0}\omega_{\pi,t-1}(\gamma_0)
+i\gamma_0\tau_{\pi,t-1}(\gamma_0)d_{\pi,t-1}(\gamma_0)$, whose imaginary part
is non-zero a.s., $d_{\pi,t-1}(\gamma_0)=q_{\gamma_0}(x_{\pi,t-1}
-\alpha_{\pi,t-2}(\gamma_0))$ having a nonatomic conditional distribution
given $\mathcal F_{t-2}$. Multiplying \eqref{eq:rank-scalar} by $(v-v_+)^2$
and letting $v\to v_+$, the affine remainder contributes zero, whereas the
first term contributes $\iota_1$ times a non-zero constant. Hence
$\iota_1=0$, and, choosing instead $w$ with $m_\pi^\Tr w=0$ and
$m_y^\Tr w\neq0$, the same argument gives $\iota_2=0$.
 
With $\iota_1=\iota_2=0$, \eqref{eq:rank-degenerate} reduces to
$\iota_3(\overline\imath_t+s^\Tr z)=0$ for every $z\in\mathbb R^3$, so that
$\iota_3s=0$. It therefore suffices that $s\neq0$. Writing $a=1-\rho_{i,0}$, a
direct computation gives
$\phi_0^\Tr M^{-1}=\Delta_0^{-1}(\phi_{\pi,0},\ \psi_0\phi_{\pi,0}+\phi_{y,0})$
and, since $Le_3=(0,-\varphi_0)^\Tr$,
\[
s^\Tr e_3
=
1-\frac{a\varphi_0(\psi_0\phi_{\pi,0}+\phi_{y,0})}{\Delta_0}
=
1-\frac{\Delta_0-1}{\Delta_0}
=
\frac{1}{\Delta_0}
\neq0 .
\]
Hence $\iota_3=0$, so that $\iota=0$, a contradiction.

\underline{Proposition~\ref{prop:separation} ($i$).} The proof, organized in six steps, is a generalization of the identification argument of Remark~\ref{rem:tail-sep}.

\emph{Step 1.}
Define
\begin{equation}\label{eq:ct-xit-z}
\begin{aligned}
c_t(\theta;z)
\coloneqq{}&
C(\vartheta)B(\vartheta_0)W_t(\gamma_0;z)
-B(\vartheta)W_t(\gamma;z),\\
\xi_t(\theta,R;z)
\coloneqq{}&
(C(\vartheta)R_0-RC(\vartheta))u_{t-1}(z)
+c_t(\theta;z)-Rc_{t-1}(\theta).
\end{aligned}
\end{equation}
By construction,
$\xi_t(\theta,R)=\xi_t(\theta,R;\varepsilon_{t-1})$ a.s.. Moreover, for $j\in\{\pi,y\}$, both
$\nu_{j,t-1}(\gamma_0;z)$ and
$\nu_{j,t-1}(\gamma;z)$ are affine in $z$, with gradient $m_j$. By Lemma~\ref{lem:negR}, $\iinf_\gamma r_{j,t}(\gamma)>0$ a.s.. Hence the
denominators in
$x_{j,t}^e(\gamma_0;z)$ and $x_{j,t}^e(\gamma;z)$ are strictly positive
for every $z\in\mathbb R^3$.  It thus follows that, conditional on ${\cal F}_{t-2}$, each of the three components of
$z\mapsto\xi_t(\theta,R;z)$ is a rational function of $z$ whose
denominator does not vanish on $\mathbb R^3$. Since, by hypothesis,
$\xi_t(\theta,R) = \xi_t(\theta,R;\varepsilon_{t-1})=0$ a.s. and, by Assumption~\ref{ass:density}, the conditional law of
$\eps_{t-1}$ has full support on $\mathbb R^3$, continuity implies
that, $\xi_t(\theta,R;z)=0$ for every $z\in\mathbb R^3$ and almost every realization of ${\cal F}_{t-2}$. 
Indeed, as argued in Remark~\ref{rem:tail-sep}, $\xi_t(\theta,R;z)=0$ is an identity of rational functions over $\mathbb C^3$, which allows us in the next step to examine the complex zeros of the denominators. In particular, conditional on ${\cal F}_{t-2}$, $\xi_t(\theta,R;z)$ is a linear combination of
$x_{j,t}^e(\cdot;z)$, together with terms that remain finite at their poles. Hence, and as shown below, if one $x_{j,t}^e(\cdot;z)$ has a genuine pole at some
$z^\ast$, the identity $\xi_t(\theta,R;z)=0$ can hold only if another
forecast term has a pole at the same location and the corresponding
singular parts cancel. 

\emph{Step 2.}
By Lemma~\ref{lem:negR} and the preceding discussion, conditional on
$\mathcal F_{t-2}$ we have $r_{j,t-1}(\gamma;z)>0$ for every
$z\in\mathbb R^3$, so that the denominator $r_{j,t-1}(\gamma;z)^2$ of the
fractional term in Eq.~\eqref{eq:xejt-gamma-z} vanishes only at complex $z$.
Exactly as in the proof of Lemma~\ref{lem:rank}, its zeros are the $z$ with
$\nu_{j,t-1}(\gamma;z)=\pm i\tau_{j,t-2}(\gamma)$, where
$\tau_{j,t-2}(\gamma)\coloneqq\{q_\gamma r_{j,t-2}(\gamma)/\gamma\}^{1/2}>0$;
in terms of $z$, using $x_{j,t-1}(z)=\overline x_{j,t-1}+m_j^\Tr z$, these are
the solutions of
\begin{equation}\label{eq:pole-locus}
m_j^\Tr z=a_j(\gamma),
\qquad
a_j(\gamma)\coloneqq
\alpha_{j,t-2}(\gamma)-\overline x_{j,t-1}\pm i\tau_{j,t-2}(\gamma) .
\end{equation}
At either of these locations the numerator of the fractional term equals, up
to a non-zero factor,
$(q_\gamma\omega_{j,t-2}(\gamma)\pm i\gamma\tau_{j,t-2}(\gamma)d_{j,t-2}(\gamma))^2$,
whose bracket has imaginary part $\pm\gamma\tau_{j,t-2}(\gamma)d_{j,t-2}(\gamma)$
and is therefore non-zero unless $d_{j,t-2}(\gamma)=0$, an event of
probability zero under Assumption~\ref{ass:density}. The affine term
$\alpha_{j,t-1}(\gamma;z)$ contributing no poles, the solutions of
\eqref{eq:pole-locus} are a.s. genuine double poles of
$x^e_{j,t}(\gamma;\cdot)$.

Finally, $M^{-1}L$ having rank two, $m_\pi$ and $m_y$ are linearly
independent, so the two equations in \eqref{eq:pole-locus} have different solution sets. Hence a pole of the true inflation forecast cannot be cancelled by an
output-forecast term in the identity $\xi_t(\theta,R;z)=0$, but must be matched
by a pole of the candidate inflation forecast, and comparing their locations
will identify $\gamma$.
\bk

\emph{Step 3.} Let ${\cal P}_{\pi,t}^{\pm}(\gamma)
\coloneqq
\{
z\in\mathbb C^3:
m_\pi^\Tr z
=
\alpha_{\pi,t-1}(\gamma)-\overline x_{\pi,t}
\pm i\tau_{\pi,t-1}(\gamma)
\}$ denote the two sets of
inflation pole locations, i.e. the sets of $z\in\mathbb C^3$ solving
\eqref{eq:pole-locus}. By Step~2 these are
genuine double-pole locations of $x^e_{\pi,t+2}(\gamma;\cdot)$. In view of
Eq.~\eqref{eq:ct-xit-z}, the true and candidate inflation forecasts enter
$\xi_{t+1}(\theta,R;z)$ through
$C(\vartheta)B(\vartheta_0)e_1x^e_{\pi,t+2}(\gamma_0;z)
-B(\vartheta)e_1x^e_{\pi,t+2}(\gamma;z)$, with coefficient vectors
$C(\vartheta)B(\vartheta_0)e_1=C(\vartheta)(\delta_0,\varphi_0,0)^\Tr$ and
$B(\vartheta)e_1=(\delta,\varphi,0)^\Tr$, both non-zero because
$C(\vartheta)$ is nonsingular and $\delta,\delta_0>0$.
Consider the locations in ${\cal P}^{+}_{\pi,t}(\gamma_0)$. Since the
coefficient of $x^e_{\pi,t+2}(\gamma_0;\cdot)$ is non-zero, the corresponding
term in $\xi_{t+1}(\theta,R;z)$ becomes unbounded there, and, the identity
$\xi_{t+1}(\theta,R;z)=0$ holding throughout, this divergence must be cancelled by
another term with poles at the same locations. By Step~2 an output-forecast
term cannot do so, therefore ${\cal P}^{+}_{\pi,t}(\gamma_0)$ coincides with
${\cal P}^{+}_{\pi,t}(\gamma)$ or with ${\cal P}^{-}_{\pi,t}(\gamma)$. The
second is impossible, as $\tau_{\pi,t-1}(\gamma)$ and
$\tau_{\pi,t-1}(\gamma_0)$ are both strictly positive. Since $m_\pi\neq0$,
equality of the two sets forces equality of the corresponding constants in
\eqref{eq:pole-locus}, that is, a.s.,
$\alpha_{\pi,t-1}(\gamma)+i\tau_{\pi,t-1}(\gamma)
=\alpha_{\pi,t-1}(\gamma_0)+i\tau_{\pi,t-1}(\gamma_0)$, and, two complex
numbers being equal exactly when their real and imaginary parts agree
separately, by stationarity, $\alpha_{\pi,t}(\gamma)=\alpha_{\pi,t}(\gamma_0)$ a.s. for arbitrary $t$. Subtracting the recursions
$\alpha_{\pi,t}(g)=q_g\alpha_{\pi,t-1}(g)+gx_{\pi,t}$, $g\in\{\gamma,\gamma_0\}$,
therefore gives
$(\gamma-\gamma_0)(x_{\pi,t}-\alpha_{\pi,t-1}(\gamma_0))=0$ a.s.. But
$x_{\pi,t}-\alpha_{\pi,t-1}(\gamma_0)=q_{\gamma_0}^{-1}d_{\pi,t}(\gamma_0) \neq 0$ a.s. by Step~2. Hence $\gamma=\gamma_0$ and $W_{t+1}(\gamma)=W_{t+1}(\gamma_0)$ a.s..

\emph{Step 4.} Set $N\coloneqq C(\vartheta)B(\vartheta_0)-B(\vartheta)$. By
Step~3, $c_t(\theta)=NW_t(\gamma_0)$, so that the identity
$\xi_t(\theta,R;z)=0$ becomes
$(C(\vartheta)R_0-RC(\vartheta))u_{t-1}(z)+NW_t(\gamma_0;z)-RNW_{t-1}(\gamma_0)=0$,
where, by columns,
$NW_t(\gamma_0;z)=Ne_1x^e_{\pi,t}(\gamma_0;z)+Ne_2x^e_{y,t}(\gamma_0;z)
+Ne_3i_{t-1}(z)$. Conditional on ${\cal F}_{t-2}$, the terms
$(C(\vartheta)R_0-RC(\vartheta))u_{t-1}(z)$ and $Ne_3i_{t-1}(z)$ are affine in
$z$ and $RNW_{t-1}(\gamma_0)$ does not depend on $z$, so that only the two
forecasts can generate poles. By Step~2 there are inflation-pole locations at which the output forecast
remains finite. At such a location $x^e_{\pi,t}(\gamma_0;\cdot)$ has a genuine
double pole while every other term remains finite, so that $Ne_1\neq0$ would
render at least one component of the identity unbounded, a contradiction.
Hence $Ne_1=0$ and, by the same argument at an output-pole location at which
the inflation forecast remains finite, $Ne_2=0$; that is, the first two
columns of $C(\vartheta)B(\vartheta_0)-B(\vartheta)$ vanish. Writing
$n_3\coloneqq Ne_3$, it follows that $N=n_3e_3^\Tr$ and
$c_t(\theta)=n_3i_{t-1}$, so that the identity reduces to
$(C(\vartheta)R_0-RC(\vartheta))u_{t-1}+n_3i_{t-1}-Rn_3i_{t-2}=0$ a.s.

\emph{Step 5.} It remains to show that $n_3=0$. By the reduced form, the true interest rate can be written as
$i_{t-1}(z)=c_0^\ast i_{t-2}+\kappa_0^\Tr x^e_t(\gamma_0;z)+s^\Tr u_{t-1}(z)$,
for some constant $c_0^\ast$, where
$\kappa_0^\Tr\coloneqq(1-\rho_{i,0})\phi_0^\Tr H_0$ collects the loading of the
policy rate on expectations.  By Step~2 one may choose an inflation-pole location at which the output
forecast remains finite; there $x^e_{\pi,t}(\gamma_0;\cdot)$ has a genuine
double pole while every other term stays finite, so that the identity forces
$n_3(\kappa_0^\Tr e_\pi)=0$. However, a direct computation gives
$\kappa_0^\Tr e_\pi=(1-\rho_{i,0})\{\phi_{\pi,0}\delta_0
+\varphi_0(\phi_{\pi,0}\psi_0+\phi_{y,0})\}/\Delta_0>0$, since
$|\rho_{i,0}|<1$, $\phi_{\pi,0}>0$, $\delta_0>0$,
$\varphi_0,\psi_0,\phi_{y,0}\geq0$ and $\Delta_0\geq1$.
$\kappa_0^\Tr e_\pi>0$. Hence $n_3=0$ and,
with Step~4, $C(\vartheta)B(\vartheta_0)=B(\vartheta)$, so that
$c_t(\theta)=0$ a.s..

\emph{Step 6.} By Steps~4 and~5, $\gamma=\gamma_0$ and
$C(\vartheta)B(\vartheta_0)=B(\vartheta)$ and thus $c_t(\theta)=0$ a.s., so
that the identity $\xi_t(\theta,R)=0$ reduces to $Du_{t-1}=0$ a.s., where
$D\coloneqq C(\vartheta)R_0-RC(\vartheta)$. Since
$u_{t-1}=R_0u_{t-2}+\varepsilon_{t-1}$ with $\varepsilon_{t-1}$ independent of
$u_{t-2}$, the stationary variance
$\Omega_u\coloneqq\operatorname{\sf var}(u_{t-1})$ satisfies
$\Omega_u\succeq\Sigma_0\succ0$ by Assumption~\ref{ass:density}. Hence
$0=\operatorname{\sf var}[Du_{t-1}]=D\Omega_uD^\Tr$ forces $D=0$, that is
$C(\vartheta)R_0=RC(\vartheta)$. Together with Steps~3-5 this shows that
$\gamma=\gamma_0$, $C(\vartheta)B(\vartheta_0)=B(\vartheta)$ and
$C(\vartheta)R_0=RC(\vartheta)$, and consequently $c_t(\theta)=0$ a.s.,
which proves part~($i$).

\underline{Proposition~\ref{prop:separation} ($ii$).}
 
Write $C=C(\vartheta)=(c_{jk})_{j,k=1}^3$. Since
$C=A(\vartheta)A(\vartheta_0)^{-1}$, the identity $A(\vartheta)=CA(\vartheta_0)$
holds automatically; combined with $B(\vartheta)=CB(\vartheta_0)$, established
in Part~(i), and diagonality of $C\Sigma_0C^\Tr$, where
$\Sigma_0=\operatorname{diag}(v_\pi,v_y,v_i)\succ0$, the rows of $C$ are
pairwise $\Sigma_0$-orthogonal. We show $C=I_3$.
 
If $\rho_{i,0}\neq0$, or $\rho_{i,0}=0$ with $\delta_0\phi_{\pi,0}\neq1$, or
$\rho_{i,0}=0$, $\delta_0\phi_{\pi,0}=1$ with $\mu_0\neq0$, then in each case
one of these nondegeneracy conditions forces a single entry, or pair of
entries, of $C$, respectively $c_{23}$, again $c_{23}$, or the pair
$c_{12}=c_{32}$, to vanish directly from the relevant column of
$B(\vartheta)=CB(\vartheta_0)$ together with the corresponding row of
$A(\vartheta)=CA(\vartheta_0)$; orthogonality of the rows of $C$ and the
zero/unit-diagonal pattern of $A(\vartheta_0)$ then propagate $C=I_3$ by
elementary substitution. None of these three configurations uses the sign
restrictions. We omit the routine computations and give the argument in
full only for the remaining, genuinely delicate configuration.

\underline{Case: $\rho_{i,0}=\mu_0=0$, $\delta_0\phi_{\pi,0}=1$.} Write
$f\coloneqq\varphi_0\geq0$, $p\coloneqq\phi_{\pi,0}=1/\delta_0$,
$\kappa\coloneqq\psi_0p+\phi_{y,0}$, $\Delta_0\coloneqq1+f\kappa$. Here the
two identities do \emph{not} pin down $C$ directly: they give $\mu=\rho_i=0$,
$\delta=\delta_0$, $\phi_\pi=p$, and, for some $a,b,e\in\mathbb R$,
\[
C=I_3+(ae_1+ee_3)v^\Tr+be_2w^\Tr,
\qquad
v\coloneqq(-fp,1,-f)^\Tr,
\quad
w\coloneqq(p,\kappa,1)^\Tr,
\]
with $\varphi=f+b\Delta_0$. Orthogonality reduces, with
$G\coloneqq v_y+f^2(p^2v_\pi+v_i)>0$ and
$D\coloneqq\kappa v_y-f(p^2v_\pi+v_i)$, to the nonlinear system
\[
a(v_y+bD)=-bp\,v_\pi,
\qquad
e(v_y+bD)=-b\,v_i,
\qquad
aeG=f(pe\,v_\pi+a\,v_i).
\]
If $b=0$, the first two equations give $a=e=0$. Suppose $b\neq0$. Then
$v_y+bD\neq0$, since otherwise the first equation gives $bp\,v_\pi=0$;
solving the first two equations directly gives
$a=-bp\,v_\pi/(v_y+bD)$ and $e=-b\,v_i/(v_y+bD)$, both nonzero.
Dividing, $a\,v_i=pe\,v_\pi$; substituting into the third equation gives
$aG=2fp\,v_\pi$, so that
\[
a=\frac{2fp\,v_\pi}{G},
\qquad
e=\frac{2f v_i}{G},
\qquad
b=-\frac{2f v_y}{N},
\qquad
N\coloneqq G+2fD=2v_y\Delta_0-G,
\]
the last equality following from $v_i+eD=v_iN/G$ (with $N\neq0$, since
$N=0$ would give $e v_y=0$, contradicting $e\neq0$). Since $a\neq0$ and
$p,v_\pi,G>0$, necessarily $f\neq0$, hence $f>0$. By the matrix
determinant lemma applied to the rank-two form of $C$,
$\det C=G/N=-\varphi/f$; since $\det C=\Delta(\vartheta)/\Delta_0$, this
gives
\[
\varphi\,\Delta_0=-\varphi_0\,\Delta(\vartheta).
\]
Since $\rho_i=\rho_{i,0}=0$, the sign restrictions give
$\Delta(\vartheta),\Delta_0\geq1$: the left-hand side is nonnegative, as
$\varphi\geq0$, while the right-hand side is strictly negative, as
$\varphi_0=f>0$, a contradiction. Hence $a=b=e=0$, so $C=I_3$. \hfill $\square$

{\bf Proof of Proposition~\ref{prop:const}.}
We prove
\begin{equation}\label{eq:Qn-uniform}
\ssup_{\theta\in\Theta}|Q_n(\theta)-Q(\theta)|=o_{a.s.}(1), \quad Q(\theta) \coloneqq \mmin\limits_{R \in \mathbb R^{3\times 3}}Q(\theta,R).
\end{equation}
Strong consistency then follows from the standard `argmin' theorem, since
\(Q\) is uniquely minimised at \(\theta_0\) by
Proposition~\ref{prop:separation}.
Since \(B_1(\vartheta)\) and \(B_2(\vartheta)\) are continuous and
\(\Theta\) is compact, their norms are uniformly bounded.
Moreover, Lemma~\ref{lem:invariant} gives
\(\beta_{j,t}(\gamma)^2\leq 1\). Hence,
for some \(K<\infty\),
\begin{equation}\label{eq:ut-envelope}
\ssup_{\theta\in\Theta}\|u_t(\theta)\|^2
\leq
K(
\|Y_t\|^2+\|x_{t-1}\|^2+i_{t-1}^2
+\sum_{j\in\{\pi,y\}}
\ssup_{\gamma\in\Gamma}|\alpha_{j,t-1}(\gamma)|^2).
\end{equation}
Proposition~\ref{prop:ergod}~($iii$) and
Lemma~\ref{lem:mom}~($i$) thus imply
\(\Ex[\ssup_{\theta\in\Theta}\|u_1(\theta)\|^2]<\infty\).
For \((k,l)=(0,0),(0,1),(1,1)\), define
\(v_t^{kl}(\theta)=\operatorname{\sf vec}
(u_{t-k}(\theta)u_{t-l}(\theta)^\Tr)\).
By Proposition~\ref{prop:ergod}~($i$), \(v_t^{kl}\) is stationary and
ergodic. The map \(\theta\mapsto v_t^{kl}(\theta)\) is continuous (a.s.), and \eqref{eq:ut-envelope}, stationarity, and the Cauchy-Schwarz
inequality give
\(\Ex[\ssup_{\theta\in\Theta}\|v_1^{kl}(\theta)\|]<\infty\).
Consequently, \citet[Theorem~2.2.1]{straumann2005estimation} yields
\begin{equation}\label{eq:Skl-uniform}
\ssup_{\theta\in\Theta}
\|S_{kl,n}(\theta)-S_{kl}(\theta)\|
=o_{a.s.}(1),
\qquad
(k,l)=(0,0),(0,1),(1,1),
\end{equation}
where
\(S_{kl}(\theta)=
\Ex[u_{1-k}(\theta)u_{1-l}(\theta)^\Tr]\).
Since
\(u_t(\theta)=C(\vartheta)u_t+c_t(\theta)\), where
\(c_t(\theta)\) is \(\mathcal F_{t-1}\)-measurable,
\(S_{11}(\theta)\succeq
C(\vartheta)\Sigma_0C(\vartheta)^\Tr\).
Continuity, compactness, and nonsingularity of \(C(\vartheta)\) (recall that $\operatorname{\sf det}C(\vartheta) = \Delta(\vartheta)/\Delta(\vartheta_0)$) therefore
give
\(\iinf_{\theta\in\Theta}\lambda_{\min}(S_{11}(\theta))>0\).
Completing the square in \(R\), this also justifies profiling: write
\(R^\st(\theta)\coloneqq S_{01}(\theta)S_{11}(\theta)^{-1}\) and consider
\begin{equation}
\begin{split}\label{eq:complete-square}
\Ex[\eps_t(\theta,R)\eps_t(\theta,R)^\Tr]
= \ &
\Omega(\theta)
+
(R-R^\st(\theta))S_{11}(\theta)(R-R^\st(\theta))^\Tr,\\
\Omega(\theta)\coloneqq \ &
S_{00}(\theta)-S_{01}(\theta)S_{11}(\theta)^{-1}S_{01}(\theta)^\Tr .
\end{split}
\end{equation}
The second term in \eqref{eq:complete-square} is positive semi-definite, its
\(j\)-th diagonal entry being
\(\|S_{11}(\theta)^{1/2}(R-R^\st(\theta))^\Tr e_j\|^2\), which vanishes if and
only if \((R-R^\st(\theta))^\Tr e_j=0\). Every diagonal entry of
\(\Omega(\theta,R)\) is therefore minimised at the same \(R^\st(\theta)\), so
that \(Q(\theta)=\mmin_{R}Q(\theta,R)=Q(\theta,R^\st(\theta))\), the minimizer
being unique because \(S_{11}(\theta)\succ0\); in particular, uniqueness of the
minimizer of \(Q(\theta,R)\) transfers to \(Q(\theta)\).
The decomposition also implies
\(\iinf_{\theta\in\Theta}e_j^\Tr\Omega(\theta)e_j>0\) for
\(j\in \{\pi,y,i\}\). It follows from \eqref{eq:Skl-uniform} that
\(\widehat\Omega_n(\theta)\) converges uniformly to \(\Omega(\theta)\)
and hence
\begin{equation}\label{eq:variance-uniform}
\ssup_{\theta\in\Theta}
|
\sum_{j=\pi,y,i}\log\widehat\sigma_{j,n}^2(\theta)
-
\sum_{j=\pi,y,i}\log
(e_j^\Tr\Omega(\theta)e_j)
|
=o_{a.s.}(1).
\end{equation}
Finally, \(\psi\geq0\) and Assumption~\ref{ass:para} imply
\(\Delta(\vartheta)\geq1\) on \(\Theta\). The Jacobian term
\(-2\operatorname{\sf log}|\Delta(\vartheta)|\) is therefore continuous
and identical in the sample and population criteria. Combining this with
\eqref{eq:variance-uniform} proves \eqref{eq:Qn-uniform}, and hence
\(\theta_n\to\theta_0\) almost surely. Since $\widehat R_n(\theta)$ converges (a.s.) uniformly to $R^\st(\theta) = S_{01}(\theta)S_{11}(\theta)^{-1}$, which is continuous, we get
\[
\|\widehat R_n(\theta_n)-R_0\|
\leq
\ssup_{\theta\in\Theta}
\|\widehat R_n(\theta)-R^\st(\theta)\|
+
\|R^\st(\theta_n)-R^\st(\theta_0)\|
=o_{a.s.}(1).
\]
Thus \(\widehat R_n(\theta_n)\to R_0\) almost surely.\hfill $\square$

\subsection{Results of Section \ref{sec:distnlse}}\label{sec:A2a}

The proof makes use of the following two results:

\begin{lemma}\label{lem:mom}
    Let $k \geq 1$ and denote by $\alpha_t^{(m)}(\gamma) \coloneqq \frac{{\sf d}^m}{{\sf d} \gamma^m}\alpha_t(\gamma)$ $($similarly for $r_t$ and $\beta_t$$)$ the $m$-th derivative with respect to $\gamma$.
    \begin{enumerate}
        \item[$(i)$] If $\Ex[\|v_t\|^k] < \infty$, then
    $\Ex[\ssup\limits_{\gamma \in \Gamma}|\alpha_t^{(m)}(\gamma)|^k] < \infty$,  $m \in \{0,1,2,3\}.$
        \item[$(ii)$] If $\Ex[\|v_t\|^{2k}] < \infty$, then
    $\Ex[\ssup\limits_{\gamma \in \Gamma}|r_t^{(m)}(\gamma)|^k] < \infty,$ $m \in \{0,1,2,3\}.$
        \item[$(iii)$] If $\Ex[\|v_t\|^{4k+\zeta}] < \infty$ for some $\zeta>0$, then 
    $\Ex[\ssup\limits_{\gamma \in \Gamma}|\beta_t^{(m)}(\gamma)|^k] < \infty$, $m \in \{1,2\},$
    and, if in addition, $\Ex[\|v_t\|^{6k+\zeta}] < \infty$  for some $\zeta>0$, then $\Ex[\ssup\limits_{\gamma \in \Gamma}|\beta_t^{(3)}(\gamma)|^k] < \infty.$
    \end{enumerate}    
\end{lemma} 

\begin{lemma}\label{lem:pd}
Let \(A\coloneqq
\Ex[\nabla_\theta f_t(\theta_0)\nabla_\theta f_t(\theta_0)^\Tr]\).

\emph{(a)} If \(\Ex[\|v_t\|^{4+\zeta}]<\infty\) for some
\(\zeta>0\), then \(A\) is a finite positive definite \(3\times3\)
matrix.

\emph{(b)} If \(\Ex[\|v_t\|^{8+\zeta}]<\infty\) for some
\(\zeta>0\), then, for any \(\theta_1,\theta_2\in\Theta\),
\[
\left\|
\frac{1}{n}\nabla_\theta^2 Q_n(\theta_2)
-
\frac{1}{n}\nabla_\theta^2 Q_n(\theta_1)
\right\|
\leq
B_n\|\theta_2-\theta_1\|,
\]
where \(B_n=O_p(1)\). 
\end{lemma}

{\bf Proof of Proposition~\ref{prop:norm}.} By Proposition~\ref{prop:const},
\(\theta_n\to_{a.s.}\theta_0\). Since
\(\theta_0\in\operatorname{\sf int}(\Theta)\), the first-order condition holds
with probability approaching one, and the mean-value expansion in
Eq.~\eqref{eq:mvt} applies. By Lemma~\ref{lem:pd}~(a),
\(A=\Ex[\nabla_\theta f_t(\theta_0)\nabla_\theta
f_t(\theta_0)^\Tr]\) is finite and positive definite. Moreover,
\[
\frac{1}{\sqrt n}\nabla_\theta Q_n(\theta_0)
=
-\frac{2}{\sqrt n}\sum_{t=1}^n
\nabla_\theta f_t(\theta_0)u_t.
\]
Let ${\cal F}_t \coloneqq \sigma(\{u_j,\eps_j : j \leq t\})$ and
${\cal G}_t \coloneqq \sigma({\cal F}_{t-1},\eps_t)$. Then
$\nabla_\theta f_t(\theta_0)$ is ${\cal G}_t$-measurable, and, since $u_t$ is
independent of $\eps_t$ by Assumption~\ref{ass:density},
$\Ex[u_t \mid {\cal G}_t] = \Ex[u_t \mid {\cal F}_{t-1}] = 0$. Hence
$\{\nabla_\theta f_t(\theta_0)u_t, {\cal G}_t\}$ is a homoskedastic martingale
difference sequence and the CLT for martingale difference sequences yields
$\frac{1}{\sqrt n}\nabla_\theta Q_n(\theta_0) \to_d {\cal N}(0,4\sigma_u^2 A).$
Next,
\[
\frac{1}{n}\nabla_\theta^2Q_n(\theta_0)
=
\frac{2}{n}\sum_{t=1}^n
\nabla_\theta f_t(\theta_0)\nabla_\theta f_t(\theta_0)^\Tr
-
\frac{2}{n}\sum_{t=1}^n u_tH_t(\theta_0)
\to_p 2A,
\]
by the ergodic theorem and the martingale-difference property. Since
\(\bar\theta_n\to_p\theta_0\), Lemma~\ref{lem:pd}~(b) implies $\frac{1}{n}\nabla_\theta^2Q_n(\bar\theta_n)
-
\frac{1}{n}\nabla_\theta^2Q_n(\theta_0)
=o_p(1),$
and hence
\(n^{-1}\nabla_\theta^2Q_n(\bar\theta_n)\to_p2A\). The claim follows from Slutsky's theorem. \hfill $\square$

{\bf Proof of Corollary \ref{cor:rate}.} The idea is to show that under $\delta_0=0$
\begin{align*}
D_n(\theta) \coloneqq Q_n(\theta)-Q_n(\theta_0)
= \,& \eta^\Tr \sum_{t=1}^n z_t(\gamma)z_t(\gamma)^\Tr \eta
- 2\,\eta^\Tr \sum_{t=1}^n z_t(\gamma)u_t > 0
\end{align*}
for $\eta  \coloneqq (\delta,\psi-\psi_0)^\Tr$ and 
$z_t(\gamma)\coloneqq (h_{t-1}(\gamma),y_t)^\Tr$,
uniformly outside any shrinking neighbourhood of $\eta_0$ whose radius goes to
zero slower than $1/\sqrt{n}$. This, in turn, forces the NLS minimizer to lie
within an $1/\sqrt{n}$ neighbourhood of $\eta_0$. The proof, given in the
appendix, builds on a uniform LLN and CLT for the sample moments of
$z_t(\gamma)$, uniformly in the unidentified nuisance parameter $\gamma$,
which follow from the stochastic Lipschitz bounds of Lemma~\ref{lem:mom}; see
\citet[Lem 1]{andrews:92} and \citet[Thm 2]{hansen:1996b}. Following \cite{saikkonen:95} and \cite{seo:11}, we show that for $r_n \rightarrow \infty$ such that $r_n = o(\sqrt{n})$ and ${\cal B}_{n}(\nu) \coloneqq \{\eta \in \mathbb{R}^2: \|\eta\|>\nu/r_n\}$, $\eta \coloneqq (\delta,\psi-\psi_0)^\Tr$, for some $\nu > 0$, we have
$$
\iinf\limits_{\gamma \in \Gamma, \eta \in {\cal B}_{n}(\nu)} D_n(\theta)>0.
$$
This implies that the NLS estimators of $(\delta,\psi)$ are with probability approaching one inside ${\cal B}_n(\nu)$; implying, because $\nu$ is arbitrary, that they are $\sqrt n$-consistent estimators of $(0,\psi_0)^\Tr$. In order to complete the proof, define $$u_{1,t}(\gamma,\delta) \coloneqq -(\delta-\delta_0)h_{t-1}(\gamma),\quad u_{2,t}(\gamma,\psi) \coloneqq u_t - \delta_0(h_{t-1}(\gamma)-h_{t-1}(\gamma_0))-(\psi-\psi_0)y_t,$$ so that
$Q_n(\theta) = Q_{1,n}(\theta)+Q_{2,n}(\gamma,\psi)$, where
$$Q_{1,n}(\theta) \coloneqq 2\sum_{t=1}^nu_{2,t}(\theta)u_{1,t}(\theta)+\sum_{t=1}^nu_{1,t}(\theta)^2, \quad Q_{2,n}(\gamma,\psi)\coloneqq \sum_{t=1}^nu_{2,t}(\gamma,\psi)^2.$$
Note that $Q_{1,n}(\gamma,\delta_0,\psi)=0$ and thus $Q_{n}(\gamma,\delta_0,\psi) = Q_{2,n}(\gamma,\psi).$ Moreover, if $\delta_0 = 0$, then $Q_{2,n}(\gamma,\psi) = Q_{2,n}(\gamma_0,\psi)$. So,
$$
D_n(\theta)= Q_{1,n}(\theta)+Q_{2,n}(\gamma_0,\psi)-Q_{2,n}(\gamma_0,\psi_0)$$
with $Q_{2,n}(\gamma_0,\psi)-Q_{2,n}(\gamma_0,\psi_0)=(\psi-\psi_0)^2\sum_{t=1}^ny_t^2-2(\psi-\psi_0)\sum_{t=1}^ny_tu_t$ and
$$
Q_{1,n}(\theta) = -2\delta\sum_{t=1}^n h_{t-1}(\gamma)(u_t-(\psi-\psi_0)y_t) + \delta^2\sum_{t=1}^n h_{t-1}(\gamma)^2.
$$
Now, with $\eta = (\delta, \psi-\psi_0)^\Tr$ and $z_t(\gamma)=(h_{t-1}(\gamma),y_t)^\Tr$, we get
$$
D_n(\theta) = \eta^\Tr \sum_{t=1}^n z_t(\gamma)z_t(\gamma)^\Tr \eta - 2\eta \sum_{t=1}^n z_t(\gamma)u_t, 
$$
or,  
$$
\frac1{\xi_n}D_n(\theta) = \xi_n \frac{\eta^\Tr Z_n(\gamma)\eta}{\|\eta\|^2}-2\frac{\eta^\Tr S_n(\gamma)}{\|\eta\|}, \quad \xi_n \coloneqq \sqrt{n}\|\eta\|.
$$
Note that, uniformly in $\gamma \in \Gamma$, we have
$Z_n(\gamma) \rightarrow_p \Ex[Z_n(\gamma)].$ Since for each $\gamma \in \Gamma$, $\Ex[Z_n(\gamma)]\succ 0$, it follows from continuity of $Z_n(\gamma)$ and compactness of $\Gamma$ that there is a $c \in (0,\infty)$ such that, with probability approaching one, $$\iinf\limits_{\gamma\in \Gamma}\lambda_{min}\left(\frac1{n}\sum_{t=1}^n z_t(\gamma)z_t(\gamma)^\Tr\right)\geq c.$$ Moreover, by \citet[Thm 2]{hansen:1996b} with $q =2$, $S_n(\cdot)$ is stochastically equicontinous. By the Cramer-Wold device and the CLT for MDS, we obtain {\it fidi}-convergence, implying $S_n(\gamma) \rightsquigarrow {\mathbb S}(\gamma)$ in $\ell^\infty(\Gamma)$, where ${\mathbb S}$ is a Gaussian process with $\cov[{\mathbb S}(\gamma_1),{\mathbb S}(\gamma_2)] = \sigma_u^2\Ex[z(\gamma_1)z(\gamma_2)^\Tr]$. This implies that $S_n(\gamma) = O_p(1)$ uniformly in $\gamma \in \Gamma$. Therefore, with probability approaching one,
\begin{align*}
\iinf\limits_{\gamma \in \Gamma, \eta \in {\cal B}_{n}(\nu)} \frac1{\xi_n}D_n(\theta) \geq\,& c \iinf\limits_{\eta \in {\cal B}_{n}(\nu)}\xi_n-O_p(1) = c \iinf\limits_{\eta \in {\cal B}_{n}(\nu)}\sqrt{n}\|\eta\|-O_p(1)\\
\geq \,& \sqrt n\frac{c\nu }{r_n}-O_p(1) \rightarrow +\infty.
\end{align*}
This completes the proof. \hfill $\square$

\subsubsection{Results of Section~\ref{sec:roots}}

Define
\[
D(\beta;\lambda) \coloneqq \psi^2(\delta\beta^2\rho+1)+(1-\rho^2)(1-\delta\beta^2\rho)(\sigma_u/\sigma_\eps)^2,
\]
so that $F(\beta;\lambda) = \delta\beta^2 + \psi^2\rho(1-\delta^2\beta^4)/D(\beta;\lambda)$. To begin with, we need the following lemma:

\begin{lemma}\label{lem:Flambda}
Suppose Assumption~\textup{\ref{ass:para}} holds. Then $(i)$ $F(\beta;\lambda)$
is continuously differentiable in $(\beta,\lambda)$, and
$\iinf_{\beta\in[0,1]}D(\beta;\lambda)>0$. If, in addition, $\psi_0\neq0$, then
$(ii)$ there exists $c\in(0,\infty)$ such that
$\iinf_{\beta\in[0,1]}\|F_\lambda(\beta,\lambda_0)\|\ge c$. If, in addition,
$\rho_0>0$, then $(iii)$ $G(0)<0<G(1)$; $(iv)$ the roots of $G(\cdot;\lambda_0)$
in $[0,1]$ coincide, with multiplicities, with those of a polynomial of degree
exactly four; and $(v)$ they admit exactly one of the following multiplicity
structures: three roots, all simple; two roots, one simple and one double; one
root, simple; or one root of multiplicity three, at which
$G_{\beta\beta\beta}(\beta_0)\eqqcolon6c$ satisfies $c>0$.
\end{lemma}

{\bf Proof of Corollary~\ref{cor:roots}.}
By Assumption~\ref{ass:para} and Lemma~\ref{lem:Flambda}$(i)$, the denominator of
$F$ is bounded away from zero on $[0,1]\times{\cal B}$, where ${\cal B}$ is a
compact neighbourhood of $\lambda_0$ with $\PP\{\lambda_n\in{\cal B}\}\to1$;
hence $F$ and $G$ are infinitely differentiable there. Write
$\Delta_n\coloneqq\lambda_n-\lambda_0=O_p(n^{-1/2})$ and recall
$\sqrt n\Delta_n\to_d\mathbb Z\sim{\cal N}(0,\Omega)$, $\Omega\succ0$. Let
$\beta_{1,0},\ldots,\beta_{r_0,0}$ denote the distinct roots of
$G(\cdot;\lambda_0)$, all in $(0,1)$ by Lemma~\ref{lem:Flambda}$(iii)$, with
multiplicities $m_1,\ldots,m_{r_0}$; by Lemma~\ref{lem:Flambda}$(v)$ their
configuration is one of the cases \textnormal{(1)}-\textnormal{(4)}, treated in
turn below. Finally, note that $B\neq0$ whenever $B$ denotes $G_\lambda$
evaluated at a root, by Lemma~\ref{lem:Flambda}$(ii)$ and $G_\lambda=-F_\lambda$;
since $\Omega\succ0$, $B^\Tr\mathbb Z$ is then a nondegenerate centred Gaussian
variable.

In a first step, we show that the roots of $G_n$ converge to the true roots at a rate governed by their multiplicity.  The mean value theorem in $\lambda$ and compactness of $[0,1]\times{\cal B}$ give
\begin{equation}\label{eq:unifG}
\ssup_{\beta\in[0,1]}
|\partial_\beta^jG_n(\beta)-\partial_\beta^jG(\beta)|
=O_p(n^{-1/2}),
\qquad j=0,1,2,3.
\end{equation}
Fix $\varepsilon>0$ such that the intervals
${\cal I}_i\coloneqq(\beta_{i,0}-\varepsilon,\beta_{i,0}+\varepsilon)$ are
disjoint subsets of $(0,1)$. As $G$ vanishes on $[0,1]$ only at the
$\beta_{i,0}$, we have $\iinf\{|G(\beta)|:\beta\notin\bigcup_i{\cal I}_i\}>0$, so
by \eqref{eq:unifG} all roots of $G_n$ in $[0,1]$ lie in $\bigcup_i{\cal I}_i$
with probability approaching one. In particular, $r_n$ is determined by the roots
near the $\beta_{i,0}$ alone. By Taylor's theorem (see also proof of Lemma~\ref{lem:Flambda}~($v$)), $G(\beta)=(\beta-\beta_{i,0})^{m_i}g_i(\beta)$, with $g_i(\beta_{i,0})\neq0$. Since each $g_i$ is continuous with $g_i(\beta_{i,0})\neq0$, we may choose $\varepsilon$ so that, for some $K_i>0$,
$|G(\beta)|\geq K_i|\beta-\beta_{i,0}|^{m_i}$ on ${\cal I}_i$, $i=1,\ldots,r_0$.  Any root
$\beta^\st_n\in{\cal I}_i$ of $G_n$ thus satisfies
$K_i|\beta^\st_n-\beta_{i,0}|^{m_i}\leq|G(\beta^\st_n)-G_n(\beta^\st_n)|
=O_p(n^{-1/2})$, so that
\begin{equation}\label{eq:loc}
n^\frac1{2m_i}(\beta^\st_n-\beta_{i,0})=O_p(1).
\end{equation}
All expansions below are taken over shrinking $n^\frac1{2m_i}$-neighbourhoods of $\beta_{i,0}$.

\underline{Part (a).} Every $\beta_{i,0}$ is simple, so $G_\beta(\beta_{i,0})\neq0$. By
Eq.~\eqref{eq:unifG}, $G_{n,\beta}\coloneqq \partial_\beta G_n$ is bounded away from zero and strictly monotone there on ${\cal I}_i$. Since $G$, and hence $G_n$, changes sign across ${\cal I}_i$, there is exactly one root $\beta^\st_{i,n}\in{\cal I}_i$. This gives $r_n\to_pr_0$ and $\beta^\st_{i,n}\to_p\beta_{i,0}$.
Expanding $0=G_n(\beta^\st_{i,n})$ around $(\beta_{i,0},\lambda_0)$ by
the mean value theorem in each argument separately,
\[
\sqrt n(\beta^\st_{i,n}-\beta_{i,0})
=-G_\beta(\bar\beta_{i,n};\lambda_n)^{-1}
G_\lambda(\beta_{i,0};\bar\lambda_n)^\Tr\sqrt n\Delta_n
\]
for intermediate values $\bar\beta_{i,n}$ and $\bar\lambda_n$. By consistency
and continuity in conjunction with Slutsky's theorem this yields $\sqrt n(\beta^\st_{i,n}-\beta_{i,0})\to_dJ_i^\Tr\mathbb Z$,
jointly over $i$.

\underline{Part (b).} Let $\beta_0$ be the simple and $\beta_0^\dagger$ the double root, so that
\begin{equation}\label{eq:double}
G(\beta_0^\dagger)=G_\beta(\beta_0^\dagger)=0,
\qquad
G_{\beta\beta}(\beta_0^\dagger)=2a\neq0,
\qquad
B\coloneqq G_\lambda(\beta_0^\dagger).
\end{equation}
Write ${\cal I}$ and ${\cal I}^\dagger$ for the corresponding neighbourhoods from
above. Part (a) applies directly at $\beta_0$, yielding a unique root $\beta^\st_n$ nearby
with
\begin{equation}\label{eq:simplelim}
\sqrt n(\beta^\st_n-\beta_0)=J_n^\Tr\sqrt n\Delta_n,
\qquad J_n\to_pJ\coloneqq J(\beta_0),
\end{equation}
unconditionally, since this root exists irrespective of the behaviour near
$\beta_0^\dagger$. Near $\beta_0^\dagger$, Eq.~\eqref{eq:double} and a second-order expansion give,
uniformly over $|\beta-\beta_0^\dagger|\lesssim n^{-1/4}$,
\begin{equation}\label{eq:quad}
G_n(\beta)=a(\beta-\beta_0^\dagger)^2+B^\Tr\Delta_n+o_p(n^{-1/2}).
\end{equation}
Next, since $G_{\beta\beta}(\beta_0^\dagger)=2a\neq0$ and $G_{\beta\beta}$ is
continuous, we can choose ${\cal I}^\dagger$ so that, for some $\kappa>0$,
$a^{-1}G_{\beta\beta}\geq\kappa$ uniformly on ${\cal I}^\dagger$. By
Eq.~\eqref{eq:unifG} the same holds for $a^{-1}G_{n,\beta\beta}$ with probability
approaching one, so that $G_n$ is strictly convex (concave) on ${\cal I}^\dagger$
if $a>0$ ($a<0$). As $G_\beta(\beta_0^\dagger)=0$, $G_\beta$ is strictly monotone
on ${\cal I}^\dagger$ and hence has opposite signs at its endpoints; by
\eqref{eq:unifG} so does $G_{n,\beta}$, which therefore has a unique zero
$\widehat\beta_n$, say, in ${\cal I}^\dagger$. Because
$G_{n,\beta}(\beta_0^\dagger)=G_{n,\beta}(\beta_0^\dagger)-G_\beta(\beta_0^\dagger)
=O_p(n^{-1/2})$ and $G_{n,\beta\beta}$ is bounded away from zero, the mean value
theorem yields $\widehat\beta_n-\beta_0^\dagger=O_p(n^{-1/2})$ and hence
$G_n(\widehat\beta_n)=B^\Tr\Delta_n+O_p(n^{-1})$. Finally, two applications of the
mean value theorem together with $G(\beta_0^\dagger)=G_\beta(\beta_0^\dagger)=0$
give $a^{-1}G(\beta)=\tfrac12a^{-1}G_{\beta\beta}(\bar\beta)(\beta-\beta_0^\dagger)^2
\geq\tfrac{\kappa}{2}\varepsilon^2>0$ at the two endpoints of ${\cal I}^\dagger$,
so that, by \eqref{eq:unifG}, $G_n$ has the sign of $a$ there with probability
approaching one. By convexity, it follows that for every fixed $\varrho>0$ there
are, with probability approaching one, exactly two roots in ${\cal I}^\dagger$ on
$\{a^{-1}B^\Tr\sqrt n\Delta_n<-\varrho\}$ and none on
$\{a^{-1}B^\Tr\sqrt n\Delta_n>\varrho\}$. Since
$a^{-1}B^\Tr\sqrt n\Delta_n\to_da^{-1}B^\Tr\mathbb Z$, whose law is continuous
and symmetric,
\[
\PP\{r_n=3\}\to\tfrac12,
\quad
\PP\{r_n=1\}\to\tfrac12,
\]
and $\PP\{r_n=2\}\to0$.

On $\{r_n=3\}$, let
$\beta_n^{\st-}<\beta_n^{\st+}$ be the two roots in ${\cal I}^\dagger$. By Eq.~\eqref{eq:quad},
\begin{equation}\label{eq:split}
n^{1/4}(\beta_n^{\st\pm}-\beta_0^\dagger)
=\pm\sqrt{-a^{-1}B^\Tr\sqrt n\Delta_n}+o_p(1).
\end{equation}
With \eqref{eq:simplelim} and the above, this gives the stated joint limit
conditional on $\{r_n=3\}$.

Finally, let $6c\coloneqq G_{\beta\beta\beta}(\beta_0^\dagger)$,
$D\coloneqq G_{\beta\lambda}(\beta_0^\dagger)$ and
$E\coloneqq\frac12G_{\lambda\lambda}(\beta_0^\dagger)$, so that expanding in both
arguments,
\[
G_n(\beta_0^\dagger+x)
=ax^2+cx^3+B^\Tr\Delta_n+xD^\Tr\Delta_n+\Delta_n^\Tr E\Delta_n
+R,
\]
with remainder $R = O(x^4+x^2\|\Delta_n\|+x\|\Delta_n\|^2+\|\Delta_n\|^3)$. Evaluating at $x_n^\pm\coloneqq\beta_n^{\st\pm}-\beta_0^\dagger=O_p(n^{-1/4})$,
the remainder is $O_p(n^{-1})$; writing $u_n\coloneqq(x_n^++x_n^-)/2$ and
$v_n\coloneqq(x_n^+-x_n^-)/2>0$, subtracting and adding the two resulting
equations yields
\begin{align}
0&=4au_nv_n+c(6u_n^2v_n+2v_n^3)+2v_nD^\Tr\Delta_n+O_p(n^{-1}),
\label{eq:diff}\\
0&=a(u_n^2+v_n^2)+B^\Tr\Delta_n+c(u_n^3+3u_nv_n^2)+u_nD^\Tr\Delta_n
\nonumber\\
&\quad \quad+\Delta_n^\Tr E\Delta_n+O_p(n^{-1}).
\label{eq:sum}
\end{align}
By Eq.~\eqref{eq:split}, $v_n^{-1}=O_p(n^{1/4})$ conditionally on $r_n=3$, so
dividing \eqref{eq:diff} by $2v_n$ gives
\begin{equation}\label{eq:un}
0=2au_n+c(3u_n^2+v_n^2)+D^\Tr\Delta_n+O_p(n^{-3/4}).
\end{equation}
Since $v_n^2=O_p(n^{-1/2})$, $\Delta_n=O_p(n^{-1/2})$, Eq.~\eqref{eq:un} becomes
$2au_n=-3cu_n^2+O_p(n^{-1/2})$. As $u_n=o_p(1)$ implies $u_n^2=o_p(u_n)$ and
$a\neq0$, we obtain $u_n=O_p(n^{-1/2})$, which verifies the $O_p(1)$ claim.
Substituting into \eqref{eq:sum} gives
$v_n^2=-a^{-1}B^\Tr\Delta_n+O_p(n^{-1})$, in agreement with \eqref{eq:split};
inserting this and $u_n^2=O_p(n^{-1})$ into \eqref{eq:un},
\begin{equation}\label{eq:midpoint}
\sqrt n\,u_n
=\frac1{2a}\Bigl(\frac ca B-D\Bigr)^\Tr\sqrt n\Delta_n+O_p(n^{-1/4}).
\end{equation}
If $cB\neq aD$, Step~$0$ gives the stated conditional limit; if $cB=aD$, the
leading vector vanishes and $\sqrt n\,u_n=o_p(1)$.

\underline{Part (c).}
Let $\beta_0$ be the unique root, of multiplicity three, so that
$G(\beta_0)=G_\beta(\beta_0)=G_{\beta\beta}(\beta_0)=0$ and
$G_{\beta\beta\beta}(\beta_0)=6c_3$ with $c_3>0$ by
Lemma~\ref{lem:Flambda}$(v)$. Set $B\coloneqq G_\lambda(\beta_0)$.

Consider the quartic $P(\beta;\lambda)\coloneqq G(\beta;\lambda)D(\beta;\lambda)$; see also the proof of Lemma~\ref{lem:Flambda}. Write $P_n\coloneqq P(\cdot;\lambda_n)$,
$P_0\coloneqq P(\cdot;\lambda_0)$, and let $\kappa_n\coloneqq\kappa(\lambda_n)$
and $\kappa_0\coloneqq\kappa(\lambda_0)$ denote the respective leading
coefficients. Because, as shown in the proof of Lemma~\ref{lem:Flambda}, $\kappa > 0$, it follows, by continuity and consistency, $\kappa_n>0$ with probability
approaching one. Since $\beta_0$ has multiplicity three, the remaining root
$\widetilde\beta_0$ of $P_0$ is real, and, by the fundamental theorem of algebra,
$P_0(\beta)=\kappa_0(\beta-\beta_0)^3(\beta-\widetilde\beta_0)$. Because, by
Lemma~\ref{lem:Flambda}, $G(0) < 0$ and $D(0) > 0$, we have $\kappa_0\beta_0^3\widetilde\beta_0=P_0(0)<0$, so that $\widetilde\beta_0<0$ (in fact it must be $\widetilde\beta <-1$); being
the only root of $P_0$ other than $\beta_0$, it is simple. By the implicit function theorem, $P_n$ therefore has, with probability
approaching one, a unique simple real root $\widetilde\beta_n<0$ near
$\widetilde\beta_0$.  Since $P_n$ is quartic, the factor theorem gives
\begin{equation}\label{eq:quartfactor}
P_n(\beta_0+x)
=
\kappa_n(\beta_0+x-\widetilde\beta_n)q_n(x),
\end{equation}
where $q_n(x)=x^3+b_{2,n}x^2+b_{1,n}x+b_{0,n}$ is the cubic factor and the $b_{j,n}$ are the coefficients of the remaining cubic factor. At
$\lambda_0$ this factor equals $x^3$, and hence smooth dependence on $\lambda$
gives $b_{j,n}=O_p(n^{-1/2})$, $j=0,1,2$. Moreover, setting $x=0$ in \eqref{eq:quartfactor} and using
$P_n(\beta_0)=D(\beta_0;\lambda_n)G(\beta_0;\lambda_n)$, together with
$\kappa_0(\beta_0-\widetilde\beta_0)=c_3D(\beta_0;\lambda_0)$, yields
$b_{0,n}=c_3^{-1}B^\Tr\Delta_n+O_p(n^{-1}).$
To determine how many of the three remaining roots are real, we consider the
discriminant of the cubic factor in \eqref{eq:quartfactor}. Let ${\cal D}_n$
denote this discriminant. Since $b_{j,n}=O_p(n^{-1/2})$,\footnote{For a real cubic $x^3+bx^2+cx+d$, the discriminant is
$b^2c^2-4c^3-4b^3d-27d^2+18bcd$. Since here
$b_{j,n}=O_p(n^{-1/2})$, all terms are $o_p(n^{-1})$ except
$-27b_{0,n}^2$, which explains the factor $27$.}
${\cal D}_n
=
-27b_{0,n}^2+o_p(n^{-1}),$
and therefore
\[
n{\cal D}_n
\to_d
-27(c_3^{-1}B^\Tr\mathbb Z)^2<0.
\]
Since $B^\Tr\Omega B>0$, $\PP\{{\cal D}_n<0\}\to1$. Since a real cubic has exactly one real root if
and only if its discriminant is negative, the cubic factor in
\eqref{eq:quartfactor} has exactly one real root with probability approaching
one. Hence the cubic has exactly one real root with probability approaching one.
Since $\widetilde\beta_n<0$, this is the only remaining real root of $P_n$.
Moreover, $G_n(0)<0<G_n(1)$ with probability approaching one, so continuity
implies that $G_n$ has a root in $(0,1)$. Since $P_n$ and $G_n$ have the same
roots on $[0,1]$, the unique real root of the cubic must lie in $(0,1)$.
Consequently, $r_n\to_p1$.

 Let $\beta^\st_n$ denote the unique root of $G_n$
in $(0,1)$ with probability approaching one. Putting $x_n\coloneqq\beta^\st_n-\beta_0
=O_p(n^{-1/6})$ by \eqref{eq:loc}, a third-order expansion gives, uniformly over
$|x|\leq Mn^{-1/6}$,
\[
G_n(\beta_0+x)=c_3x^3+B^\Tr\Delta_n
+xG_{\beta\lambda}(\beta_0)^\Tr\Delta_n+O(x^4)+O_p(x^2\|\Delta_n\|)
+O_p(\|\Delta_n\|^2),
\]
where at this rate all terms after the first two are $o_p(n^{-1/2})$. Hence
$\sqrt n\,c_3x_n^3=-B^\Tr\sqrt n\Delta_n+o_p(1)\to_d-B^\Tr\mathbb Z$, and
continuity gives
$n^{1/6}(\beta^\st_n-\beta_0)\to_d-(c_3^{-1}B^\Tr\mathbb Z)^{1/3}$. 
\hfill$\square$

\subsubsection{Results of Section~\ref{sec:inf}}\label{app:inf}

{\bf Proof of Corollary \ref{cor:SEs}.} The claim follows from Proposition~\ref{prop:norm} and \citet[Thm 7.4]{newmc:94}. \hfill $\square$

\begin{remark}\label{rem:prof} As in \textup{\cite{hansen:17} (}see also \textup{\citealp{mm:25})}, we can reduce the numerical complexity by concentrating out $\delta$ and $\psi$. That is, set 
 $(\delta_n,\psi_n)^\Tr \coloneqq (\delta_n(\gamma_n),\psi_n(\gamma_n))^\Tr$ and let $\gamma_n$ the estimator that minimises the profiled objective
    $$Q^\st_n(\gamma) \coloneqq \sum_{t = 1}^{n}(\pi_{t} - \delta_n(\gamma) h_{t-1}(\gamma)-\psi_n(\gamma)y_{t})^2,$$
    where, for a given $\gamma$, the $2\times 1$ OLS estimator is given by
    $$(\delta_n(\gamma),\psi_n(\gamma))^\Tr \coloneqq  \left[\sum_{t=1}^n (h_{t-1}(\gamma),y_t)^\Tr(h_{t-1}(\gamma),y_t)\right]^{-1}\sum_{t=1}^n(h_{t-1}(\gamma),y_t)^\Tr\pi_t.$$
\end{remark}

{\bf Proof of Corollary \ref{cor:sup}.} Note that $F_n(\gamma) = t_n(\gamma)^2$, where, by Frisch-Waugh-Lovell,
\[
t_n(\gamma) = \frac{\frac1{\sqrt n}\sum_{t=1}^n u_t \tilde h_{t-1,n}(\gamma)}{\hat \sigma_u(\gamma) (\frac1{n}\sum_{t=1}^n \tilde h_{t-1,n}(\gamma)^2)^{1/2}}.
\]
Here, $\tilde h_{t,n}(\gamma) \coloneqq h_t(\gamma) - \tilde b_n(\gamma)^\Tr z_t$, $\tilde b_n(\gamma) \coloneqq (\sum_{t=1}^n z_tz_t^\Tr)^{-1}\sum_{t=1}^n z_th_{t-1}(\gamma)$ are the residuals upon partialling out $z_t = (1,y_t)^\Tr$ and
\[
\hat\sigma_u^2(\gamma) \coloneqq \frac1{n}\sum_{t=1}^n(\pi_t-x_t(\gamma)^\Tr b_n(\gamma))^2, \quad b_n(\gamma) \coloneqq (\sum_t x_t(\gamma)x_t(\gamma)^\Tr)^{-1}\sum_t x_t(\gamma)\pi_t,
\]
where $x_t(\gamma) = (h_{t-1}(\gamma),z_t^\Tr)^\Tr$.
Let $\tilde h_t(\gamma)$ denote the population residual $\tilde h_{t}(\gamma) \coloneqq h_t(\gamma) - \tilde b(\gamma)^\Tr z_t$, $\tilde b(\gamma) \coloneqq \Ex[z_tz_t^\Tr]^{-1}\Ex[z_th_{t-1}(\gamma)]$ and define the infeasible $t$-statistic:
\[
\tilde t_n(\gamma) \coloneqq  \frac{\frac1{\sqrt n}\sum_{t=1}^n u_t \tilde h_{t-1}(\gamma)}{\sigma_u \omega(\gamma)}, \quad \omega^2(\gamma) \coloneqq \Ex[\tilde h_t(\gamma)^2].
\]
Set $b(\gamma) \coloneqq \Ex[x_t(\gamma)x_t(\gamma)^\Tr]^{-1}\Ex[x_t(\gamma)\pi_t]$ and let us first prove the following:
\begin{align*}
    \textup{($i$)}\; \ssup\limits_{\gamma \in \Gamma}\|\tilde b_n(\gamma)-\tilde b(\gamma)\| = o_p(1)\quad \textup{($ii$)}\; \ssup\limits_{\gamma \in \Gamma}\|b_n(\gamma)-b(\gamma)\| = o_p(1)
\end{align*}
{\it Proof of $(i)$ and $(ii)$.} Next, 
\begin{align*}
\|\tilde b_n(\gamma)-\tilde b(\gamma)\| \leq \,& \|(\frac1{n}\sum_{t=1}^nz_tz_t^\Tr)^{-1}-\Ex[z_tz_t^\Tr]^{-1}\| \|\frac1{n}\sum_{t=1}^n z_th_{t-1}(\gamma)\| \\
\,& +  \|(\frac1{n}\sum_{t=1}^nz_tz_t^\Tr)^{-1}\| \|\frac1{n}\sum_{t=1}^n z_th_{t-1}(\gamma)-\Ex[z_th_{t-1}(\gamma)]\|.
\end{align*}
By the LLN, $\frac1{n}\sum_{t=1}^nz_tz_t^\Tr \rightarrow_p \Ex[z_tz_t^\Tr] \succ 0$. Because, by Lemma~\ref{lem:mom}, $\Ex[\ssup_\gamma h_t(\gamma)^2] < \infty$, the LLN for stationary ergodic processes and \citet[Thm 2]{hansen:1996b} yield $\ssup_\gamma\|\frac1{n}\sum_{t=1}^n z_th_{t-1}(\gamma)-\Ex[z_th_{t-1}(\gamma)]\|=o_p(1)$.  The proof of part ($ii$) is similar. 

Next, use ($i$) and ($ii$) to show uniformly in $\gamma \in \Gamma$ 
\[
\textup{(a) }\; |\frac1{n}\sum_{t=1}^n \tilde h_{t-1,n}(\gamma)^2-\omega(\gamma)^2| = o_p(1) \quad \textup{(b) }\; |\frac1{\sqrt n}\sum_{t=1}^n u_t (\tilde h_{t-1,n}(\gamma)-\tilde h_{t-1}(\gamma))| = o_p(1).
\]
{\it Proof of parts $(a)$ and $(b)$.} Note that $| \tilde h_{t,n}(\gamma)^2-\omega(\gamma)^2| \leq  |\tilde h_{t,n}(\gamma)^2-\tilde h_t(\gamma)^2|+ |\tilde h_{t}(\gamma)^2-\omega(\gamma)^2|$. Because, by Lemma~\ref{lem:mom}, $\Ex[\ssup_\gamma h_t(\gamma)^4] < \infty$, the LLN for stationary ergodic processes and \citet[Thm 2]{hansen:1996b} yield  $\ssup_\gamma|\frac1{n}\sum_{t=1}^n \tilde h_{t-1}(\gamma)^2-\omega^2(\gamma)| = o_p(1)$. Moreover, $|\tilde h_{t,n}(\gamma)^2-\tilde h_t(\gamma)^2| \leq |\tilde h_{t,n}(\gamma)-\tilde h_t(\gamma)||\tilde h_{t,n}(\gamma)+\tilde h_t(\gamma)|$, where
$|\tilde h_{t,n}(\gamma)-\tilde h_t(\gamma)| \leq |\tilde b_n(\gamma)-\tilde b(\gamma)| \| z_t \|.$ By part ($i$) and the LLN,
\[
\ssup\limits_{\gamma \in \Gamma}\frac1{n}\sum_{t=1}^n(\tilde h_{t,n}(\gamma)-\tilde h_t(\gamma))^2 \leq \|\tilde b_n(\gamma)-\tilde b(\gamma)\|^2 \|\frac1{n}\sum_{t=1}^nz_tz_t^\Tr\| = o_p(1).
\]
Part ($a$) follows from Cauchy-Schwarz. Turning to part ($b$), note that
$$\ssup\limits_{\gamma \in \Gamma}\|\frac1{\sqrt n}\sum_{t=1}^n u_t (\tilde h_{t-1,n}(\gamma)-\tilde h_{t-1}(\gamma))\| \leq  \ssup\limits_{\gamma \in \Gamma}\|\tilde b_n(\gamma)-b(\gamma)\|\| \frac1{\sqrt n}\sum_{t=1}^n z_tu_t\| = o_p(1),
$$
using that, by the CLT for MDS, $\frac1{\sqrt n}\sum_{t=1}^n z_tu_t = O_p(1)$. This proves part ($b$). 

Now, by continuity of $\gamma \mapsto \omega(\gamma)$ and compactness of $\Gamma$, Lemma~\ref{lem:pd} ($a$) yields $\iinf_\gamma \omega(\gamma)>0$. Therefore, ($a$), ($b$), and the uniform consistency of $\hat\sigma_u(\gamma)$ imply that, uniformly in $\gamma$, $t_n(\gamma) = \tilde t_n(\gamma) + o_p(1)$. It remains to be proven that
 $\tilde t_n(\gamma) \rightsquigarrow {\mathbb U}(\gamma)$ in $\ell^\infty(\Gamma)$. This is, however, a direct consequence of the multivariate CLT for MDS ({\it fidi}-convergence) and \citet[Thm 2]{hansen:1996b}. This proves the claim. \hfill $\square$

{\it Validity of multiplier bootstrap.} We follow \cite{hansen:17} and propose the use of a Gaussian multiplier bootstrap to obtain critical values: For $b \in \{1,\dots,B\}$, let $\{\pi_{1,b},\dots,\pi_{n,b}\}$, be a draw of standard normal variates. Use $\pi_{t,b}$ in place of $\pi_t$ to compute the bootstrap equivalent $sup{\sf F}^b$, say, of {\it sup}{\sf F}. For $B$ large enough, we can then approximate the bootstrap $p$-value using $p_{n,B} \coloneqq \frac1{B}\sum_{b=1}^B1\{sup{\sf F} \leq sup{\sf F}^b\}$, and reject at a significance level $\alpha \in (0,1)$ if $p_{n,B} \leq \alpha$. As shown in the appendix, $p_{n,B}$ is consistent for the true $p$-value as $n$ and $B$ diverge. For each $b \in \{1,\dots,B\}$,
\[
t_{n,b}(\gamma) \coloneqq \frac{\frac1{\sqrt n}\sum_{t=1}^n \pi_{t,b} \tilde h_{t-1,n}(\gamma)}{\hat\sigma_{u,b}(\gamma)(\frac1n\sum_{t=1}^n \tilde h_{t-1,n}(\gamma)^2)^{1/2}},
\]
where $\hat\sigma_{u,b}(\gamma)^2$ is the residual variance from the regression of $\pi_{t,b}$ on $z_t$; $\pi_{t,b} \stackrel{{\sf IID}}{\sim} {\mathcal N}(0,1)$ are bootstrap multiplieres independent of the data. Since $\ssup_\gamma|\hat\sigma_{n,b}(\gamma)-1| = o_{\mathbb P}(1)$, following the proof of the corollary, $\ssup_\gamma|t_{n,b}(\gamma) -\tilde t_{n,b}(\gamma)| = o_{\mathbb P}(1)$, with 
\[
\tilde t_{n,b}(\gamma) \coloneqq \frac1{\sqrt n}\sum_{t=1}^n \zeta_{t,b}(\gamma), \quad \zeta_{t,b}(\gamma) \coloneqq \frac{\pi_{t,b} \tilde h_{t-1}(\gamma)}{\omega(\gamma)}.
\]
For each $n$, conditionally on the data, $\tilde t_{n,b}(\gamma)$ is thus Gaussian with  $\Ex^\st[\tilde t_{n,b}(\gamma)] = 0$ and $\cov^\st[\tilde t_{n,b}(\gamma_1),\tilde t_{n,b}(\gamma_2)] = \frac{\tilde h_{t-1}(\gamma_1)\tilde h_{t-1}(\gamma_2)}{\omega(\gamma_1)\omega(\gamma_2)}$, where $\Ex^\st[\cdot]$  and $\cov^\st[\cdot]$ denote expectation and covariance  based on the empirical measure ${\mathbb P}$ induced by the data. Next, recall that $\cov[{\mathbb U}(\gamma_1),{\mathbb U}(\gamma_2)] = \Ex[\tilde h_{t-1}(\gamma_1)\tilde h_{t-1}(\gamma_2)]/(\omega(\gamma_1)\omega(\gamma_2))$. Moreover, 
\[
\ssup\limits_{\gamma_1,\gamma_2 \in \Gamma} |\frac1{n}\sum_{t=1}^n(\cov^\st[\zeta_{t,b}(\gamma_1),\zeta_{t,b}(\gamma_2)]-\cov[{\mathbb U}(\gamma_1),{\mathbb U}(\gamma_2)])| = o_p(1),
\]
which follows from the two-parameter case of \citet[Thm 2]{hansen:17} with $q > 2$, as $\Ex[\ssup_{\gamma_1,\gamma_2}|\tilde h_{t}(\gamma_1)\tilde h_{t}(\gamma_2)|^q]\leq \Ex[\ssup_{\gamma} |\tilde h_{t}(\gamma)|^{2q}]<\infty$ for some $q \in (2,4]$ by Lemma~\ref{lem:mom}. We therefore obtain, $\tilde t_{n,b}(\gamma) \rightsquigarrow_{\mathbb P} {\mathbb U}(\gamma)$ in $\ell^\infty(\Gamma)$, which proves the claim.

{\bf Proof of Corollary \ref{cor:CI}.}  First, we show
$$
\sqrt{n}(G_n(\beta)-G(\beta)) \rightsquigarrow {\mathbb G}(\beta) \quad \text{ in }\; \ell^\infty([0,1]).
$$
Because $\lambda \mapsto G(\beta;\lambda)$ is differentiable at $\lambda_0$, by the mean value theorem,
$$
G_n(\beta)-G(\beta) = G_{\lambda}(\beta; \bar\lambda_{n,\beta})^\Tr(\lambda_n-\lambda_0)=-F_{\lambda}(\beta; \bar\lambda_{n,\beta})^\Tr(\lambda_n-\lambda_0),
$$
where $\bar\lambda_{n,\beta}\coloneqq \lambda_0+t_{n,\beta}(\lambda_n-\lambda_0)$ for some (random) $t_{n,\beta} \in [0,1]$ that depends on $\beta \in [0,1]$. Clearly, $\ssup_\beta \|\bar\lambda_{n,\beta}-\lambda_0\| \leq \|\lambda_n-\lambda_0\| = O_p(n^{-1/2})$. Moreover, $G_\lambda(\beta,\lambda)$ is continuos on the compact set $[0,1] \times B$, where $B$ is a compact ball around $\lambda_0$ such that $\PP\{\lambda_n \in B\} \rightarrow 1$; thus, by Heine-Cantor, $G(\beta,\lambda)$ is uniformly continuous on the compact set $[0,1] \times B$, so that
$\ssup_\beta\|G_{\lambda}(\beta; \bar\lambda_{n,\beta})-G_{\lambda}(\beta,\lambda_0)\| = o_p(1),$
and therefore uniformly in $\beta$, $\sqrt{n}(G_n(\beta)-G(\beta)) = -F_{\lambda}(\beta)^\Tr \sqrt{n}(\lambda_n-\lambda_0)+o_p(1).$ Thus, $\sqrt{n}(G_n(\beta)-G(\beta)) \rightsquigarrow {\mathbb G}(\beta)$ in $\ell^\infty([0,1])$, where $\cov[{\mathbb G}(\beta_1),{\mathbb G}(\beta_2)] = F_\lambda(\beta_1)^\Tr \Omega F_\lambda(\beta_2)$. Next, for each $\beta \in [0,1]$
$$
 G(\beta) \in [G_n(\beta) \pm c_\alpha n^{-1/2}s_n(\beta)] \Leftrightarrow  T_n(\beta) \coloneqq \frac{\sqrt{n}|G_n(\beta)-G(\beta)|}{s_n(\beta)} \leq c_\alpha.
$$
Next, the statement $\{\forall \beta: T_n(\beta) \leq c_\alpha\}$ is equivalent to $\{\ssup_\beta T_n(\beta) \leq c_\alpha\}$. Because $\PP\{T \leq c_\alpha\}=1-\alpha$, $\ssup_\beta|s_n(\beta)-s(\beta)|=o_p(1)$ and, as a result of $\Omega \succ 0$ and Assumption~\ref{ass:para2}, $\iinf_\beta s(\beta)\geq \sqrt{\lambda_{min}(\Omega)}\iinf_\beta \|F_\lambda(\beta)\|>0$, it follows, by the continuous mapping theorem, $\ssup_\beta T_n(\beta) \rightarrow_d T$. This proves the claim. \hfill $\square$

{\it Proof of multiplier bootstrap.} Since the limiting distribution is unknown and non-pivotal, to obtain $c_\alpha$, we suggest to use a multiplier bootstrap based on the first order approximation:
\[
\sqrt{n}(G_n(\beta)-G(\beta)) = \frac1{\sqrt n}\sum_{t=1}^n m_t(\beta;\kappa_0)+o_p(1), 
\]
where
\[
 m_t(\beta;\kappa) \coloneqq  -F_\lambda(\beta;\lambda)^\Tr\phi_t(\kappa), \quad \kappa \coloneqq (\lambda^\Tr,\gamma)^\Tr,
\]
and $\frac1{\sqrt n}\sum_t \phi_t(\kappa_0)$ is the influence function of $\sqrt{n}(\lambda_n-\lambda_0)$. Thus, for a draw  $b \in \{1,\dots,B\}$ of standard Gaussian multipliers $\{\xi_{1,b},\dots,\xi_{n,b}\}$, and, conditionally on the data, we have
\[
\frac1{\sqrt n}\sum_{t=1}^n \xi_{t,b} m_t(\beta;\kappa_0) \rightsquigarrow_{\mathbb P} {\mathbb G}(\beta) \quad \textup{in} \quad \ell^\infty([0,1]),
\]
where $ \rightsquigarrow_{\mathbb P}$ means that weak convergence holds in probability with respect to the data-generating measure. In practice, neither $\kappa_0$ nor $\phi_t(\kappa_0)$ is observed. As shown in below, we can however construct $\phi_{t,n}(\kappa_n)$ such that
$\frac{1}{n}\sum_{t=1}^n\|\phi_{t,n}(\kappa_n)-\phi_t(\kappa_0)\|^2=o_p(1)$, and define the feasible linearization
$m_{t,n}(\beta)\coloneqq -F_\lambda(\beta;\lambda_n)^\top\phi_{t,n}(\kappa_n)$. We then simulate the critical value from the studentized bootstrap statistic
\[
T_{n,b} \coloneqq \ssup_{\beta\in[0,1]}\frac{\big|\frac{1}{\sqrt n}\sum_{t=1}^n \xi_{t,b} m_{t,n}(\beta)\big|}{s_n(\beta)},
\qquad
s_n(\beta)^2\coloneqq \frac{1}{n}\sum_{t=1}^n m_{t,n}(\beta)^2,
\]
and take $c_{\alpha,n,B}$ as the empirical $1-\alpha$ quantile of $\{T_{n,b}\}_{b=1}^B$. As verified in the appendix, if $B \rightarrow \infty$ and $n \rightarrow \infty$, $c_{\alpha,n,B}\rightarrow_p c_\alpha$, yielding the uniform bands in Corollary~\ref{cor:CI}.

First we show that if $\ssup_{\beta}\frac1{n}\sum_{t=1}^n (m_{t}(\beta)-m_{t,n}(\beta,\kappa_n))^2 = o_p(1)$, then ${\mathbb G}_{n,b}(\beta) \rightsquigarrow_{\mathbb P} \mathbb{G}(\beta)$ in $\ell^\infty([0,1])$. To see this, define first by ${\mathbb G}_{n,b}^\circ(\beta) \coloneqq \frac1{\sqrt n} \sum_{t=1}^n \xi_{t,b}m_t(\beta)$ and ${\mathbb G}_{n,b}(\beta) \coloneqq \frac1{\sqrt n} \sum_{t=1}^n \xi_{t,b}m_{t,n}(\beta,\kappa_n)$ the infeasible and feasible multiplier bootstrap processes, respectively. Conditionally on the data, ${\mathbb G}_{n,b}^\circ(\beta)$ is mean zero Gaussian with covariance kernel
\[
\cov^\st[{\mathbb G}_{n,b}^\circ(\beta_1),{\mathbb G}_{n,b}^\circ(\beta_2)] = \frac1{n}\sum_{t=1}^n m_t(\beta_1)m_t(\beta_2).
\]
By the LLN for stationary ergodic processes and \citet[Thm 2]{hansen:1996b} in conjunction with $\Ex[\ssup_\beta |m_t(\beta)|^{2q}]<\infty$ for some $q \in (2,4]$, the preceding display converges uniformly in $\beta_1,\beta_2 \in [0,1]$ to the covariance kernel of ${\mathbb G}(\beta)$, thus showing that ${\mathbb G}_{n,b}^\circ(\beta) \rightsquigarrow_{\mathbb P} {\mathbb G}(\beta)$ in $\ell^\infty([0,1])$. Next, recall that by construction the Gaussian multipliers $\xi_{t,b}$ are independent of the data, uncorrelated with unit variance such that
\[
\var^\st[{\mathbb G}_{n,b}(\beta)-{\mathbb G}_{n,b}^\circ(\beta)] = \frac1{n}\sum_{t=1}^n (m_{t}(\beta)-m_{t,n}(\beta,\kappa_n))^2.
\]
By hypothesis, the preceding display is, uniformly in $\beta$, $o_p(1)$, so that, by Markov's inequality $\ssup_\beta |{\mathbb G}_{n,b}(\beta)-{\mathbb G}_{n,b}^\circ(\beta)| = o_p(1)$, which verifies the claim.  Next, we have to find $m_{t,n}(\beta,\kappa_n)$ that satisfies the hypothesis. In doing so, note that
\begin{align*}
\frac1{2}(m_{t,n}(\beta,\kappa_n)-m_t(\beta))^2 \leq \,&  \| F_\lambda(\beta;\lambda_n)-F_\lambda(\beta) \|^2 \phi_t(\kappa_0)^2\\
\,& + \|F_\lambda(\beta;\lambda_n)\|^2 (\phi_{t,n}(\kappa_n)-\phi_t(\kappa_0))^2
\end{align*}
By Lemma~\ref{lem:Flambda} and consistency of $\lambda_n$, we have $\ssup_\beta\| F_\lambda(\beta;\lambda_n)-F_\lambda(\beta) \| = o_p(1)$, while, by the LLN for stationary ergodic processes, $\frac1{n}\sum_t^n \phi_t(\kappa_0)^2 = O_p(1)$. Because $\ssup_{\beta,\lambda \in [0,1]\times B}\|F_\lambda(\beta;\lambda)\|^2 < \infty$, it remains to find $\phi_{t,n}(\kappa_n)$ such that 
$$
\frac1{n}\sum_{t=1}^n(\phi_{t,n}(\kappa_n)-\phi_t(\kappa_0))^2 = o_p(1).
$$
It can be shown, that a suitable choice is
$$
  \phi_{t,n}(\kappa) \coloneqq  \begin{bmatrix}
  \phi_{t,n,\delta}(\theta)\\
    \phi_{t,n,\psi}(\theta)\\
    \phi_{t,n,\rho}(\rho) \\
    \phi_{t,n,\sigma_u^2}(\kappa) \\
    \phi_{t,n,\sigma_\eps^2}(\kappa)
\end{bmatrix}, \textup{ with } A_n^{-1}\dot f_{t}(\theta_n) u_{t}(\theta_n) = (\phi_{t,n,\gamma}(\theta_n),\phi_{t,n,\delta}(\theta_n),\phi_{t,n,\psi}(\theta_n))^\Tr,
$$
and $u_t(\theta) = \pi_t - f_t(\theta)$,
$$\phi_{t,n,\rho}(a,\rho) = \frac{(y_{t-1}-\bar y)\eps_t(a,\rho)}{\frac1{n}\sum_{t=1}^n (y_{t-1}-\bar y)^2}, \quad \eps_t(a,\rho) \coloneqq y_t - a-\rho y_{t-1},$$
while
$$\phi_{t,n,\sigma_u^2}(\theta) \coloneqq u_t(\theta)^2-\hat\sigma_u^2, \quad \phi_{t,n,\sigma_\eps^2}(\rho) \coloneqq \eps_t(\rho)^2-\hat\sigma_\eps^2.$$

\subsection{Proofs of auxiliary results}\label{sec:appaux}

{\bf Proof of Lemma~\ref{lemA:attract}.}

 \textbf{C1} (\underline{Step 1.}) Fix \(x^{(0)},x^{(1)}\in\mathbb R^2\) such that $x_j^{(0)}\neq x_j^{(1)},$ $j\in\{\pi,y\},$
and fix arbitrary \(i^{(0)},i^{(1)}\in\mathbb R\). Throughout,
parenthesized superscripts are interpreted modulo two, so that
\[
x^{(k)} \coloneqq x^{(k \operatorname{\sf mod} 2)},
\qquad k\in\mathbb Z,
\]
and analogously for all other quantities carrying parenthesized
superscripts. Here, \(k \operatorname{\sf mod} 2\) equals \(0\) if \(k\) is even and
\(1\) if \(k\) is odd. Define $v\coloneqq \frac{x^{(0)}-x^{(1)}}{1+q}$ and note that, by construction, $v_j\neq0,$ $j\in\{\pi,y\}.$ Since $x^{(s)}$ has period two, we look for a two-periodic solution of each learning recursion $(\alpha_s,\omega_s,r_s)$. Thus, we assume that
$\alpha_s$ (and, below, $r_s,\omega_s$) takes only two values ($\alpha_s = \alpha^{(0)}$ for even $s$ and $\alpha_s = \alpha^{(1)}$ for odd $s$) and solve for these values by substituting this two-periodic form into the respective recursions. In particular, the mean recursion requires
$\alpha^{(s)} = q\alpha^{(s-1)}+\gamma x^{(s)}.$
Consequently, \(\alpha^{(0)}\) and \(\alpha^{(1)}\) satisfy 
\[
\alpha^{(0)}
=
q\alpha^{(1)}+\gamma x^{(0)},
\qquad
\alpha^{(1)}
=
q\alpha^{(0)}+\gamma x^{(1)}.
\]
Solving these equations gives
\[
\alpha^{(0)}
=
\frac{x^{(0)}+q x^{(1)}}{1+q},
\qquad
\alpha^{(1)}
=
\frac{q x^{(0)}+x^{(1)}}{1+q}.
\]
The prediction errors associated with these values are
$\nu^{(s)}\coloneqq x^{(s)}-\alpha^{(s-1)}.$
It follows that
\[
\nu^{(0)}
=
x^{(0)}-\alpha^{(1)}
=
\frac{x^{(0)}-x^{(1)}}{1+q}
=
v,
\]
whereas $\nu^{(1)} = x^{(1)}-\alpha^{(0)} = -v.$
The recursion for \(r_j\) requires $r_j^{(s)}
= q r_j^{(s-1)} + \gamma(\nu_j^{(s)})^2.$
Since $\nu^{(0)}=-\nu^{(1)}=v$, the two values \(r_j^{(0)}\) and \(r_j^{(1)}\) satisfy
\[
r_j^{(0)}
=
q r_j^{(1)}+\gamma v_j^2,
\qquad
r_j^{(1)}
=
q r_j^{(0)}+\gamma v_j^2.
\]
Subtracting the two equations yields $(1+q)(r_j^{(0)}-r_j^{(1)})=0,$ so that $r_j^{(0)}=r_j^{(1)} \eqqcolon r_j,$ say, for which we get $r_j=q r_j+\gamma v_j^2 = v_j^2$.
Since \(1-q=\gamma\), Therefore,
\[
r^{(0)}=r^{(1)}=r \coloneqq (v_\pi^2,v_y^2)^\Tr > 0.
\]
Similarly, we get for the first autocovariance recursion
\[
\omega_j^{(s)}
=
q\omega_j^{(s-1)}
+
\gamma q\,\nu_j^{(s)}\nu_j^{(s-1)}.
\]
the following solution 
\[
\omega^{(0)}=\omega^{(1)}
=
\omega
=
-qr.
\]
For every \(j\in\{\pi,y\}\) and \(s\in\{0,1\}\),
\[
\begin{aligned}
r_j
-\frac{|\omega_j|}{\sqrt q}
-\frac{\gamma}{2q^2}
(x_j^{(s)}-\alpha_j^{(s)})^2
&=
v_j^2-\sqrt q\,v_j^2-\frac{\gamma}{2}v_j^2
\\
&=
(1-\sqrt q-\frac{\gamma}{2})v_j^2
=
\frac12(1-\sqrt q)^2v_j^2
>0.
\end{aligned}
\]
The last inequality follows because \(q\in(0,1)\) and \(v_j\neq0\).
Hence the two alternating values of the learning state lie in the
interior of \({\sf S}_0\). Furthermore, $\beta_j^2 \coloneqq \frac{\omega_j^2}{r_j^2}=q^2 = (1-\gamma)^2$.

The corresponding values of \(h = (h_\pi,h_y)^\Tr\) are therefore
\[
h_j^{(s)} = \alpha_j^{(s)}
+
\beta_j^2 (x_j^{(s)}-\alpha_j^{(s)})=
\alpha_j^{(s)}+q^3\nu_j^{(s)}.
\]
The shock values \(u^{(0)}\) and \(u^{(1)}\) supporting the prescribed
sequences \(x^{(s)}\) and \(i^{(s)}\) are uniquely recovered from the
structural equations. Specifically,
\[
{\cal L}u^{(s)}
=
\begin{bmatrix}
M[x^{(s)}-Hh^{(s-1)}-g\,i^{(s-1)}]
\\ 
i^{(s)}-\rho_i i^{(s-1)}
-(1-\rho_i)\phi^\Tr x^{(s)}
\end{bmatrix},
\]
where
\[
{\cal L}
\coloneqq
\begin{bmatrix}
L\\
e_3^\Tr
\end{bmatrix}
\]
is nonsingular. The corresponding controls are
$
w^{(s)}
=
u^{(s)}-Ru^{(s-1)}.$

Define
\[
S^{(s)}
\coloneqq
(
x^{(s)\Tr},
\alpha^{(s)\Tr},
\omega^\Tr,
r^\Tr,
u^{(s)\Tr},
i^{(s)}
)^\Tr,
\qquad s\in\{0,1\}.
\]
Then \(S^{(0)}\) and \(S^{(1)}\) satisfy the control model and
alternate under the controls \(w^{(1)}\) and \(w^{(0)}\). Both states
belong to \({\sf X}_0\). Finally, set (we could have chosen also $S^{(1)}$)
\[
Y^\star
\coloneqq
\pi(S^{(0)})
\in{\sf Y}_0.
\]
The next step is to show that, from any initial state and for every
sufficiently large horizon, an admissible control sequence can be
chosen whose terminal state is arbitrarily close to \(Y^\star\).

\textbf{C1} (\underline{Step 2.})  Let \(Y_0\in{\sf Y}_0\) be arbitrary and set
\(S_0=\iota(Y_0)\in{\sf X}_0\).  For each terminal horizon
\(k\ge1\), define $p_s^{[k]}\coloneqq (s-k)\operatorname{\sf mod}2,$ $s=0,\ldots,k,$ so that \(p_k^{[k]}=0\).  Starting from \(S_0\), recursively choose
\(u_s^{[k]}\) as the unique solution of
\[
{\cal L}u_s^{[k]}
=
\begin{bmatrix}
M[
x^{(p_s^{[k]})}
-Hh_{s-1}^{[k]}-g i_{s-1}^{[k]}
]
\\
i^{(p_s^{[k]})}
-\rho_i i_{s-1}^{[k]}
-(1-\rho_i)\phi^\Tr x^{(p_s^{[k]})}
\end{bmatrix},
\qquad
\widetilde w_s^{[k]}
\coloneqq u_s^{[k]}-Ru_{s-1}^{[k]}.
\]
Since \({\cal L}\) is nonsingular, this defines a \(k\)-step control
vector
\(w^{[k]}
=( w_1^{[k]},\ldots, w_k^{[k]})\), and the
corresponding controlled path satisfies
\[
x_s^{[k]}=x^{(p_s^{[k]})},
\qquad
i_s^{[k]}=i^{(p_s^{[k]})},
\qquad s=1,\ldots,k.
\]
Moreover, forward invariance implies \(S_s^{[k]}\in{\sf X}_0\).

To control the remaining coordinates, define
\[
\Delta_s^{[k]}
\coloneqq
\alpha_s^{[k]}-\alpha^{(p_s^{[k]})}.
\]
Subtracting the mean recursion of the two-periodic solution from the
actual recursion gives
\[
\Delta_s^{[k]}=q\Delta_{s-1}^{[k]},
\qquad
\Delta_s^{[k]}
=
q^s\bigl(\alpha_0-\alpha^{(p_0^{[k]})}\bigr).
\]
Since \(p_0^{[k]}\in\{0,1\}\), the initial discrepancy is bounded
uniformly in \(k\).  Writing
\[
\nu_s^{[k]}=x_s^{[k]}-\alpha_{s-1}^{[k]},
\]
we therefore have
\[
\nu_s^{[k]}-\nu^{(p_s^{[k]})}
=-\Delta_{s-1}^{[k]}
=O(q^{s-1}),
\]
uniformly over \(k\ge s\).  Using
\[
\nu^{(0)}=v,\qquad \nu^{(1)}=-v,\qquad
r=v^{\circ2},\qquad \omega=-qr,
\]
the stable \(r\)- and \(\omega\)-recursions consequently yield, for
any fixed \(\varrho\in(q,1)\),
\[
\|r_s^{[k]}-r\|
+
\|\omega_s^{[k]}-\omega\|
\le C(Y_0)\varrho^s,
\qquad s\le k,
\]
where the one-time initial discrepancy in the \(\omega\)-recursion is
absorbed into \(C(Y_0)\).  Since \(r_j=v_j^2>0\), continuity of the
remaining transformations and nonsingularity of \({\cal L}\) then
give
\[
\|
{\cal G}_Y^{(s)}
\bigl(
Y_0;
w_1^{[k]},\ldots, w_s^{[k]}
\bigr)
-
\pi(S^{(p_s^{[k]})})
\|
\le C(Y_0)\varrho^s.
\]
Evaluating the preceding bound at \(s=k\) and using
\(p_k^{[k]}=0\), we obtain
\[
\|
{\cal G}_Y^{(k)}
\bigl(Y_0;w_1^{[k]},\ldots,w_k^{[k]}\bigr)
-Y^\star
\|
\leq C(Y_0)\varrho^k
\longrightarrow0.
\]
Since \(f_\eps(w)>0\) for every \(w\in\mathbb R^3\), the constructed
controls \(w_1^{[k]},\ldots,w_k^{[k]}\) belong to \({\cal O}\).
Therefore, for every open neighbourhood \(U\) of \(Y^\star\), there
exists \(T\in\mathbb Z_{>0}\) such that, for every \(k\geq T\),
\[
{\cal G}_Y^{(k)}
(Y_0;w_1^{[k]},\ldots,w_k^{[k]})\in U.
\]
Hence \(Y^\star\) is steadily attracting. \hfill $\square$

{\bf Proof of Lemma~\ref{lemA:control}.} Over any finite horizon \(m\in\mathbb Z_{>0}\), let
\[
c_t\coloneqq(\nu_t^\Tr,i_t)^\Tr, \quad \nu_t \coloneqq x_t-\alpha_{t-1}.
\]
Since $S_{t-1}$ depends only on
\((w_1,\ldots,w_{t-1})\), \(c_t = c(w_1,\ldots,w_t)\). Hence the
Jacobian ${\cal J}_m \coloneqq \partial(c_1,\ldots,c_m)/\partial(w_1,\ldots,w_m)$ of $(w_1,\ldots,w_m)\longmapsto(c_1,\ldots,c_m)$
is block lower triangular, with diagonal blocks
\begin{equation}
\frac{\partial c_t}{\partial w_t}=J_w
\coloneqq
\begin{bmatrix}
M^{-1}L\\
(1-\rho_i)\phi^\Tr M^{-1}L+e_3^\Tr
\end{bmatrix}.
\label{eq:rank-Jw}
\end{equation}
Since ${\sf det}(J_w) = {\sf det}(M^{-1}){\sf det}({\cal L}) =\Delta^{-1}\neq0$, \(J_w\) is nonsingular, so the two control sequences are equivalent given $S_{t-1}$. Indeed, by the chain rule,
\[
\frac{\partial Y_m}{\partial(w_1,\ldots,w_m)}
=
\frac{\partial Y_m}{\partial(c_1,\ldots,c_m)}{\cal J}_m.
\]
Since ${\cal J}_m$ is nonsingular, right multiplication by it
does not change rank. Thus, the controllability condition may equivalently be verified
using \((\nu_t,i_t)\) as local control coordinates.

To proceed, recall that for each \(j \in \{\pi,y\}\),
\[
x_{j,t}=\alpha_{j,t-1}+\nu_{j,t},
\qquad
\alpha_{j,t}=\alpha_{j,t-1}+\gamma\nu_{j,t},
\]
\[
r_{j,t}=qr_{j,t-1}+\gamma\nu_{j,t}^2,
\quad
\omega_{j,t}
=
q\omega_{j,t-1}
+\gamma\nu_{j,t}d_{j,t-1},
\quad
d_{j,t}=x_{j,t}-\alpha_{j,t}= q\nu_{j,t}.
\]
Starting from \(Y^\star\), choose $\nu_{j,1}=0,$ $\nu_{j,2}=1,$ and $\nu_{j,3}=0.$
All forcing terms in the \(\omega\)-recursion vanish for $t\leq 4$, so
$\omega_{j,4}=q^4\omega_{j,0}.$ Writing \(d_{j,0}=x_{j,0}-\alpha_{j,0}\), direct differentiation of
the four-step recursion gives
$$
\frac{\partial
(x_{j,4},\alpha_{j,4},\omega_{j,4},r_{j,4})}
{\partial
(\nu_{j,1},\dots,\nu_{j,4})} \Bigg|_{\nu_1=0,\nu_2=1,\nu_3=0} = \begin{bmatrix}
\gamma&\gamma&\gamma&1\\
\gamma&\gamma&\gamma&\gamma\\
\gamma q^3(d_{j,0}+1)&0&\gamma q(q+\nu_{j,4})&0\\
0&2\gamma q^2&0&2\gamma\nu_{j,4}
\end{bmatrix},
$$
with determinant $-2\gamma^3q^4 (q^2d_{j,0}-\gamma q-\nu_{j,4}).$ Because we start at \(Y^\star=\pi(S^{(0)})\), the proof of {\bf C1} (Step 1) implies $\omega_{j,0}=-qv_j^2\neq0,$
and hence \(\omega_{j,4}=q^4\omega_{j,0}\neq0\).  If we choose
$\nu_{j,4} \notin \{0,\,q^2d_{j,0}-\gamma q\},$ then the determinant is nonzero and \(d_{j,4}=q\nu_{j,4}\neq0\).

After the first four controls, consider $h_{j,4}
=
\alpha_{j,4}
+
\frac{\omega_{j,4}^2}{r_{j,4}^2}
(x_{j,4}-\alpha_{j,4}).$ Note that \(h_{j,4}\) is already determined by
\((\nu_{j,1},\ldots,\nu_{j,4})\). We add a fifth control
\(\nu_{j,5}\), because \(h_4\)
enters the effective state only at date \(5\), through $\eta_5
=
\Sigma V^\Tr
\begin{bmatrix}
h_4\\ i_4
\end{bmatrix}.$

We therefore factor the five-step map as
\[
(\nu_{j,1},\ldots,\nu_{j,5})
\longmapsto
(x_{j,4},\alpha_{j,4},\omega_{j,4},r_{j,4},\nu_{j,5})
\longmapsto
(x_{j,5},\alpha_{j,5},\omega_{j,5},r_{j,5},h_{j,4}).
\]
Put differently, \(\nu_{j,1},\ldots,\nu_{j,4}\) determine
\(x_{j,4},\alpha_{j,4},\omega_{j,4},r_{j,4}\), and hence also
\(h_{j,4}\). Together with \(\nu_{j,5}\), these date-\(4\)
quantities determine
\(x_{j,5},\alpha_{j,5},\omega_{j,5},r_{j,5}\). Now, applying the chain rule gives
\begin{align}
&
{\sf det}
\frac{\partial
(x_{j,5},\alpha_{j,5},\omega_{j,5},r_{j,5},h_{j,4})}
{\partial
(\nu_{j,1},\ldots,\nu_{j,5})}
\nonumber\\
&\quad=
{\sf det}
\frac{\partial
(x_{j,4},\alpha_{j,4},\omega_{j,4},r_{j,4})}
{\partial
(\nu_{j,1},\ldots,\nu_{j,4})}
\,
\frac{q^2\omega_{j,4}}{r_{j,4}^2}
(
2\gamma\nu_{j,5}d_{j,4}
-q\omega_{j,4}
).
\label{eq:rank-J5}
\end{align}
Since \(d_{j,4}\neq0\), the last factor is a nonconstant affine
function of \(\nu_{j,5}\).  It is therefore nonzero whenever
\[
\nu_{j,5}
\neq
\frac{q\omega_{j,4}}{2\gamma d_{j,4}}.
\]
Thus, for each \(j\in\{\pi,y\}\), the map
\[
(\nu_{j,1},\ldots,\nu_{j,5})
\longmapsto
(x_{j,5},\alpha_{j,5},\omega_{j,5},r_{j,5},h_{j,4})
\]
has rank five at the selected control values.

Given \(i_1,i_2,i_3\), the map
\[
(\nu_{1},\ldots,\nu_{5},i_4,i_5)
\mapsto
z^\dagger
\coloneqq
(x_5^\Tr,\alpha_5^\Tr,\omega_5^\Tr,r_5^\Tr,
 h_4^\Tr,i_4,i_5)^\Tr
\]
has, a block-triangular
Jacobian whose diagonal blocks are the two nonsingular matrices in
\eqref{eq:rank-J5} and \(I_2\) (because of $i_4$, $i_5$).  It is therefore invertible at the selected controls.

Now, re-order for convenience the coordinates of \(z^\dagger = (\alpha_5^\Tr,\omega_5^\Tr,r_5^\Tr,x_5^\Tr,i_5,h_4^\Tr,i_4)^\Tr\). The structural equations imply
\[
u_5
=
{\cal L}^{-1}
\begin{bmatrix}
M(x_5-Hh_4-g\,i_4)\\
i_5-\rho_i i_4-(1-\rho_i)\phi^\Tr x_5
\end{bmatrix},
\qquad
\eta_5
=
\Sigma V^\Tr
\begin{bmatrix}h_4\\ i_4\end{bmatrix}.
\]
Consequently,
\begin{equation}
\frac{\partial Y_5}{\partial z^\dagger}
=
\begin{bmatrix}
I_6&0&0\\
0&W&\ast\\
0&0&\Sigma V^\Tr
\end{bmatrix},
\qquad
W
\coloneqq
{\cal L}^{-1}
\begin{bmatrix}
M&0\\
-(1-\rho_i)\phi^\Tr&1
\end{bmatrix}.
\label{eq:rank-final}
\end{equation}
Since
\[
{\sf det}(W)={\sf det}(M)=\Delta\neq0
\]
and, by the reduced singular-value decomposition of \(B\),
\[
(\Sigma V^\Tr)(\Sigma V^\Tr)^\Tr
=
\Sigma^2\succ0,
\]
the matrix \(\Sigma V^\Tr\) has full row rank \(r\).  Hence
\eqref{eq:rank-final} has rank
\[
6+3+r=9+r={\sf dim}({\sf Y}_0).
\]
Since the change from
\((w_1,\ldots,w_5)\) to
\((\nu_1,i_1,\ldots,\nu_5,i_5)\) is invertible, the same
rank holds in the original control coordinates:
\[
\operatorname{\sf rank}({\cal C}_5)=9+r.
\]
Moreover, under Assumption~\ref{ass:density},
\(f_\eps(w)>0\) for every \(w\in\mathbb R^3\), so the corresponding
five-step control sequence is element of ${\cal O}=\mathbb R^3$. This finishes the proof. \hfill $\square$

{\bf Proof of Lemma~\ref{lem:negR}.} Fix \(j\in\{\pi,y\}\) and an integer \(l\ge1\). The stationary
infinite-past representation of the variance recursion gives
\[
r_{j,t}(\gamma)
=
\gamma\sum_{i=0}^{\infty}
(1-\gamma)^i\nu_{j,t-i}(\gamma)^2.
\]
Hence, uniformly over \(\gamma\in\Gamma\),
\[
r_{j,t}(\gamma)
\ge
c_l\sum_{i=0}^l\nu_{j,t-i}(\gamma)^2,
\qquad
c_l
\coloneqq
\ubar\gamma(1-\bar\gamma)^l>0.
\]
Define
$z_{j,t}^{(l)}(\gamma)
\coloneqq(
\nu_{j,t-l}(\gamma),\ldots,\nu_{j,t}(\gamma))^\Tr$ and note that
\begin{equation}
r_{j,t}(\gamma)
\ge
c_l\|z_{j,t}^{(l)}(\gamma)\|^2.
\label{eq:Rmoment-lower}
\end{equation}
We first derive a small-ball bound for fixed \(\gamma\). Let $\mathcal F_t
\coloneqq
 \sigma(\varepsilon_s:s\le t) $  and recall from the proof of Lemma~\ref{lem:invariant} that the conditional density of $\nu_{j,t}$ is bounded by $\bar f_j$. Successive conditioning consequently
shows that, conditional on \(\mathcal F_{t-l-1}\), the joint density of
\(z_{j,t}^{(l)}(\gamma)\) is bounded by
\(\bar f_j^{\,l+1}\). Thus, for every \(r>0\),
\begin{equation}
\ssup_{\gamma\in\Gamma}
\PP\{
\|z_{j,t}^{(l)}(\gamma)\|\le r
\}
\le
C_l r^{l+1},
\label{eq:Rmoment-fixed-gamma}
\end{equation}
where \(C_l<\infty\) does not depend on \(t\) or \(\gamma\). Next, define
$L_{j,t}^{(l)}
\coloneqq
\ssup_{\gamma\in\Gamma}
\|
\partial_\gamma z_{j,t}^{(l)}(\gamma)
\|.$ From the definition of $\nu_{j,t}(\gamma)$, $\Ex[\|x_0\|^2]<\infty$, and the earlier arguments in the proof of Lemma~\ref{lem:filter} we get that
$\Ex[L_{j,t}^{(l)}]<\infty.$ Now, fix \(0<r<1\) and \(K\ge1\). On the event
\(\{L_{j,t}^{(l)}\le K\}\), the map
\(\gamma\mapsto z_{j,t}^{(l)}(\gamma)\) is \(K\)-Lipschitz. Cover
\(\Gamma\) by a deterministic grid
\(\gamma_1,\ldots,\gamma_N\) with mesh width at most \(r/K\), where, by a standard argument, the
number of grid points $N$, say, satisfies $N\le C\frac{K}{r}.$
If
\[
\iinf_{\gamma\in\Gamma}
\|z_{j,t}^{(l)}(\gamma)\|
\le r
\quad\text{and}\quad
L_{j,t}^{(l)}\le K,
\]
then, for some grid point \(\gamma_i\),
$\|z_{j,t}^{(l)}(\gamma_i)\|\le2r.$
Consequently, by~\eqref{eq:Rmoment-fixed-gamma},
\[
\begin{aligned}
\PP\{
\iinf_{\gamma\in\Gamma}
\|z_{j,t}^{(l)}(\gamma)\|
\le r,\,
L_{j,t}^{(l)}\le K
\}
\le
\sum_{i=1}^N
\PP\{
\|z_{j,t}^{(l)}(\gamma_i)\|\le2r
\}
\le
C_l K r^l.
\end{aligned}
\]
Moreover, Markov's inequality gives
\[
\PP\{L_{j,t}^{(l)}>K\}
\le
\frac{\Ex[L_{j,t}^{(l)}]}{K}.
\]
Choosing \(K=r^{-l/2}\), we obtain
\begin{equation}
\PP\left\{
\iinf_{\gamma\in\Gamma}
\|z_{j,t}^{(l)}(\gamma)\|
\le r
\right\}
\le
C_l r^{l/2}.
\label{eq:Rmoment-uniform-gamma}
\end{equation}

By~\eqref{eq:Rmoment-lower},
\[
\{\ubar r_t\le\varepsilon\}
\subset
\bigcup_{j\in\{\pi,y\}}
\left\{
\iinf_{\gamma\in\Gamma}
\|z_{j,t}^{(l)}(\gamma)\|
\le
\sqrt{\varepsilon/c_l}
\right\}.
\]
It follows from~\eqref{eq:Rmoment-uniform-gamma} that, for all
sufficiently small \(\varepsilon>0\),
\begin{equation}
\PP\{\ubar r_t\le\varepsilon\}
\le
C_l\varepsilon^{l/4}.
\label{eq:Rmoment-tail}
\end{equation}
For \(0<\varepsilon_0<1\), we get
\[
\Ex[\ubar r_t^{-\tau}]
\le
\varepsilon_0^{-\tau}
+
\tau\int_0^{\varepsilon_0}
\varepsilon^{-\tau-1}
\PP\{\ubar r_t\le\varepsilon\}
\,d\varepsilon.
\]
Choose \(l>4\tau\). By~\eqref{eq:Rmoment-tail}, the integrand near zero
is bounded by a constant times $\varepsilon^{l/4-\tau-1}$,which is integrable because \(l/4>\tau\). This proves the claim. \hfill $\square$

{\bf Proof of Lemma \ref{lem:mom}.} Put $q_\gamma \coloneqq 1-\gamma$. By Proposition~\ref{prop:ergod}, the moment conditions imposed on $\eps_t$ imply the corresponding moment conditions
on $\pi_t$. Note also that the claim of the Lemma~\ref{lem:mom} extends directly component-wise to the general model.

($i$) Using
\[
\alpha_t(\gamma)
=
\gamma\sum_{i=0}^{\infty}q_\gamma^i\pi_{t-i},
\]
we obtain, by termwise differentiation,
\[
\dot\alpha_t(\gamma)
=
\sum_{i=0}^{\infty}
q_\gamma^{i-1}\{1-(i+1)\gamma\}\pi_{t-i},
\]
\[
\ddot\alpha_t(\gamma)
=
\sum_{i=0}^{\infty}
q_\gamma^{i-2}i\{(i+1)\gamma-2\}\pi_{t-i},
\]
and
\[
\dddot\alpha_t(\gamma)
=
\sum_{i=0}^{\infty}
q_\gamma^{i-3}i\{3(i-1)-(i^2-1)\gamma\}\pi_{t-i}.
\]
Since $\Gamma=[\ubar\gamma,\bar\gamma]\subset(0,1)$, the preceding
series and their derivatives converge absolutely and uniformly in
$\gamma\in\Gamma$. Moreover, there are constants
$c_m\in(0,\infty)$ such that
\[
\ssup_{\gamma\in\Gamma}
|\alpha_t^{(m)}(\gamma)|
\leq
c_m\sum_{i=0}^{\infty}
(1+i)^m(1-\ubar\gamma)^i|\pi_{t-i}|,
\qquad
m\in\{0,1,2,3\}.
\]
Using (by Hölder's inequality)
\(|\sum_i a_i b_i|^k
\leq
(\sum_i a_i)^{k-1}
\sum_i a_i|b_i|^k,\)
$a_i\geq0,$ stationarity, and
\(\sum_{i=0}^{\infty}(1+i)^m(1-\ubar\gamma)^i<\infty,\)
we obtain
\(
\Ex[
\ssup_{\gamma\in\Gamma}
|\alpha_t^{(m)}(\gamma)|^k
]
<\infty,\) \(m\in\{0,1,2,3\}.
\)

($ii$) Write \(x_t(\gamma)\coloneqq\pi_t-\alpha_{t-1}(\gamma).\) The stationary variance recursion is
\[
r_t(\gamma)
=
q_\gamma r_{t-1}(\gamma)+\gamma x_t(\gamma)^2
=
\gamma\sum_{i=0}^{\infty}q_\gamma^i x_{t-i}(\gamma)^2.
\]
Lemma~\ref{lem:invariant} and the $c_r$-inequality therefore yield, respectively, a.s. the following lower and upper bounds 
\[
0<R_t = \iinf_{\gamma\in\Gamma}r_t(\gamma)\leq \ssup_{\gamma\in\Gamma}r_t(\gamma)
\leq
2\bar\gamma\sum_{i=0}^{\infty}
(1-\ubar\gamma)^i
(|\pi_{t-i}|^2
+
\ssup_{\gamma\in\Gamma}
|\alpha_{t-i-1}(\gamma)|^2). 
\]
The uniform $k$-moment bound for $r_t(\gamma)$ follows from part ($i$)
and $\Ex[|\pi_t|^{2k}]<\infty$. Differentiating the recursion gives
\begin{align*}
\dot r_t(\gamma)
={}&
q_\gamma\dot r_{t-1}(\gamma)
+x_t(\gamma)(x_t(\gamma)-2\gamma\dot\alpha_{t-1}(\gamma))
-r_{t-1}(\gamma),
\\
\ddot r_t(\gamma)
={}&
q_\gamma\ddot r_{t-1}(\gamma)
+2[
\gamma(\dot\alpha_{t-1}(\gamma)^2
-x_t(\gamma)\ddot\alpha_{t-1}(\gamma))
-2x_t(\gamma)\dot\alpha_{t-1}(\gamma)
-\dot r_{t-1}(\gamma)
],
\end{align*}
and
\begin{align*}
\dddot r_t(\gamma)
= 
q_\gamma\dddot r_{t-1}(\gamma)
+ {}& 6\dot\alpha_{t-1}(\gamma)
(\dot\alpha_{t-1}(\gamma)
+\gamma\ddot\alpha_{t-1}(\gamma))
\\
-{}& 2x_t(\gamma)
(3\ddot\alpha_{t-1}(\gamma)
+\gamma\dddot\alpha_{t-1}(\gamma))
-3\ddot r_{t-1}(\gamma).
\end{align*}
Because $q_\gamma\leq1-\ubar\gamma<1$, each derivative admits a
stationary infinite sum representation; e.g. $\dot r_t(\gamma) = \sum_{j \geq 0} q_\gamma^j \xi_{t-j}^{(1)}(\gamma)$, with $\xi_{t}^{(1)})(\gamma) \coloneqq x_t(\gamma)(x_t(\gamma)-2\gamma\dot\alpha_{t-1}(\gamma))
-r_{t-1}(\gamma)$, and similarly for $\ddot r_t(\gamma)$ and $\dddot r_t(\gamma)$. The forcing terms contain
only lower-order derivatives of $r_t(\gamma)$ and $\alpha_t(\gamma)$. Hence, using part ($i$), the weighted-sum inequality from part ($i$), and Cauchy-Schwarz, yields
\(
\Ex[
\ssup_{\gamma\in\Gamma}
|r_t^{(m)}(\gamma)|^k
]
<\infty,\) \(m\in\{0,1,2,3\},\) provided $\Ex[|\pi_t|^{2k}]<\infty$.

($iii$) Let \(\omega_t(\gamma)\coloneqq r_t(\gamma)\beta_t(\gamma).\) Since \(\pi_{t-1}-\alpha_{t-1}(\gamma)
=q_\gamma x_{t-1}(\gamma),\) the covariance recursion can be written as
$\omega_t(\gamma) = q_\gamma \omega_{t-1}(\gamma)
+\gamma q_\gamma x_t(\gamma)x_{t-1}(\gamma),$
and therefore
\[
\omega_t(\gamma)
=
\gamma q_\gamma
\sum_{j=0}^{\infty}
q_\gamma^j
x_{t-j}(\gamma)x_{t-j-1}(\gamma).
\]
The same argument as in part ($ii$) shows that, for
$j\in\{0,1,2,3\}$ and $k\geq1$,
$\Ex[\ssup_{\gamma\in\Gamma}|\omega_t^{(j)}(\gamma)|^k]<\infty$, whenever $\Ex[|\pi_t|^{2k}]<\infty.$

By Lemma~\ref{lem:filter}, $\Ex[R_t^{-\tau}]<\infty$ for every $\tau>0.$ Moreover, the bound for the stationary candidate filter gives $\ssup_{\gamma\in\Gamma}|\beta_t(\gamma)| \leq 1$. Differentiating the identity
$\omega_t(\gamma)=r_t(\gamma)\beta_t(\gamma)$ gives
\begin{align*}
\dot\beta_t(\gamma)
={}&
\frac{\dot \omega_t(\gamma)-\dot r_t(\gamma)\beta_t(\gamma)}{r_t(\gamma)},
\\
\ddot\beta_t(\gamma)
={}&
\frac{\ddot \omega_t(\gamma)-\ddot r_t(\gamma)\beta_t(\gamma)-2\dot r_tv\dot\beta_t(\gamma)}{r_t(\gamma)},
\\
\dddot\beta_t(\gamma)
={}&
\frac{\dddot \omega_t(\gamma)-\dddot r_t(\gamma)\beta_t(\gamma)
-3\ddot r_t(\gamma)\dot\beta_t(\gamma)
-3\dot r_t(\gamma)\ddot\beta_t(\gamma)}{r_t(\gamma)}.
\end{align*}
For $j\in\{1,2,3\}$, put $Q_{j,t}
\coloneqq
\ssup_{\gamma\in\Gamma}
(|\omega_t^{(j)}(\gamma)|+|r_t^{(j)}(\gamma)|)$
The preceding identities imply, for a finite constant
$c\in(0,\infty)$,
\begin{align*}
\ssup_{\gamma\in\Gamma}|\dot\beta_t(\gamma)|
&\leq
R_t^{-1}Q_{1,t},
\\
\ssup_{\gamma\in\Gamma}|\ddot\beta_t(\gamma)|
&\leq
c\left\{
R_t^{-1}Q_{2,t}
+R_t^{-2}Q_{1,t}^2
\right\},
\\
\ssup_{\gamma\in\Gamma}|\dddot\beta_t(\gamma)|
&\leq
c\left\{
R_t^{-1}Q_{3,t}
+R_t^{-2}Q_{1,t}Q_{2,t}
+R_t^{-3}Q_{1,t}^3
\right\}.
\end{align*}

Suppose first that $\Ex[\|v_t\|^{4k+\zeta}]<\infty$ for some
$\zeta>0$. Choose $p>1$ sufficiently close to one that
$4kp\leq4k+\zeta$, and let $p'>1$ be such that $1/p+1/p' = 1$. By
H\"older's inequality, $\Ex[R_1^{-\tau}]<\infty$, part ($ii$),
and the moment bound of the auto-covariances,
\begin{align*}
\Ex[R_t^{-2k}Q_{1,t}^{2k}]
&\leq
\Ex[R_t^{-2kp'}]^{1/p'}
\Ex[Q_{1,t}^{2kp}]^{1/p}
<\infty,
\\
\Ex[R_t^{-k}Q_{2,t}^{k}]
&\leq
\Ex[R_t^{-kp'}]^{1/p'}
\Ex[Q_{2,t}^{kp}]^{1/p}
<\infty.
\end{align*}
The corresponding, weaker bounds for
$R_t^{-k}Q_{1,t}^k$ are immediate. Consequently,
\[
\Ex\left[
\ssup_{\gamma\in\Gamma}
|\beta_t^{(m)}(\gamma)|^k
\right]
<\infty,
\qquad
m\in\{1,2\}.
\]

Finally, suppose that
$\Ex[\|v_t\|^{6k+\zeta}]<\infty$ for some $\zeta>0$. Choose
$p>1$ sufficiently close to one that $6kp\leq6k+\zeta$. Applying
H\"older's inequality once more, together with Cauchy-Schwarz for the
mixed term $Q_{1,t}Q_{2,t}$, gives
$\Ex[R_t^{-k}Q_{3,t}^k]<\infty,$ $\Ex[R_t^{-2k}(Q_{1,t}Q_{2,t})^k]<\infty,$ and $\Ex[R_t^{-3k}Q_{1,t}^{3k}]<\infty.$ It follows that \(\Ex[
\ssup_{\gamma\in\Gamma}
|\dddot\beta_t(\gamma)|^k]
<\infty.
\)
This completes the proof. \hfill $\square$

{\bf Proof of Lemma \ref{lem:pd}.} \underline{Lemma \ref{lem:pd} ($a$).} Finiteness of $A$ follows from Lemma~\ref{lem:mom} under the stated
moment condition. It remains to prove that $A$ is non-singular. By
stationarity, write
$A=\Ex[g_{t+1}g_{t+1}^\Tr]$, where
$g_{t+1}\coloneqq
(\delta_0\dot h_t(\gamma_0),h_t(\gamma_0),y_{t+1})^\Tr$.
Since $A$ is positive semi-definite, it is singular if and only if there
exists $\iota\coloneqq(\iota_1,\iota_2,\iota_3)^\Tr\neq0$ such that
$\Ex[(\iota^\Tr g_{t+1})^2]=0$, or equivalently,
\begin{equation}\label{eq:A-degenerate}
\iota_1\delta_0\dot h_t(\gamma_0)
+
\iota_2h_t(\gamma_0)
+
\iota_3y_{t+1}
=
0
\qquad a.s.
\end{equation}
We show that the existence of such a vector $\iota$ yields a contradiction.

Conditional on $\mathcal F_t$, the first two terms in
\eqref{eq:A-degenerate}, as well as $a+\rho y_t$, are fixed, whereas
$y_{t+1}=a+\rho y_t+\eps_{y,t+1}$ and $\eps_{y,t+1}$ is independent of
$\mathcal F_t$. Since the left-hand side of \eqref{eq:A-degenerate}
vanishes a.s., its conditional variance is zero. Hence
$\iota_3^2\sigma_\eps^2=0$. Since $\sigma_\eps^2>0$, it follows that
$\iota_3=0$.

Condition on $\mathcal F_{t-1}$ and let $z\in\mathbb R$ denote a possible
realization of $\pi_t$. As in Remark~\ref{rem:tail-sep}, write
\[
h_t(\gamma;z)
=
\alpha_t(\gamma;z)
+
\frac{N(\gamma;z)}{r_t(\gamma;z)^2},
\]
where
$N(\gamma;z)\coloneqq
q_\gamma\omega_t(\gamma;z)^2\nu_t(\gamma;z)$,
$q_\gamma\coloneqq1-\gamma$, and
$\nu_t(\gamma;z)\coloneqq z-\alpha_{t-1}(\gamma)$.

Differentiating with respect to $\gamma$, holding $z$ fixed, gives
\begin{equation}\label{eq:hdot-rational}
\dot h_t(\gamma;z)
=
\dot\alpha_t(\gamma;z)
+
\frac{
\dot N(\gamma;z)r_t(\gamma;z)
-
2N(\gamma;z)\dot r_t(\gamma;z)
}{
r_t(\gamma;z)^3
},
\end{equation}
where
\[
\begin{aligned}
\dot N(\gamma;z)
={}&
-\omega_t(\gamma;z)^2\nu_t(\gamma;z)
+
2q_\gamma\omega_t(\gamma;z)
\dot\omega_t(\gamma;z)\nu_t(\gamma;z)-
q_\gamma\omega_t(\gamma;z)^2\dot\alpha_{t-1}(\gamma).
\end{aligned}
\]
The derivative recursions are given in the proof of
Lemma~\ref{lem:mom}. As functions of $z$,
$\dot\alpha_t(\gamma;\cdot)$ and $\dot\omega_t(\gamma;\cdot)$ are affine,
$\dot r_t(\gamma;\cdot)$ is quadratic, and
$\dot N(\gamma;\cdot)$ has degree at most three. Thus
$\dot h_t(\gamma;\cdot)$ admits a rational representation with
denominator $r_t(\gamma;\cdot)^3$.

Since $r_t(\gamma_0;z)>0$ for every real $z$, both
$h_t(\gamma_0;\cdot)$ and $\dot h_t(\gamma_0;\cdot)$ are continuous on
$\mathbb R$. By the full-support assumption on the conditional
distribution of $\pi_t$, equation~\eqref{eq:A-degenerate}, with
$\iota_3=0$, implies, for almost every realization of
$\mathcal F_{t-1}$,
\[
\iota_1\delta_0\dot h_t(\gamma_0;z)
+
\iota_2h_t(\gamma_0;z)
=
0
\qquad
\text{for every }z\in\mathbb R.
\]
Fix such a realization. Writing both terms over the common denominator
$r_t(\gamma_0;z)^3$ yields a polynomial of degree at most seven that
vanishes for every real $z$. It must therefore be the zero polynomial,
so the equality holds as an identity of rational functions over
$\mathbb C$. The two roots of $r_t(\gamma;\cdot)$ are $z_\pm(\gamma)
=
\alpha_{t-1}(\gamma)
\pm
i\tau_{t-1}(\gamma),$ $
\tau_{t-1}(\gamma)
\coloneqq
\sqrt{q_\gamma r_{t-1}(\gamma)/\gamma}.$
Write $z_\pm\coloneqq z_\pm(\gamma_0)$. Since
$\tau_{t-1}(\gamma_0)>0$, the roots are distinct and
$r_t(\gamma_0;z)=\gamma_0(z-z_+)(z-z_-)$. By the argument in
Remark~\ref{rem:tail-sep}, $N(\gamma_0;z_\pm)\neq0$ a.s., so
$h_t(\gamma_0;\cdot)$ has genuine poles of order two at $z_+$ and
$z_-$. Equation~\eqref{eq:hdot-rational} shows that
$\dot h_t(\gamma_0;\cdot)$ has poles of order at most three at the same
two points.

We show that the pole at $z_+$ is genuinely of order three. Since
$r_t(\gamma_0;z_+)=0$, evaluating the numerator of the rational term in
\eqref{eq:hdot-rational} at $z_+$ gives $-2N(\gamma_0;z_+)\dot r_t(\gamma_0;z_+).$ From the derivative recursion in Lemma~\ref{lem:mom} and $\nu_t(\gamma_0;z_+)=i\tau_{t-1}(\gamma_0)$, it follows that
$\operatorname{\sf Im}\dot r_t(\gamma_0;z_+)
=
-2\gamma_0\tau_{t-1}(\gamma_0)
\dot\alpha_{t-1}(\gamma_0)$. Moreover, the derivative recursion for $\alpha_{t-1}$ gives
$\dot\alpha_{t-1}(\gamma_0)
=
\pi_{t-1}
-
\alpha_{t-2}(\gamma_0)
+
q_{\gamma_0}\dot\alpha_{t-2}(\gamma_0)$.
Conditional on $\mathcal F_{t-2}$, the last two terms are fixed, whereas
$\pi_{t-1}$ has a nonatomic distribution. Hence
$\dot\alpha_{t-1}(\gamma_0)\neq0$ a.s., and therefore
$\dot r_t(\gamma_0;z_+)\neq0$ a.s. Together with
$N(\gamma_0;z_+)\neq0$ a.s., this shows that
$\dot h_t(\gamma_0;\cdot)$ has a genuine pole of order three at $z_+$.

Multiplying the rational identity by $(z-z_+)^3$ and letting
$z\to z_+$, the term involving $h_t(\gamma_0;\cdot)$ converges to zero
because it has a pole of order two only, whereas the term involving
$\dot h_t(\gamma_0;\cdot)$ converges to $\iota_1\delta_0a_1$, where
\[
a_1
\coloneqq
\lim_{z\to z_+}
(z-z_+)^3\dot h_t(\gamma_0;z)
=
-\frac{
2N(\gamma_0;z_+)\dot r_t(\gamma_0;z_+)
}{
\gamma_0^3(z_+-z_-)^3
}
\neq0
\qquad a.s.
\]
Since $\delta_0\neq0$, it follows that $\iota_1=0$. The rational identity
then reduces to $\iota_2h_t(\gamma_0;z)=0$ for every real $z$. Since
$h_t(\gamma_0;\cdot)$ has a genuine double pole at $z_+$, it is not
identically zero, and hence $\iota_2=0$.

Thus $\iota_1=\iota_2=\iota_3=0$, contradicting $\iota\neq0$. Therefore
$A$ is non-singular and, being positive semi-definite, $A\succ0$. \hfill $\square$

\underline{Lemma \ref{lem:pd} ($b$).} Note that \[
\frac{1}{n}\nabla_\theta^2 Q_n(\theta)
=
\frac{2}{n}\sum_{t=1}^n g_t(\theta), \quad H_t(\theta)
\coloneqq
\nabla_\theta^2 f_t(\theta)
=
\begin{bmatrix}
\delta\ddot h_{t-1}(\gamma) & \dot h_{t-1}(\gamma) & 0 \\
\dot h_{t-1}(\gamma) & 0 & 0 \\
0 & 0 & 0
\end{bmatrix},
\], where $g_t(\theta) \coloneqq -\dot f_t(\theta)\dot f_t(\theta)^\Tr + (\pi_t-f_t(\theta))H_t(\theta)$. Before proceeding, note that $\dot f_t(\theta)$ and $H_t(\theta)$ are affine in $(\delta,\psi)^\Tr$; i.e. the complications arise only due to the nonlinear $\gamma$-functionals. Then, for $\theta_1$, $\theta_2 \in \Theta$,
\begin{align*}
g_t(\theta_1) - g_t(\theta_2) = \,& \dot f_t(\theta_1)\dot f_t(\theta_1)^\Tr-\dot f_t(\theta_2)\dot f_t(\theta_2)^\Tr \\
\,& + [(\pi_t-f_t(\theta_1))H_t(\theta_1)-(\pi_t-f_t(\theta_2))H_t(\theta_2)]  \\
\,& \eqqcolon A_t(\theta_1,\theta_2)+B_t(\theta_1,\theta_2),
\end{align*}
say. By $|A_t(\theta_1,\theta_2)|_2\leq (|\dot f_t(\theta_1)|_2+|\dot f_t(\theta_2)|_2)|\dot f_t(\theta_1)-\dot f_t(\theta_2)|_2$, the mean-value theorem yields
\[
|A_t(\theta_1,\theta_2)|_2 \leq 2\ssup\limits_{\theta \in \Theta} |\dot f_t(\theta)|_2\ssup\limits_{\theta \in \Theta} |H_t(\theta)|_2 |\theta_1-\theta_2|_2.
\]
Next, $B_t(\theta_1,\theta_2) = (f_t(\theta_2)-f_t(\theta_1))H_t(\theta_2)+(\pi_t-f_t(\theta_1))(H_t(\theta_1)-H_t(\theta_2))$
\begin{align*}
|B_t(\theta_1,&\theta_2)|_2\\
\leq & \, [\ssup\limits_{\theta \in \Theta} |\dot f_t(\theta)|_2\ssup\limits_{\theta \in \Theta} |H_t(\theta)|_2 + (|\pi_t|+\ssup\limits_{\theta \in \Theta} | f_t(\theta)|_2)\ssup\limits_{\theta \in \Theta} |\nabla_\theta H_t(\theta)|_2] \\ & \times |\theta_1-\theta_2|_2.
\end{align*}
Hence, $|g_t(\theta_1)-g_t(\theta_2)|_2 \leq C_t |\theta_1-\theta_2|_2$, where
\[
C_t \coloneqq 3\ssup\limits_{\theta \in \Theta} |\dot f_t(\theta)|_2\ssup\limits_{\theta \in \Theta} |H_t(\theta)|_2 + (|\pi_t|+\ssup\limits_{\theta \in \Theta} | f_t(\theta)|_2)\ssup\limits_{\theta \in \Theta} |\nabla_\theta H_t(\theta)|_2 = 3C_{1,t}+C_{2,t},
\]
say, where $C_{1,t}$, $C_{2,t}$ are implicitly defined. It thus remains to be shown that $B_n \coloneqq \frac1{n}\sum_{t=1}^n C_t = O_p(1).$ By Markov's inequality, it suffices to show that $\Ex[C_t] < \infty$. By Cauchy-Schwarz,
$$\Ex[C_{1,t}]  \leq \Ex^{1/2}[\ssup\limits_{\theta \in \Theta} |\dot f_t(\theta)|_2^2]\Ex^{1/2}[\ssup\limits_{\theta \in \Theta} |H_t(\theta)|_2^2] < \infty,$$ where the strict equality follows, by Lemma~\ref{lem:mom}, from the assumed finite $8$ moments. Next, by Hölder's inequality with $p = 4$, $q = 4/3$ ($1/p+1/q = 1$), we get
\begin{align*}
\Ex[C_{2,t}] \leq   \Ex^{1/4}[(|\pi_t|+\ssup\limits_{\theta \in \Theta} | f_t(\theta)|_2)^4]\Ex^{3/4}[\ssup\limits_{\theta \in \Theta} |\nabla_\theta H_t(\theta)|_2^{4/3}].
\end{align*}
Lemma~\ref{lem:mom} directly shows that the first factor is finite. Note that $\nabla_\theta H_t(\theta)$ involves up to the third derivatives of the recursions; in particular, we need a uniform $q = 4/3$-moment bound of $\dddot\beta_t(\gamma)$. By Lemma~\ref{lem:mom}, this requires $6\times 4/3 = 8$ moments. Thus, $\Ex[C_t]<\infty$, completing the proof.  \hfill $\square$

{\bf Proof of Lemma~\ref{lem:Flambda}.} \underline{Part ($i$).} Under Assumption~\ref{ass:para}, $1\pm\delta\beta^2\rho>0$, $1-\rho^2>0$, and $\sigma_u^2/\sigma_\eps^2\in(0,\infty)$; hence $D(\beta;\lambda)>0$ uniformly in $(\beta,\lambda)$, and $F(\beta;\lambda)$, being a ratio of polynomials
with nondegenerate denominator, is continuously differentiable on $[0,1]\times B$. This proves
$(i)$.

\underline{Parts ($ii$) and ($iii$).}  For $(ii)$, it suffices to show that some partial derivative of $F$ is
uniformly bounded away from zero. If $\rho_0=0$,
\[
\Bigl|\frac{\partial}{\partial\rho}F(\beta;\lambda_0)\Bigr|_{\rho_0=0}
=
\frac{\psi_0^2(1-\delta_0^2\beta^4)}{\psi_0^2+(\sigma_u/\sigma_\eps)^2}.
\]
If $\rho_0\neq0$, a direct computation (quotient rule applied to
$\psi^2\rho(1-\delta^2\beta^4)/D$, noting the numerator does not depend on
$\sigma_u^2$) gives
\[
\frac{\partial}{\partial\sigma_u^2}F(\beta;\lambda_0)
=
-\,\psi_0^2\,\rho_0\,
\frac{(1-\delta_0^2\beta^4)(1-\rho_0^2)(1-\delta_0\beta^2\rho_0)}
     {D(\beta;\lambda_0)^2\,\sigma_\eps^2}.
\]
In either case, since $|\delta_0|<1$ implies
$1-\delta_0^2\beta^4\ge1-\delta_0^2>0$ uniformly on $\beta\in[0,1]$, and all
remaining factors are bounded and of fixed sign on $[0,1]\times B$ by
Assumption~\ref{ass:para}, the displayed derivative is bounded away from zero
uniformly in $\beta\in[0,1]$ whenever $\psi_0\neq0$. This proves $(ii)$. For $(iii)$, evaluating the two displays above at $\beta=0$ and $\beta=1$ and
simplifying gives
\[
F(0;\lambda_0)=\frac{\psi_0^2\rho_0}{\psi_0^2+(1-\rho_0^2)(\sigma_u/\sigma_\eps)^2},
\]
and
\[
1-F(1;\lambda_0)
=
\frac{(1-\delta_0)(1-\rho_0)[\psi_0^2+(\sigma_u/\sigma_\eps)^2(1-\delta_0\rho_0)(1+\rho_0)]}
     {D(1;\lambda_0)}.
\]
Because $|\delta_0|<1$ together with $\rho_0>0$ and
$\psi_0\neq0$, every factor on the right-hand side of each display is
strictly positive. Hence $F(0;\lambda_0)>0$ and
$F(1;\lambda_0)<1$, i.e.\ $G(0)=-F(0;\lambda_0)<0$ and
$G(1)=1-F(1;\lambda_0)>0$.

\underline{Part $(iv)$.} Set $P(\beta)\coloneqq G(\beta;\lambda_0)D(\beta;\lambda_0)$. By
$(i)$, $G(\cdot;\lambda_0)$ and $P$ have the same roots in $[0,1]$; they also
have the same multiplicities, since Leibniz' rule gives, for a root $\beta^\st$
of multiplicity $m$, $\partial^j_\beta P(\beta^\st)=0$ for $j<m$ and
\begin{equation}\label{eq:leibniz}
\partial^m_\beta P(\beta^\st)=\partial^m_\beta G(\beta^\st;\lambda_0)\,
D(\beta^\st;\lambda_0)\neq0 .
\end{equation}
Writing $D(\beta;\lambda_0)=A_0+A_2\beta^2$ with
$A_0\coloneqq\psi_0^2+(1-\rho_0^2)(\sigma_u/\sigma_\eps)^2$ and
$A_2\coloneqq\delta_0\rho_0\{\psi_0^2-(1-\rho_0^2)(\sigma_u/\sigma_\eps)^2\}$,
a direct computation gives
\[
P(\beta)=-\psi_0^2\rho_0+A_0\beta-A_0\delta_0\beta^2+A_2\beta^3+\kappa\beta^4,
\qquad
\kappa\coloneqq\psi_0^2\rho_0\delta_0^2-A_2\delta_0
=\delta_0^2\rho_0(1-\rho_0^2)(\sigma_u/\sigma_\eps)^2 .
\]
Since $\delta_0\neq0$ and $\rho_0\in(0,1)$ by Assumption~\ref{ass:para},
$\kappa>0$, so $P$ has degree exactly four.

\underline{Part $(v)$.} Let $\beta_1<\cdots<\beta_{r_0}$ denote the distinct roots in
$[0,1]$, with multiplicities $m_1,\ldots,m_{r_0}$; by $(iii)$ neither endpoint
is a root, and by $(iv)$ their number is finite with $\sum_i m_i\le4$. Writing
$G(\beta;\lambda_0)=(\beta-\beta_i)^{m_i}g_i(\beta)$ with $g_i$ continuous and
$g_i(\beta_i)\neq0$ locally, $G$ changes sign at $\beta_i$ if and only if $m_i$
is odd; as $G(0)<0<G(1)$ by $(iii)$, the number of odd $m_i$ is odd. It remains
to enumerate the partitions with $\sum_i m_i\le4$ and an odd number of odd
parts. For $r_0\ge4$, necessarily $r_0=4$ and $m_i\equiv1$, giving four odd
parts; this is excluded. For $r_0=3$ the candidates $(1,1,1)$ and $(1,1,2)$ have
three and two odd parts, leaving $(1,1,1)$. For $r_0=2$ the candidates
$(1,1),(1,2),(1,3),(2,2)$ have two, one, two and zero odd parts, leaving
$(1,2)$. For $r_0=1$ the candidates $(1),(2),(3),(4)$ have one, zero, one and
zero odd parts, leaving $(1)$ and $(3)$. This yields the four structures, which
are mutually exclusive because they differ in $r_0$ or in the multiplicities.
Finally, let $\beta_0$ have multiplicity three. As $P$ has real coefficients and
degree four, its remaining root $\beta^\dagger$ is real and
$P(\beta)=\kappa(\beta-\beta_0)^3(\beta-\beta^\dagger)$. Evaluating at zero,
$\kappa\beta_0^3\beta^\dagger=P(0)=G(0;\lambda_0)D(0;\lambda_0)<0$ by $(i)$ and
$(iii)$; since $\kappa>0$ and $\beta_0>0$, this gives $\beta^\dagger<0$. Hence
$\partial^3_\beta P(\beta_0)=6\kappa(\beta_0-\beta^\dagger)>0$, and
\eqref{eq:leibniz} with $m=3$ yields
$6c=\partial^3_\beta G(\beta_0;\lambda_0)=6\kappa(\beta_0-\beta^\dagger)/
D(\beta_0;\lambda_0)>0$. \hfill$\square$

\section{Additional Monte Carlo simulations}\label{sec:addMC}

We simulate the model of Eqs.~\eqref{eq:pc}-\eqref{eq:var-shock}, where
$$\vartheta=(\mu,\varphi,\delta,\psi,\phi_\pi,\phi_y,\rho_i)^\Tr
=(0.85,1,0.95,0.15,1.5,0.5,0.6)^\Tr, \qquad \gamma_0=0.05.$$ The shock process
is a VAR(1), with
\[
R=\begin{bmatrix}0.30&0.05&0.08\\0.10&0.40&0.00\\0.05&0.00&0.20\end{bmatrix},
\]
and $\eps_t \sim {\mathcal N}(0,\Sigma)$, $\Sigma=0.01\ I_3$.
Beliefs are
initialised at $\fra=(0.15,0.15)^\Tr$ with $r_{j,0}=\beta_{j,0}=0$, as in
Remark~\ref{rem:initial}, and we set $u_0=i_0=0$.

The QMLE minimises the profiled criterion~\eqref{eq:fiml-objective} over
$\theta=(\gamma,\vartheta^\Tr)^\Tr$, with $R$ concentrated out by 
least squares and the innovation variances by their empirical counterparts. To obtain starting values, we first evaluate it
on the grid $\gamma\in\{0.01,0.02,\ldots,0.12\}$, optimising over the remaining
parameters at each grid point, and then use the minimiser as the starting value
for a quasi-Newton search over the full parameter vector. Results are based on $M=1{,}000$ replications for each of $n\in\{250,500,1{,}000,2{,}000\}$.

\begin{table}[ht]
\centering\small
\begin{tabular}{l c cc cc cc cc}
\toprule
& & \multicolumn{2}{c}{$n=250$} & \multicolumn{2}{c}{$n=500$}
  & \multicolumn{2}{c}{$n=1{,}000$} & \multicolumn{2}{c}{$n=2{,}000$}\\
\cmidrule(lr){3-4}\cmidrule(lr){5-6}\cmidrule(lr){7-8}\cmidrule(lr){9-10}
& true & mean & sd & mean & sd & mean & sd & mean & sd\\
\midrule
$\gamma$    & 0.05 & 0.0496 & 0.0188 & 0.0506 & 0.0117 & 0.0509 & 0.0079 & 0.0506 & 0.0055\\
$\mu$       & 0.85 & 0.8400 & 0.1812 & 0.8579 & 0.1219 & 0.8575 & 0.0817 & 0.8563 & 0.0591\\
$\varphi$   & 1.00 & 1.0701 & 0.1542 & 1.0511 & 0.1206 & 1.0302 & 0.0907 & 1.0171 & 0.0621\\
$\delta$    & 0.95 & 0.8897 & 0.1347 & 0.9110 & 0.0951 & 0.9263 & 0.0732 & 0.9383 & 0.0524\\
$\psi$      & 0.15 & 0.1942 & 0.1327 & 0.1819 & 0.0929 & 0.1706 & 0.0665 & 0.1608 & 0.0442\\
$\phi_\pi$  & 1.50 & 1.7153 & 0.5723 & 1.6283 & 0.3723 & 1.5767 & 0.2620 & 1.5414 & 0.1695\\
$\phi_y$    & 0.50 & 0.6861 & 0.4668 & 0.6137 & 0.3031 & 0.5649 & 0.2079 & 0.5368 & 0.1339\\
$\rho_i$    & 0.60 & 0.5845 & 0.1271 & 0.5881 & 0.0925 & 0.5942 & 0.0678 & 0.5970 & 0.0469\\
\midrule
$R_{yy}$      & 0.40 & 0.4000 & 0.0774 & 0.4018 & 0.0560 & 0.4012 & 0.0377 & 0.4007 & 0.0262\\
$R_{\pi\pi}$  & 0.30 & 0.3080 & 0.0774 & 0.3091 & 0.0529 & 0.3080 & 0.0372 & 0.3041 & 0.0256\\
$R_{ii}$      & 0.20 & 0.2223 & 0.1313 & 0.2223 & 0.1016 & 0.2139 & 0.0791 & 0.2086 & 0.0553\\
\midrule
$\sigma_{y}$   & 0.10 & 0.1010 & 0.0067 & 0.1010 & 0.0048 & 0.1007 & 0.0032 & 0.1003 & 0.0022\\
$\sigma_{\pi}$ & 0.10 & 0.1014 & 0.0098 & 0.1012 & 0.0060 & 0.1009 & 0.0042 & 0.1005 & 0.0027\\
$\sigma_{i}$   & 0.10 & 0.1050 & 0.0145 & 0.1039 & 0.0110 & 0.1023 & 0.0079 & 0.1013 & 0.0052\\
\bottomrule
\end{tabular}
\caption{Monte Carlo mean and standard deviation of the QMLE of
$\theta=(\gamma,\vartheta^\Tr)^\Tr$ (upper panel), of the diagonal of the shock
coefficient matrix $R$ (middle panel), and of the innovation standard deviations
$\sigma_{\varepsilon,j}$ (lower panel), based on $M=1{,}000$ replications. The
off-diagonal elements of $R$ are estimated jointly but not reported; the
innovation covariance is restricted to be diagonal.}
\label{tab:mcmv}
\end{table}

Table~\ref{tab:mcmv} reports the results. The estimator is well behaved in all
three blocks. The reported means approach their true values monotonically in
$n$, for instance $\delta$ moves through $0.890$, $0.911$, $0.926$ and $0.938$
towards $0.95$, and $\phi_y$ through $0.686$, $0.614$, $0.565$ and $0.537$
towards $0.50$, and the standard deviations halve as $n$ quadruples throughout,
consistent with the $\sqrt n$ rate.

\section{Further empirical results}\label{app:empirical}

Figure~\ref{fig:empraw} reports the data used in the analysis, namely the CPI inflation (upper panel), together with the CBO estimate for the output gap (lower panel, blue line) and the Unit Labour Cost (lower panel, pink line).
\begin{figure}[!h]
\centering
\begin{tikzpicture}

  \begin{axis}[
    name=bottom,
    width=13cm,
    height=5cm,
    at={(0,0)},
    anchor=south west,
    xmin=1960,
    xmax=2020,
    axis lines=box,
    xtick={1960,1970,1980,1990,2000,2010,2020},
    xticklabel style={
      /pgf/number format/fixed,
      /pgf/number format/precision=0,
      /pgf/number format/set thousands separator={}
    },
    xlabel={},
    ylabel={{\it output gap}},
    y label style={color=blue},
    yticklabel style={color=blue},
    grid=none,
    table/col sep=space
  ]
    \addplot+[no marks, thick, blue] table[
      x expr=1960 + \coordindex/4,
      y=y
    ] {Figure5_dat.txt};
  \end{axis}

  \begin{axis}[
    name=top,
    width=13cm,
    height=4cm,
    at={(bottom.north west)},   
    anchor=south west,
    xmin=1960,
    xmax=2020,
    axis lines=box,
    xtick={1960,1970,1980,1990,2000,2010,2020},
    xticklabels={},             
    xlabel={},
    ylabel={$\pi$},
    grid=none,
    table/col sep=space
  ]
    \addplot+[no marks, black, thick] table[
      x expr=1960 + \coordindex/4,
      y=pi
    ] {Figure5_dat.txt};
  \end{axis}

  \begin{axis}[
    width=13cm,
    height=4cm,
    at={(bottom.south west)},   
    anchor=south west,
    xmin=1960,
    xmax=2020,
    axis y line*=right,         
    axis x line=none,           
    ylabel={{\it unit labour costs}},
    y label style={color=pink},
    yticklabel style={color=pink},
    ytick pos=right,
    grid=none,
    table/col sep=space
  ]
    \addplot+[no marks, thick, pink] table[
      x expr=1960 + \coordindex/4,
      y=ya
    ] {Figure5_dat.txt};
  \end{axis}

\end{tikzpicture}
\caption{Percentage change of the CPI (upper panel) and, in the lower panel, the output gap and the percentage change of unit labour costs .}\label{fig:empraw}
\end{figure}

Figure \ref{fig:BetaAnchoring} displays our empirical estimates of $\beta_t$ computed both with the output gap and the unit labor cost proxy. Our results are similar to those reported by \citet{Lansing2009} and \citet{JorgensenLansing2025b}, essentially by construction, and are also broadly consistent with the anchoring index proposed by \citet{naggert2023anchoring}, which we computed from the Michigan Survey of Consumers, and is reported as a benchmark in the same plot. In both cases, the evidence suggests that expectations were relatively de-anchored up to around 2000, whereas all indicators decline thereafter, pointing to a stronger degree of anchoring.

The main difference concerns the timing of the regime shift, which appears to occur somewhat later when measured by $\beta_t$. This delay is likely related to the way $\beta_t$ is estimated, since it depends on past information and therefore inherits a certain degree of persistence or inertia. In terms of the behavioural equilibria, this pattern suggests that a high-$\beta$ equilibrium is more plausible prior to the early 2000s, followed by a transition toward the low-$\beta$ equilibrium thereafter.

\begin{figure}[h!]
\centering
\begin{tikzpicture}

  \begin{axis}[
    name=bottom,
    width=13cm,
    height=6cm,
    at={(0,0)},
    anchor=south west,
    xmin=1985,
    xmax=2020,
    axis x line*=box,
    axis y line*=left,
    xtick={1985,1990,1995,2000,2005,2010,2015,2020},
    xticklabel style={
      /pgf/number format/fixed,
      /pgf/number format/precision=0,
      /pgf/number format/set thousands separator={}
    },
    xlabel={},
    ylabel={{$\beta_t$}},
    y label style={color=black},
    yticklabel style={color=black},
    grid=none,
    table/col sep=space
  ]
    \addplot+[no marks, thick, blue] table[
      x expr=1985 + \coordindex/4,
      y=bet_t_OG
    ] {Figure6_dat.txt};

    \addplot+[no marks, thick, green] table[
      x expr=1985 + \coordindex/4,
      y=bet_t_ULC
    ] {Figure6_dat.txt};
 \end{axis}

  \begin{axis}[
    width=13cm,
    height=6cm,
    at={(bottom.south west)},   
    anchor=south west,
    xmin=1985,
    xmax=2020,
    axis y line*=right,         
    axis x line=none,           
    ylabel={{\it MSC anchor index}},
    y label style={color=red},
    yticklabel style={color=red},
    ytick pos=right,
    grid=none,
    table/col sep=space
  ]
    \addplot+[no marks, thick, red] table[
      x expr=1985 + \coordindex/4,
      y=anchor
    ] {Figure6_dat.txt};
  \end{axis}

\end{tikzpicture}
\caption{Comparison between estimated $\beta_t$ from output gap (blue line) and Unit Labor Cost (green line). The red line is the anchoring index computed from \citet{naggert2023anchoring}}
\label{fig:BetaAnchoring}	
\end{figure}

\end{appendices}

\end{document}